# The Mechanical Behavior of Macroscale Single-crystal Graphene

Anirban Kundu[1], Seyed Kamal Jalali[2], Minhyeok Kim[1,3], Meihui Wang[1], Da Luo[1], Sun Hwa Lee[1], Nicola M. Pugno[2,4], Won Kyung Seong[1*], and Rodney S. Ruoff[1,3,5,6*]

[1]*Center for Multidimensional Carbon Materials (CMCM), Institute for Basic Science (IBS), Ulsan 44919, Republic of Korea.*

[2]*Laboratory for Bioinspired, Bionic, Nano, Meta Materials & Mechanics, Department of Civil, Environmental and Mechanical Engineering, University of Trento, Via Mesiano, 77, 38123 Trento, Italy.*

[3]*Department of Chemistry, Ulsan National Institute of Science and Technology (UNIST), Ulsan 44919, Republic of Korea.*

[4]*School of Engineering and Material Science, Queen Mary University of London, Mile End Road, London E1 4NS, UK.*

[5]*School of Energy and Chemical Engineering, Ulsan National Institute of Science and Technology (UNIST), Ulsan 44919, Republic of Korea.*

[6]*Department of Materials Science and Engineering, Ulsan National Institute of Science and Technology (UNIST), Ulsan 44919, Republic of Korea.*

Corresponding author: rsruoff@ibs.re.kr(RSR), wks1130@ibs.re.kr(WKS)

## Abstract

Despite extensive microscale studies, the *macroscopic* mechanical properties of monolayer graphene remain underexplored. Here, we report the Young's modulus ($E = 1.11 \pm 0.04$ TPa), tensile strength ($\sigma = 27.40 \pm 4.36$ GPa), and failure strain ($\varepsilon_f = 6.01 \pm 0.92$ %) of centimeter-scale single-crystal monolayer graphene (SCG) 'dog bone' samples with edges aligned along the zigzag (zz) direction, supported by an ultra-thin polymer (polycarbonate) film. For samples with edges along the armchair (ac) direction, we obtain $E = 1.01 \pm 0.10$ TPa, $\sigma = 20.21 \pm 3.22$ GPa, $\varepsilon_f = 3.69 \pm 0.38$ %, and for chiral samples whose edges were between zz and ac, we obtain $E = 0.75 \pm 0.12$ TPa, $\sigma = 23.56 \pm 3.42$ GPa, and $\varepsilon_f = 4.53 \pm 0.40$ %. The SCG is grown on single crystal Cu(111) foils by chemical vapor deposition (CVD). We used a home-built 'float-on-water' (FOW) tensile testing system for tensile loading measurements that also enabled in situ crack observation. The quantized fracture mechanics (QFM) analysis predicts an edge defect size from several to tens of nanometers based on chirality and notch angle. Through Weibull analysis and given that the fatal defects are confined on the edges of macroscale samples, we projected strength ranging from 13.67 to 18.43 GPa for an A4-size SCG according to their chirality. The exceptional mechanical performance of macroscale single crystal graphene (SCG) paves the way for its widespread use in a very wide variety of applications.

# Introduction

Carbon fibers have been the most mechanically robust materials on a macroscopic scale (centimeters or larger), and are widely used in fields demanding high stiffness and strength(*1, 2*) (and high specific stiffness and strength—due to the low density of carbon materials). We present the first experimental investigation of the chirality-dependent mechanical properties of centimeter-scale single-crystal graphene (SCG) and report a unique combination of mechanical properties that surpass both carbon fibers and carbon nanotubes (CNTs). High performance carbon fibers are reported to have strengths ($\sigma$) of 3.6-8 GPa, Young's moduli ($E$) of 230-588 GPa, and failure strains ($\varepsilon_f$) of 0.7-2.2% in their composite form(*3, 4*). Whereas the strength of individual fibers are reported to be as high as 18 GPa at smaller length scales (order of tens of μm), aligning with theoretical predictions(*2, 4*). Single-walled CNTs (SWCNTs) have a reported maximum engineering strength of 43 GPa at a millimeter scale (gauge length of 1.5 mm and diameter of 2.1 nm)(*5*). Our research shows SCG has remarkable performance at the centimeter scale. SCG exhibits an average strength of 27.40 GPa along the zigzag (zz) direction, an average Young's modulus of 1.11 TPa, and a notable 6.01% failure strain. These properties, observed at a scale where existing high performance carbon-based materials show limitations, position SCG as a revolutionary material with transformative potential across various applications, including aerospace, automotive, flexible electronics, and civil engineering.

Strength is a kinetic, not thermodynamic, parameter, and it is important not to confuse strength with modulus, which measures the material's stiffness (or resistance) to deformation, rather than the maximum stress it can endure. The '*ideal tensile strength*' of graphene could be taken as a well calculated value of the stress-at failure (at 0K or around room temperature, no reactive species present in such a calculation). Exfoliated (or CVD grown) graphene samples can exhibit 'ideal' strength values (130 GPa) when tested at the micron length scale(*6, 7*): *at this length scale a reasonably high fraction of samples will have no defects.* At larger length scale (millimeters, centimeters, and up) it is much more likely that graphene (*or any other material*) will have defects, referred to as "flaws" by the solid mechanics community(*8*). Such defects are 'stress concentrators', and the tensile stress on a sample can be much larger very close to flaws(*9*), thus the failure of macroscale samples of almost all materials occurs at a much lower stress than the ideal strength(*10*). This is the influence of *length scale* on the fracture strength of real materials(*11, 12*). To the best of our knowledge, high strength and failure strain of graphene have been explored only in microscale(*6, 13*).

Advances in chemical vapor deposition (CVD) of graphene(*14-16*) have enabled the production of large-scale, SCG(*17, 18*) on single crystal Cu(111) foils. The armchair (ac) and zigzag (zz) edges of SCG can be identified from the rolling marks in Cu(111) foils that are along the <112> direction(*19*). The perpendicular direction to the rolling marks is the Cu <110>, aligned with the SCG zz edge(*20*). This allowed us to identify the chirality dependent mechanical response of SCG in macroscale, which had not been experimentally explored for any length scale.

The tensile loading response of centimeter-scale SCG by using dynamic mechanical analyzer (DMA) in conjunction with a camphor-assisted transfer process(*21*), yielded $E$ values up to 737 GPa but comparatively lower strength of 3.33 GPa. Transferring large-scale graphene remains challenging, as defects can be introduced in atomically thin monolayer graphene during the multistep transfer process(*22, 23*). However, we can transfer centimeter scale graphene/polymer films onto target substrates, such as silicon wafer(*17*) or a liquid surface. We have tested the "float on water" (FOW) tensile technique(*24*) on SCG, as described further below, and it has worked very well for macroscale SCG supported by thin polymer film.

We report tensile loading of 1 cm long, 2 mm wide, SCG with graphene adhered to a 200 nm thick polycarbonate (PC) layer, and analysis of stress-strain curves to obtain the Young's modulus ($E$), and stress ($\sigma$) and strain at failure ($\varepsilon_f$). We discovered that crack initiation and subsequent propagation within graphene are controlled by the graphene crystal orientation, armchair (ac) or zigzag (zz). We found that the strength of SCG is highest along the zz direction and lower along the chiral (edges between ac and zz) and ac edges. Both zz and ac-oriented SCG exhibit ideal stiffness, whereas chiral samples show reduced stiffness. The toughness modulus of SCG (zz, ac, or chiral) is significantly higher than for any known material, and the zz-SCG's toughness modulus is *about 10x higher* than the commercially available carbon fiber. Through modeling, we anticipate the mechanical response of stacked graphene-polymer layers with a model system that mitigates the impact of edge defects. We elucidate the fracture mechanics of macroscale SCG, and the exceptionally high tensile strength values observed open avenues for its wide application in automobile and transportation (trains, planes) sectors, and especially for flexible large scale optoelectronic devices, straintronics, and new applications where the planar structure is essential.

## Results and Discussions

*Tensile Test Measurements in FOW System*

We conducted uniaxial tensile measurements on SCG films utilizing the FOW system. The primary objective of our study was to investigate the mechanical response in macroscale SCG films. We used a thin layer of Poly(bisphenol-A-carbonate) (PC) as the support material for our SCG films because the inherent ultrathin nature (~0.34 nm) and brittleness of SCG have, to date, made it challenging to prepare a freestanding macroscale sample suitable for tensile measurement. This choice was made due to three key factors: First, PC exhibit high optical transparency, a crucial property for the FOW system's accurate measurement capabilities. Second, the PC possesses considerable mechanical strength, ensuring minimal interference with the SCG film's intrinsic behavior during the tensile testing process(*25*). Third, PC's higher tensile failure strain than SCG ensures full support until its failure point. This allows us to isolate the effects of the SCG film's properties on its mechanical response.

SCG was synthesized on Cu(111) foils using thermal CVD. Raman spectroscopy of as-grown SCG on Cu (111) showed an $I_{2D}/I_G$ ratio of 2.56 and a 2D full width at half maximum (FWHM) of 35.16 cm$^{-1}$ (Fig. S12)(*17*). SCG on the SCG-PC films was verified using Raman spectroscopy (Fig. 1c). In the Raman spectra of SCG-PC, the intense G (~1584.9 cm$^{-1}$) and 2D (~2667.1 cm$^{-1}$) peaks with negligible D peak confirm the high quality of this single crystal graphene. We observed a blueshift in G (~3.3 cm$^{-1}$) and 2D (~15.1 cm$^{-1}$) band after removal of PC film, which is attributed due to the removal of strain generated at the interface between the PC and SCG. PC/chloroform solutions were spin-coated onto the as-grown SCG/Cu(111) samples, with thickness variation controlled by adjusting the spin speed. AFM was used to determine film thickness (Fig. S13 and S14). PC and SCG-PC samples were designated based on their PC film thickness, as detailed in Tables S2-S3. A "dogbone' cutter (model no. SDMP-1000 and JISK 6251-7) was then used to cut out dog bone specimens (with gauge length 10 mm, width 2 mm, see Fig. S15), from which the Cu(111) was etched away, as described in Fig. S16. Thus, the dog bone 200 nm SCG-PC samples were floated on water and attached to the FOW system components that apply tensile loading with van der Waals type adhesion clamping.

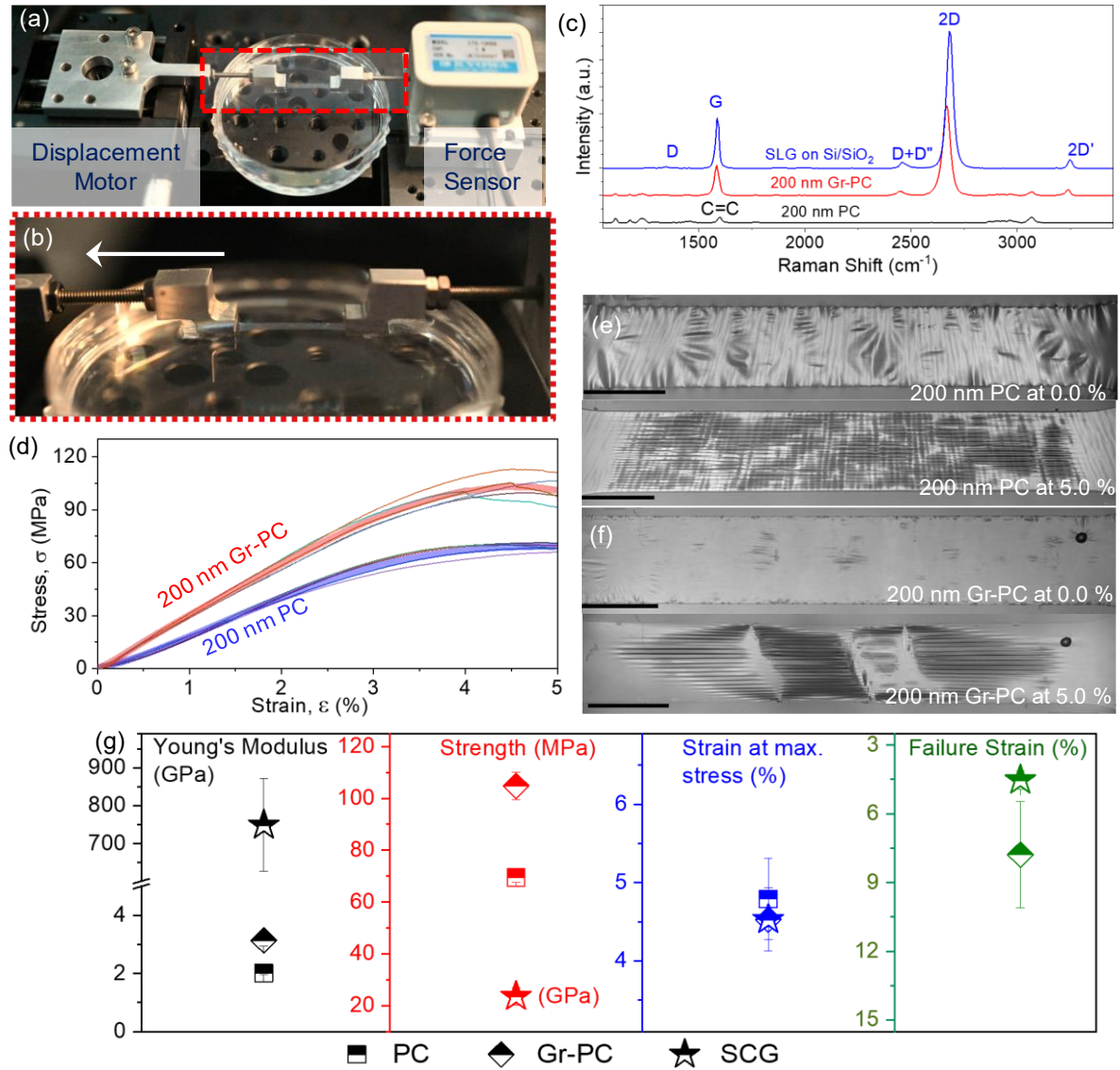

**Fig. 1: Tensile testing of macroscale single-crystal graphene using a float-on-water (FOW) measurement system.** (a,b) An experimental setup illustrating the FOW system with the dogbone-shaped SCG-PC specimen. The white arrow shows the displacement direction. (c) Raman spectra of PC, SCG-PC film, and SCG on Si/SiO$_2$ wafer. The presence of strong G and 2D bands in SCG on PC was observed, and no D peak wea detected, confirming the quality of SCG. (d) Stress-strain curves of 200 nm PC and 200 nm SCG-PC films, each recorded from five different samples. The red and blue curves present the average stress-strain curves. Representative images of (e) 200 nm PC and (f) 200 nm SCG-PC samples at 0% and 5% applied strain showcasing graphene failure at 5% strain, whereas no breakage is observed in PC at 5% applied strain. (g) Comparison of average Young's modulus, tensile strength, strain at maximum stress, and failure strain values of PC, SCG-PC samples, and SCG (extracted from SCG-PC and PC curves).

The tensile tests were done on samples floating on the water surface at room temperature under constant displacement rates of 0.002 mm/s, equivalent to a strain rate of 0.0002 (mm/mm)/s (Fig. 1a-b). The mechanical properties of similar 200-nm thick PC films (with no graphene) were characterized independently using the FOW system. Fig. 1d shows the stress-strain curves for 200 nm PC and SCG-PC films. Complete stress-strain curves are presented in Fig. S17c and S18c.

Monochromatic imaging showed wrinkles in bare PC films even at 0% strain (Fig. 1e) due to their intrinsic flexibility(*26*), while SCG-PC films were smooth (Fig. 1f), indicating uniform contact with the water surface facilitated by the SCG's flatness. At 5% strain, both films displayed horizontal lines perpendicular to the displacement direction within the gauge area, indicating stress concentration zone(s) (Fig. 1e-f, detailed analysis in Fig. S19 and Fig. S20). The higher density of such lines in the SCG-PC film suggests greater stress than the bare PC film due to SCG's higher strength, leading to earlier failure (~7.69%) compared to bare PC (>20%) films. Mechanical properties of PC and SCG-PC films are summarized in Tables S4-S5. The SCG-PC films exhibited roughly 1.5 times higher $E$ and $\sigma$ compared to bare PC films. By using these values, we extracted $E$ and $\sigma$ values for SCG using the rule of mixtures according to the following equation:

$$E_{SCG-PC} = E_{PC} \frac{t_{PC}}{t_{PC}+t_{SCG}} + E_G \frac{t_{SCG}}{t_{PC}+t_{SCG}} \qquad (1)$$

where $E_{SCG}$ and $E_{PC}$ are the Young's modulus values for SCG and PC, respectively, and $t_{SCG}$ and $t_{PC}$ are the thicknesses of graphene and PC, respectively. Fig. 1g compares the $E$, $\sigma$, strain at maximum stress, and failure strain of SCG, SCG-PC, and PC films (a total of 25 combinations as summarized in Table S6). The highest $E$ value for SCG was 970.5 GPa (Table S6), approaching microscale monolayer graphene's modulus measured by a push-to-pull (PTP) micromechanical device(*13*). In our initial investigations, the edge orientation (zigzag or armchair) of SCG was not controlled during the preparation of dog bone samples, resulting in chiral SCG edges. The maximum tensile strength we observed was 30.48 GPa (Table S6), representing a nearly tenfold increase compared to the yield strength of previously reported macroscale graphene(*21*). However, this value remains lower than the theoretical tensile strength of graphene. This is likely due to edge defects in SCG introduced during mechanical cutting, as observed by SEM (Fig. S21) and TEM (Fig. S22a-b) analyses. TEM imaging shows the presence of nanometer-sized defects or cracks along the SCG edge. The average $E$ and $\sigma$ across 25 combinations (of five chiral SCG-PC and 5 bare PC samples) were 748.9 ± 121.9

GPa and 23.56 ± 3.42 GPa, respectively (Fig. 1g). The ideal strength of monolayer graphene, approximately 130 GPa, has been observed at an applied strain reported as 25%(*6*). In our study here, the failure of chiral SCG occurred at an average strain of 4.53%, which corresponds to an estimated strength of 41.2 GPa at this strain (with $E = 1$ TPa and the third order modulus, $D = 2$ TPa)(*6*). While the average $E$ is comparable to previous macroscale results obtained using a DMA system on free-standing graphene/polymer films(*21*), increased strength and failure strain is observed showing our approach (FOW measurement system and SCG) yields significantly higher values. While the DMA test(*21*), reported a yield strain of 0.6% and the corresponding strength of 3.33 GPa, our chiral SCG samples carry substantially higher stress of 5.58 GPa at a strain of 0.6%. This improvement can be attributed to the absence of grain boundaries and adlayers in our single-crystal SCG, and utilizing an optimal polymeric substrate with a 200 nm thickness compared to 100 nm(*21*) (see the supplementary section 4 on role of polymer thickness). Note that defining the strength of a centimeter-scale SCG solely by its yield stress underestimates its capacity. Although microcrack initiation can cause a deviation from linearity, our chiral SCGs smoothly continued to bear loads at stress levels up to 23.56 GPa, demonstrating their full failure strength—7.08x greater than the previously reported yield stress. This highlights the well-known concept of *damage tolerance*, where macroscale SCG, like many materials, can redistribute stress after the initiation of damage, thereby delaying total failure(*27, 28*).

The crack initiation and propagation were visualized such as shown in supplementary Video S1. This video correlates the mechanical response with the formation and growth of a notch on the SCG-PC film. We observed that the strain at maximum stress of a SCG-PC sample denotes the failure strain of the SCG. The measured failure strain of macroscale SCG (~4.53%) was comparable to microscale measurements by the push-to-pull method (~5.8%)(*13*). Lower SCG failure strain compared to the theoretical limits (that are reported to be approximately 13–19% along the armchair and 20–26% along the zigzag direction)(*29-31*) likely originated due to the existence of critical edge defects as discussed (through QFM analysis) in the model prediction section later. The similarity between the SCG failure strain and the strain at maximum stress of bare PC (Table S4-S5) suggests that the failure strain of SCG is influenced by the strain at the point of maximum stress in the PC layer. Mechanical property variation of SCG with PC thicknesses (with different $E$ and $\sigma$) are discussed in supplementary section 4.

*Identification of Crack Initiation and Propagation in Macroscale SCG*

The transparent PC substrate in SCG-PC films allows direct visualization of crack initiation and propagation within the SCG layer (Fig. 2a-f). At 2% applied strain, stress concentration zones (SCZ) become evident, primarily at the edges (see Fig. 2c). The reason for observing the SCZs at ~2% can be attributed due to the transition from linear to non-linear response in the polymer's stress-strain curve (see Fig. S16c). With increasing strain, the SCZs intensity manifests as horizontal lines to the loading direction (Fig. 2c-d). At 3% strain, a significant increase in the density of horizontal lines at SCZs shows the initiation of macrocrack propagation (Fig. 2g-h). An observed crack appears in SCG at 3.5% strain, originating from an existing SCZ (Fig. 2h). Simultaneously, new SCZs emerge at other edges on the top of the SCG-PC film (Fig. 2i), serving as potential sites for further crack initiation. This process continues, ultimately leading to complete failure of the SCG layer. Crack identification is facilitated by the distinct contrast between SCG (bright white lines) and the PC layer (Fig. 2i-k). A crack that initiates at one edge appears to trigger the formation of a complementary crack on the opposite edge. These cracks then propagate in opposite directions, culminating in complete SCG failure (Fig. 2l). Even after SCG failure, the PC layer with fragmented SCG on its surface continues to elongate under tensile stress, and the SCG-PC film (Fig. S20) exhibits a faster breakdown compared to the bare PC films (Fig. S19) due to its higher ultimate stress.

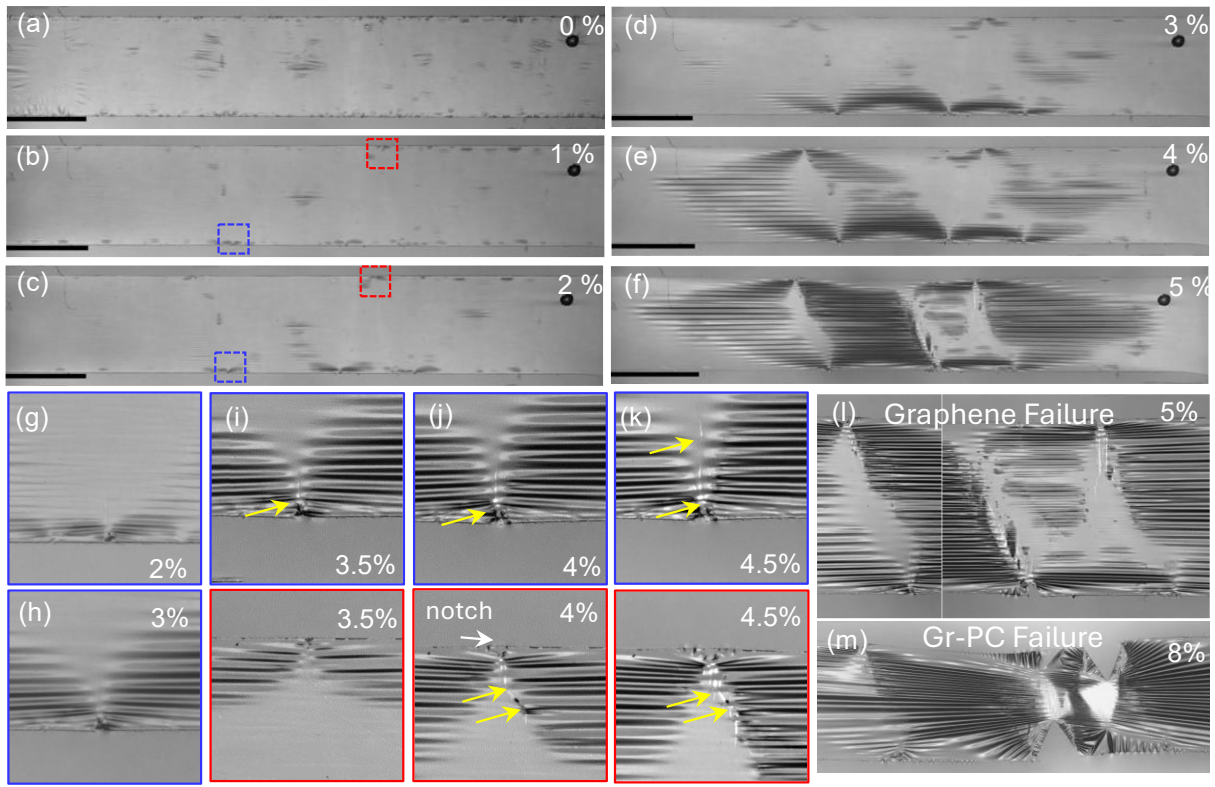

**Fig. 2: Evolution of SCG-PC films under tensile loading, illustrating crack initiation and propagation in graphene.** Monochromatic images of the SCG-PC films at various strain levels (a-f) are shown under tensile loading. No cracks are observed for (g) 2% and (h) 3% strain, and crack initiation occurs at approximately (i) 3.5% applied strain, followed by crack propagation (i-k), SCG failure (l), and complete failure of SCG-PC at around 8% strain (m). Scale bar: ~2.0 mm. The yellow and white arrows indicate the cracks in SCG and notches in the SCG-PC film, respectively. The blue and red outlines denote the bottom and top edge, respectively.

To further analyze crack characteristics, tensile testing was interrupted at various strain points, and the SCG films were transferred to a Si/SiO$_2$ wafer for SEM analysis. Chloroform treatment removed the PC layer, exposing the SCG morphology. Fig. S23 compares the effect of tensile loading on SCG at the crack initiation point and the failure strain of SCG. Smaller microcracks parallel to the main macrocrack were observed at initiation, likely contributing to its propagation until complete SCG failure. Further analysis (Fig. S23c-d) revealed that the crack angles were multiples of approximately 30°, reflecting that the crack in SCG propagates along armchair or zigzag directions. Microcracks were parallel to each other and perpendicular to the elongation direction, while larger macrocracks tended to align with the armchair or zigzag edges of graphene (Fig. S23e-f). These observations suggest that crack propagation in SCG is influenced by both stress concentration points at opposing edges and the intrinsic defects of graphene.

*Mechanical Properties of SCG along armchair (ac) and zigzag (zz) directions*

The zigzag (zz) edge of SCG grown on Cu(111) preferentially aligns with the Cu<110> direction, while the armchair (ac) edge aligns with the Cu(112) direction (see Fig. 3a)(*20*). The Cu<112> direction can be optically identified from rolling marks on the Cu foil (Fig. 3b) as confirmed by electron backscatter diffraction (EBSD) data (Fig. S35). To determine the orientation of graphene edges relative to Cu rolling marks, hexagonal SCG islands were grown on Cu(111) (Fig. 3c, Fig. S36). The angle between the edge of SCG islands and rolling marks was calculated from optical and SEM images as 90.79 ± 0.27 º and 90.14 ± 2.06 º, respectively (Fig. S36). Polarized Raman spectroscopy performed on marked (black dotted line) graphene edges on Cu(111) foil (Fig. 3c) showed an increase in the G peak intensity with the angle between incident and scattered light (varied using the analyzer), which also confirmed that the edge perpendicular to the rolling marks corresponds to the zz edge of SCG(*32*) (Fig. 3d). Polymer coated SCG/Cu(111) foils were aligned with the rolling marks and cut to prepare dogbone-shaped SCG-PC samples with zz (perpendicular) and ac (parallel) edges.

The perpendicular orientation (~89.24 ± 0.82 º) denotes the dogbone sample with zz edge, while the parallel orientation (~0.63 ± 1.27 º) denotes the ac edge; see supplementary Fig. S37. Polarized Raman spectra at these perpendicular and parallel edges are shown in Fig. S38. We performed TEM imaging at the edge of SCG parallel to the rolling marks. The diffraction pattern (Fig. S21c) and high resolution TEM image (Fig. S21d) confirms the ac direction along the edge(*33*).

Stress-strain curves were obtained from tensile loading in the FOW system along the zigzag (Fig. 3e) and armchair (Fig. 3f) direction for these different dogbone samples. The average $E$ values for ac and zz SCG-PC samples were 3.53 GPa and 3.68 GPa, respectively, while the average $\sigma$ values were 99.8 MPa and 110.7 MPa, respectively. We thus observed 1.12 times (99.8 to 110.7 MPa) enhanced strength for zz SCG-PC samples, and similar $E$ values. The ac SCG-PC samples showed a higher average failure strain of 10.10% compared to 7.79% for zz SCG-PC samples. The mechanical response of zz and ac SCG-PC samples are summarized in supplementary Table S10. The mechanical properties ($E$ and $\sigma$) of zz-SCG and ac-SCG were obtained by comparing the stress strain curves in Fig. 3e-f (and supplementary Fig. S39) with those of bare PC films (200 nm thick). $E$ values for zz- and ac-SCG were 1.11 ± 0.04 TPa and 1.01 ± 0.10 TPa, respectively. The average strength for zz-SCG was 27.40 ± 4.36 GPa, while that for ac-SCG was 20.21 ± 3.22 GPa. Chiral SCG (with edge orientations not aligned to zz or ac) samples exhibited an average strength of 23.56 ± 3.42 GPa (Fig. 3g), falling between the average strength for zz- and ac-SCG.

The mechanical properties ($E$, $\sigma$ and failure strain) of ac, zz, and chiral SCG are compared in Fig. 3g. These results align with previously reported theoretical findings from density functional theory (DFT) and molecular dynamics (MD) simulations.(*34, 35*) We compared the previously reported theoretical and experimental results with our experimental values in supplementary Table S13. Failure strains of ~ 3.69% and ~ 6.01% for SCG were observed for the ac and zz direction, respectively. Observation of higher strength and failure strain along the zz direction compared to the ac direction can be attributed to the larger bond angle variation along the zz direction.(*35, 36*)

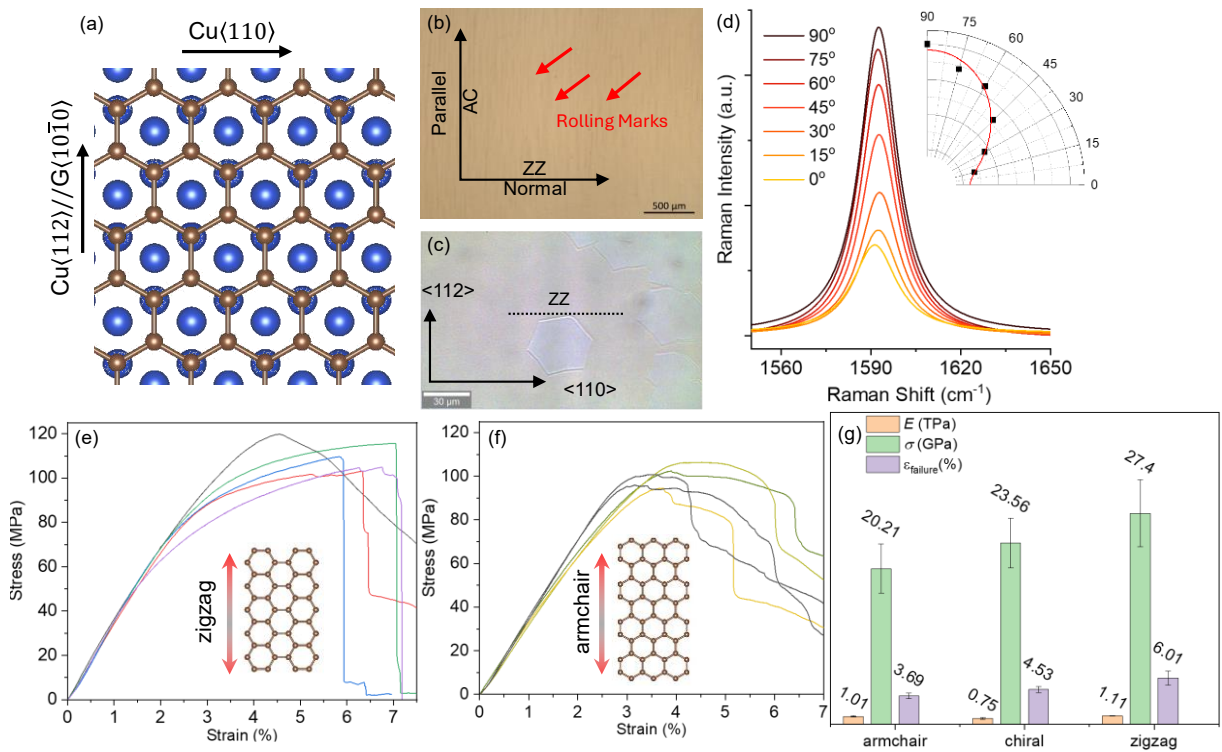

**Fig. 3: Mechanical properties of SCG along armchair (ac) and zigzag (zz) direction.** (a) Preferred orientation of SCG on Cu(111). (b) Identification of rolling marks of Cu(111) on as grown SCG on Cu(111) sample. (c) Optical image of hexagonal SCG. (d) Polarized Raman spectra of zz-SCG edge on Cu(111) as marked in (c) by black dotted line. Stress strain curve of SCG for tensile loading along (e) zigzag and (f) armchair direction. (g) Comparison of mechanical properties of SCG for armchair, chiral and zigzag direction.

*Theoretical Model Predictions*

We now consider scaling from our current dimensions of 1 cm along the tensile loading axis with an area of 0.2 cm², to an A4-sized sheet. This results in scale ratios of $R = 29.7$ for length and 3118.5 for the area. According to the Weibull scale law provided in the Section 1 of supplementary information, the strength of a reference sample, $\sigma_0$ (here our current samples), is related to the scaled strength, $\sigma$, (such as of an A4-sized sheet) by the power law, $\sigma = \sigma_0 (R)^{-\frac{1}{\alpha}}$, where $\alpha$ is the Weibull modulus, which characterizes the material's statistical strength distribution(*37*). To perform strength scaling, we fitted a Weibull distribution (Fig. S1) on the measured strengths of chiral, zz, and ac SCGs listed in Tables S6, S11, and S12. From this, the reference strengths of $\sigma_0$ were determined to be 25.00, 29.24, and 21.54 GPa for chiral, zz, and ac SCGs, respectively. The Weibull moduli $\alpha$ were 8.22, 7.35, and 7.46, corresponding to chiral, zz, and ac. When scaling by length, we assume the critical flaws are edge-confined, consistent with our experimental findings where failure initiates at the edge. For an A4-size SCG sample with $R = 29.7$, the predicted strengths for zz, chiral, and ac orientations are 18.43 GPa, 16.55 GPa, and 13.67 GPa, respectively. However, when scaling by area ($R = 3118.5$), assuming surface-distributed flaws, the predicted strengths drop to 9.78 GPa, 9.39 GPa, and 7.32 GPa, indicating a significant reduction in strength due to the increased number of critical flaws; however as noted, the flaws were always identified at the edges, and not in the interior. To estimate the size corresponding to theoretical strength, we inversely scaled down by setting $\sigma = 130$ GPa and solving for $R$. For zz SCGs, the theoretical strength would correspond to a length of 0.174 μm and an area of 350 μm². Using the same approach, ac SCGs yield lengths of 0.015 μm and areas of 30.2 μm², while chiral samples result in 0.013 μm and 26 μm². These scaling estimates, from theoretical strength to A4-size sheets, are illustrated in Fig. 4a.

We did a fracture mechanics analysis using QFM (quantized fracture mechanics), which extends linear elastic fracture mechanics by introducing a fracture quantum(*38*), as detailed in the Section 2 of supplementary information. This fracture quantum is calculated such that a crack length of zero exhibits an ideal strength of 130 GPa. Accordingly, we determine critical lengths of 6.23, 8.53, and 11.70 nm for a sharp crack at the edge of zz, chiral, and ac SCGs with the strength 27.40, 23.56, and 20.21 GPa, respectively. By substituting the sharp crack with a V-notch as the critical defect on the SCG edge(*39*) and assuming a notch angle of 30°, we find critical notch depths (perpendicular to the edge) of 13.43, 19.65, and 28.87 nm for zz, chiral, and ac SCGs, respectively. Fig. 4b illustrates the variation of the critical edge defect against the SCG strength.

Finally, we used a damage evolution model to capture the softening behavior of SCG on an ultra-thin polymer substrate, as detailed in Section 3 of supplementary information. The calibrated nonlinear stress-strain curves, derived from the constitutive laws presented in Eqs. S30-S32, are plotted in Fig. 4c. Then, we calculated the toughness values, defined as the area under the stress-strain curve, for the zigzag, armchair, and chiral SCGs, obtaining 1.53, 1.26, and 1.15 $GJ/m^3$, respectively. We also applied this model to a stack created by folding a SCG-PC sample, allowing us to predict the stack's stress-strain response, as shown in Fig. S9. We find that the folded SCG-PC sample with 100 layers of SCG-PC stacking can sustain a stress of 125.15 MPa at 4.53% strain, which is 1.80 times higher than the PC sample and 1.19 times higher than the SCG-PC sample.

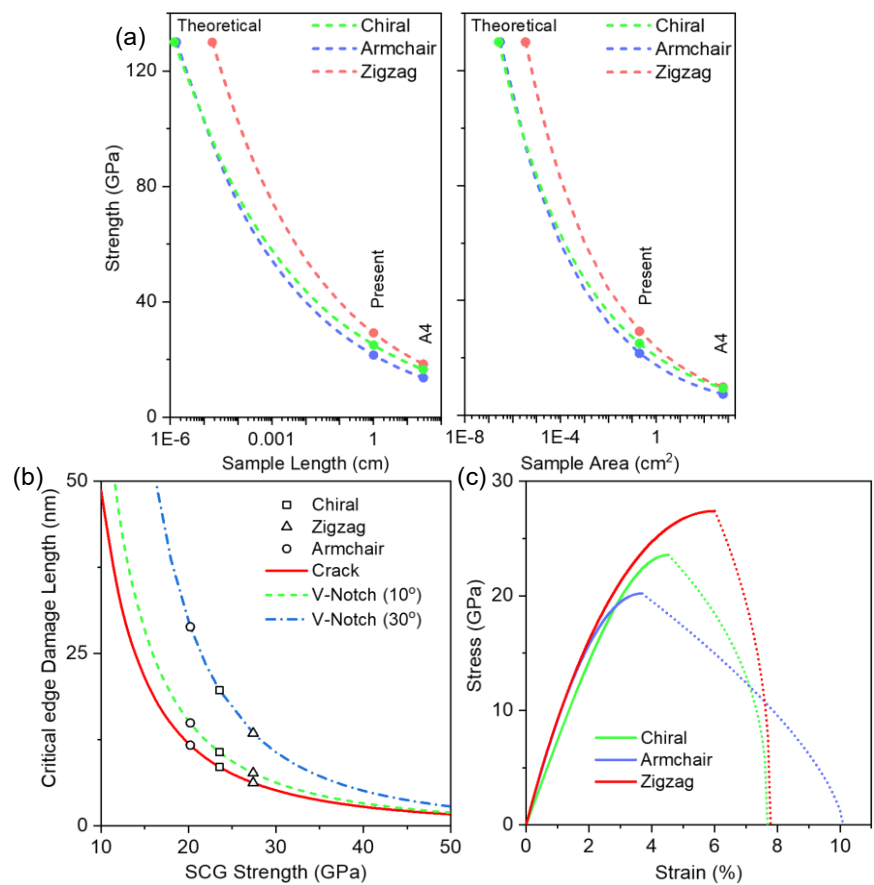

**Fig. 4: Model predictions for mechanical properties of chiral, armchair, and zigzag macroscale SCGs.** (a) Weibull strength scaling from the present sample size to A4-sized sheets, as well as projecting to the theoretical strength of 130 GPa. It is provided for both length scaling corresponding to the edge-confined defects, and area scaling, where the fatal defects are surface-distributed. (b) Estimation of critical edge damage lengths via QFM, considering either a sharp crack or a V-notch at the edge, and detailing the corresponding stress levels at which fracture occurs for each chirality. (c)

The calibrated nonlinear stress-strain curves obtained from the damage evolution model, illustrating the nonlinear softening behavior of SCGs on an ultra-thin polymer substrate.

## Conclusion

We found near-intrinsic tensile loading mechanical properties in centimeter-scale single crystal monolayer graphene (SCG) measured with a float on water testing system. We report average values of $E$ = 1.11 TPa and $\sigma$ = 27.40 GPa along the zigzag edges of SCG, approaching the theoretical limits for graphene. Loading along the armchair edges yielded $E$ = 1.01 TPa and $\sigma$ = 20.21 GPa while chiral SCG exhibits had $E$ and $\sigma$ of 0.75 TPa and 23.56 GPa, respectively. Fitting a Weibull distribution on the experimental data set of the current sample size, i.e., 10 x 2 mm, we predict that if the fatal damage is initiated solely on the edges of a macroscale sample (as we observed in all samples), the strength of an A4-size SCG can range from 13.67 to 18.43 GPa depending on chirality with zigzag as the maximum, armchair the minimum, and chiral in between. Although edge defects and supporting layer slightly impact tensile strength, our findings confirm that optimizing the support layer thickness enhances stiffness. This study establishes the *remarkable mechanical resilience of large-scale single crystal monolayer graphene*, demonstrating a failure strain of up to 6.01% for tensile loading along the zigzag edge. The developed methodology of macroscale graphene mechanics, characterized by uniform and well-controllable tensile loading, offers potential for studying the mechanical properties of other 2D materials and their heterostructures in macroscale. The SCG will be a revolutionary material due to its remarkably high strength-to-weight ratio (also, the highest specific strength of any material, ever) and ultra-thin structure. SCG's exceptional macroscale strength, ideal modulus, and large failure strain- herald a range of applications, some of which will be new. The immediate applications can be in fields like aerospace, automotive, space, urban air mobility (UAM), medical devices, civil engineering, and sporting goods—any field requiring simultaneously high strength, stiffness, and elongation.

**Materials and Methods**

Growth of Large-Area Single-Crystal Graphene: To grow single-crystal single-layer graphene (SCG), 35 mm × 50 mm Cu(111) foils were loaded into a 40 mm diameter quartz tube with a uniform heating zone around 100 mm and placed at the center of the heating zone. First, the furnace was heated to 1060 °C within 60 min in a 300 sccm Ar and 40 sccm $H_2$ atmosphere at 30 Torr pressure, followed by introducing 34 sccm 0.1 mol% $CH_4$ for 60 min while the temperature was held at 1060 °C. The sample was then rapidly cooled to room temperature under the same gaseous conditions (300 sccm Ar, 40 sccm $H_2$, and 34 sccm 0.1 mol% $CH_4$). The as-prepared SCG on Cu(111) was characterized using Raman spectroscopy and SEM to confirm the uniformity and quality of graphene.

Preparation of Polycarbonate-Graphene Film (SCG-PC): Poly(bisphenol A carbonate) [PC] from Sigma-Aldrich (average Mw ~ 45000) was dissolved in chloroform, and 0.5-5 wt% solutions were prepared. PC/chloroform solution was spin-coated at a speed of 3000 RPM on bare Cu(111) foil to prepare PC films and on Gr/Cu(111) foil to prepare SCG-PC films. The thicknesses of the PC and SCG-PC films were measured by AFM after transferring the sample onto $Si/SiO_2$ wafers.

Tensile Test Measurement Setup: The Float-on-Water (FOW) measurement setup was built by assembling two critical components: (i) a force sensor (model no. LTS-100GA from the Kyowa Electronic Instruments Co. Ltd.) and (ii) a displacement motor (model no. C863 from Physik Instrumente (PI)). The force exerted upon the samples due to displacement was recorded using a Kyowa sensor interface (Model No: PCD-430A). Monochromatic images of the thin films were captured using the Allied Vision camera (Model No: Manta G-146B).

Mechanical Cutter: PC-coated graphene films and bare PC films on Cu foil were transferred into the mechanical cutter (model no. SDMP-1000 and JISK 6251-7 from Dumbell Co. Ltd.) and cut into "dumbbell" or "dogbone" shapes (gauge length × gauge width) conforming to the international ISO-34-1 standard. The optical image of the mechanical cutter and a schematic of the shape were presented in Fig. S8.

Armchair and zigzag SCG-PC dogbone sample preparation: Rolling marks of as grown SCG-Cu(111) foil were identified using the optical microscope. The orientation of armchair (ac) and zigzag (zz) direction of SCG with respect to the rolling marks were recognized by polarized Raman spectroscopy. The variation in edge (ac or zz) orientation were further measured by optical and SEM images. PC coated graphene films were aligned with the mechanical cutter in parallel and perpendicular direction to the rolling marks of Cu(111) to obtain the ac or zz edge in the dogbone sample. The orientation of rolling marks was further measured from the optical images after preparation of dogbone samples.

Transferring SCG-PC Film onto Water Surface: The PC and SCG-PC films on Cu foil were floated on an aqueous etchant (0.5 M $(NH_4)_2S_2O_8$) to remove Cu(111). The floated films were then transferred to a water bath (repeated three times) to remove excess etchant and were further used for FOW measurements. As per requirements, the films were transferred onto Si/SiO$_2$ wafers for AFM and Raman spectroscopic characterizations. The floated SCG-PC films were transferred on the Qunatifoil Cu TEM grid (200 mesh) followed by removal of PC layer for TEM samples preparation.

Characterization: A Bruker Dimension Icon atomic force microscope (AFM) was used to measure the surface morphology and thickness of the PC and SCG-PC films studied in this work. Raman spectroscopy was performed using a WiTec micro-Raman instrument with a 488 nm (for SCG on Cu(111)) and 532 nm (for SCG/SCG-PC on Si/SiO$_2$) laser line. A Zeiss optical microscope (OM, AxioCamMRc5) was used to characterize the morphology of the specimens. A scanning electron microscope (SEM, Verios 460, FEI) was used to image the morphologies of the samples and the graphene structures. TEM imaging was done using an aberration corrected TEM (JEM-ARM300F) at an acceleration voltage of 80 kV.

## 1. Weibull Statistical Analysis for Strength Prediction of an A4-Size SCG

Here, we project the measured strength of our centimeter-scale SCG samples to an A4-size one. According to the Weibull model, when the reference sample, denoted by a subscript "0", is scaled by a factor of $R$, its strength, $\sigma_0$, is related to the scaled strength $\sigma$ via a power-law equation(*37*):

$$\sigma = \sigma_0 (R)^{-\frac{1}{\alpha}} \tag{S1}$$

where $\alpha$ is the Weibull modulus. As $\alpha \to \infty$, the size effect diminishes and $\sigma \to \sigma_0$. Conversely, as $\alpha$ decreases, the size effect becomes more pronounced. It is important to note that when $R > 1$, the strength decreases due to the greater number of defects in a larger sample. Our reference samples with a total area of $A_0 = 0.2$ cm$^2$ and a length of $l_0 = 1$ cm along the tensile load are projected to an A4-sized sample with a length $l = 29.7$ cm and an area of $A = 623.7$ cm$^2$ resulting in a scale factor of $R = l/l_0 = 29.7$ for the length and $R = A/A_0 = 3118.5$ for the area.

The reference strength, $\sigma_0$, and the Weibull modulus, $\alpha$, can be determined by fitting the Weibull statistical distribution to the measured strength of the reference samples. A larger $\alpha$ indicates a smaller dispersion in strength between samples, signifying higher quality SCG samples and reduced dependence of their strength on size. Figs. S1 (a-c) show the fitted graphs for the strength values of chiral, zigzag, and armchair SCGs, as presented in Tables S6, S11, and S12, respectively. It provides the reference strengths of $\sigma_0$ = 25.00 GPa, 29.24 GPa, and 21.54 GPa, and the Weibull modulus of $\alpha$ = 8.22, 7.35, and 7.46 for chiral, zigzag, and armchair SCGs.

The size-effect prediction based on the comparison of sample areas as the scale factor, $R$, suggests that fatal defects are distributed across the surface, and failure may initiate from any point on the sample's surface. In contrast, using the length ratio as the scale factor, $R$, implies that the fatal defects are located at the edges, with failure always initiating from the sample's edge. The predicted strength for an A4-size zigzag SCG sample (the strongest) is 18.43 GPa when scaled by length and 9.78 GPa when scaled by area. This indicates that when the critical flaw is confined to the edge, as consistently observed in our tensile tests, scaling to an A4-size zigzag sample results in a 37.0% reduction in strength, compared to a more substantial 66.6% reduction when the flaws are assumed to be distributed across the entire surface. Similarly, the strength of an A4-size armchair SCG is obtained as 13.67 GPa and 7.32 GPa, while for the A4-size chiral one they are predicted 16.55 GPa and 9.39 GPa, for the length and the area scaling respectively.

In addition, we employed the Weibull scaling law to estimate the sample size corresponding to the ideal theoretical strength of 130 GPa. When we scale based on sample length, achieving this ideal strength for the zigzag SCG requires a sample with a length as small as 0.174 µm, whereas using area as the scaling factor yields an area of 350 µm², corresponding to a circular region with a diameter of 21.11 µm. If we do the same with armchair to estimate the length and area corresponding to ideal strength, we obtain 0.015 µm and 30.2 µm² respectively, while chiral samples yield 0.013 µm and 26 µm².

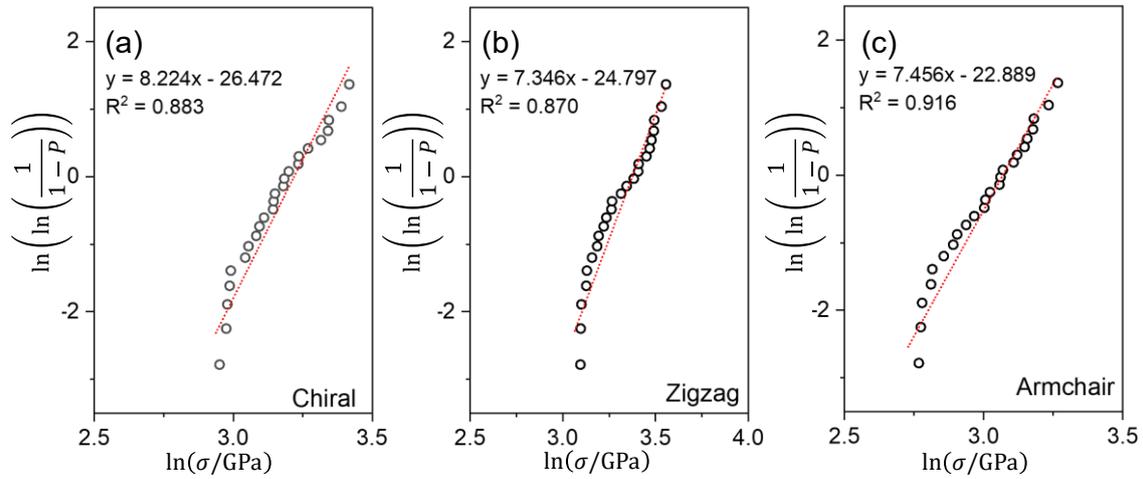

**Fig. S1: The predicted strength for an A4-sized SCG by the Weibull analysis.** Curve-fitting of Weibull distribution on the measured strength of (a) chiral, (b) zigzag, and (c) armchair SCGs respectively listed in Tables S6, S11, and S12.

## 2. Quantized Fracture Mechanics Analysis for Macroscale SCG with Edge Defects

According to quantized fracture mechanics (QFM)(*38*), the critical crack length, $l_c$, for a material with fracture toughness $K_{IC}$ and the corresponding strength $\sigma_c$ can be calculated as,

$$l_c = \frac{K_{IC}^2}{\pi Y^2 \sigma_c^2} - q \tag{S2a}$$

where $Y$ is a function of geometry and load type. The primary distinction between QFM and the classical linear elastic fracture mechanics (LEFM) in estimating $l_c$ lies in the introduction of the fracture quantum, $q$. While LEFM predicts an infinite strength for an ideal defect-free material, incorporating $q$ into the classical equation enables fracture to occur at an ideal theoretical strength $\sigma_i$ even when $l_c = 0$. To achieve this, the quantum fracture should be defined as,

$$q = \frac{K_{IC}^2}{\pi Y^2 \sigma_i^2} \tag{S2b}$$

Replacing the fracture toughness of $K_{IC} = 4.4$ MPa$\sqrt{m}$ (40), the upper limit theoretical strength of $\sigma_i = 130$ GPa for graphene, and $Y=1.122$ for defect starting from the edge into Eq. S2b, gives $q = 0.29$ nm. Then, employing Eq. S2a, one can evaluate the crack length, $l_c$, which reduces the theoretical strength, $\sigma_i$, to the strength $\sigma_c$. Relying on a nonlocal theory approach(39), it is shown that if a sharp crack on the edge of a plate under tension is replaced with a V-notch of the same length (depth perpendicular to the edge), the strength of the cracked plate, $\sigma_c$, is reduced to $\sigma_{notch}$. These two strengths are related as follows,

$$\sigma_{notch} = \sigma_i \left(\frac{\sigma_c}{\sigma_i}\right)^{2\alpha(\gamma)} \tag{S3}$$

where $\gamma$ is the notch angle and the exponent $\alpha(\gamma)$, ranging from 0.5 to 0, reflects the stress singularity. When $\gamma = 0$, the V-notch turns into a crack and $\sigma_{notch} = \sigma_c$ which demands $\alpha(0) = 0.5$, while when $\gamma \to \pi$ the edge defect tends to vanish, meaning that $\sigma_{notch} = \sigma_i$ which requires $\alpha(\pi) = 0$. For any other re-entrant angle, $\gamma$, one can solve the following eigen-equation to find the corresponding $\alpha$:

$$(1 - \alpha)\sin(2\pi - \gamma) = \sin[(1 - \alpha)(2\pi - \gamma)] \tag{S4}$$

Solving Eq. S2 for $\sigma_c$ gives,

$$\sigma_c = \left(\frac{\sigma_{notch}}{\sigma_i^{(1-2\alpha)}}\right)^{\frac{1}{2\alpha}} \tag{S5a}$$

Then, the updated version of the QFM formulation for critical length for a V-notch defect on the edge is obtained by substituting Eq. S5a into Eq. S2a as,

$$l_c = \frac{K_{IC}^2}{\pi Y^2 \left(\frac{\sigma_{notch}}{\sigma_i^{(1-2\alpha)}}\right)^{1/\alpha}} - q \tag{S5b}$$

Note that for $\gamma = 0$ ($\alpha = 0.5$) the notch and the crack are the same and Eq. S5b turns into Eq. S2a. Fig. S2 plots the predicted critical lengths, $l_c$, corresponding to the strength for a crack and 10⁰, and 30⁰ V-notches. It is seen that a V-notch defect with a longer depth than a sharp crack length is required to make an equivalent reduction in the strength and the required critical depth increases by increasing the notch angle. In other words, V-notch defects are less destructive than sharp cracks. For the zigzag SCG with the strength 27.40 GPa, the corresponding defect lengths (depth) are 6.23 nm, 7.66 nm, and 13.43 nm, for the armchair with the strength 20.21 GPa, are 11.70 nm, 14.91 nm, and 28.87 nm, and for the chiral with the strength 23.56 GPa, are 8.53 nm, 10.68 nm, and 19.65 nm, for a sharp crack and three V-notches with the angles 10⁰, and 30⁰, respectively.

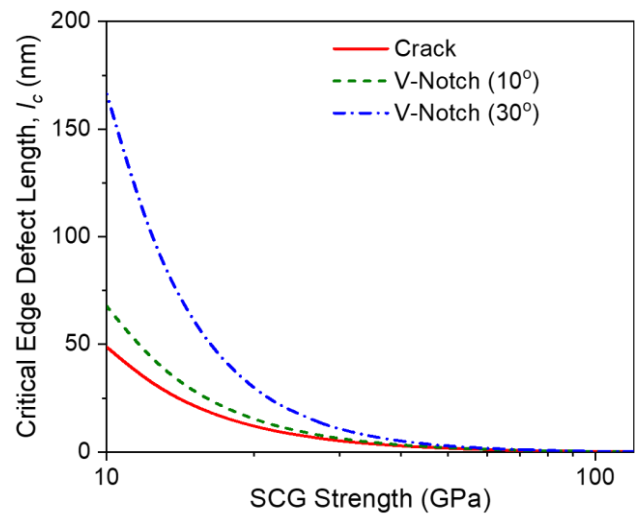

**Fig. S2: The predicted defect length on the edge of a macro-scaled SCG by QFM.** The variation of critical length of an edge defect, $l_c$, obtained by Eq. S5b for a sharp crack and 10⁰, and 30⁰ V-notches versus the corresponding reduced strength ranging from 10 to 130 GPa.

## 3. A Damage Evolution Model for Macroscale SCG

Given the macroscale (Fig. S19) and in-situ microscale (Fig. S21) observations of the failure mechanism of SCG- PC samples under tension, and the notable nonlinearity in back-calculated stress-strain of SCGs, we made a damage evolution model to describe the behavior of macroscale SCG on an ultra-thin polymeric substrate. The model aims to interpret the experimental observation of pronounced softening at strains below 6% owing to the simultaneous emergence and propagation of microcracks alongside the main macrocrack, since this notable softening at such low strains cannot be attributed to the intrinsic nonlinearity of the C-C bond.

## 3.1. Stress equilibrium in the SCG-PC bilayer system

Consider a bilayer that consists of a substrate of thickness $t_S$ and an SCG of thickness $t_G$ laid on it resulting in a volume fraction of $v_S = t_S/t_B$ and $v_G = t_G/t_B$ where $t_B = t_S + t_G$. Under a unidirectional applied tensile strain, $\varepsilon$, the force equilibrium along the loading gives the average stress carried by the SCG along the loading direction, $\bar{\sigma}_G$, as:

$$\bar{\sigma}_G(\varepsilon) = \frac{1}{v_G}(\sigma_B(\varepsilon) - v_S\sigma_S(\varepsilon)) \tag{S6}$$

where $\sigma_B$ and $\sigma_S$ are the stress in the bilayer and the substrate, respectively. Increasing the applied strain $\varepsilon$ causes a corresponding stress in each component through the components' constitutive law. Having the stress-strain curve of the bilayer and the bare substrate (without the SCG) obtained from individual tensile tests, one can back calculate the corresponding stress-strain curve of the SCG. Due to higher elastic modulus and ultimate strength of SCG compared to the substrate, it is expected that $\sigma_B(\varepsilon) > \sigma_S(\varepsilon)$, however, this reinforcement role disappears at the failure strain of SCG, $\varepsilon_{fG}$. Setting $\bar{\sigma}_G(\varepsilon_{fG}) = 0$ in Eq. S6, the failure strain of SCG must satisfy the following:

$$\sigma_B(\varepsilon_{fG}) - v_S\sigma_S(\varepsilon_{fG}) = 0 \tag{S7}$$

When $t_S \gg t_G$, then $v_S \cong 1$, and Eq. S7 simplifies to $\sigma_B(\varepsilon_{fG}) = \sigma_S(\varepsilon_{fG})$. In other words, the failure strain of SCG is where the stress-strain curves of the bare substrate and the bilayer intersect as schematically shown in Fig. S3a. Note that the term $\sigma_S(\varepsilon_{fG})$ in Eq. S7 imposes that the substrate stress-strain curve must be continued to $\varepsilon_{fG}$ which means the failure strain of the substrate needs to be greater than of the SCG, $\varepsilon_{fS} \geq \varepsilon_{fG}$. Otherwise, the substrate will fail before the SCG and since the one-atom thick macroscale SCG cannot stand alone, the failure of the substrate results in the failure of the SCG and the bilayer. It is concluded that a substrate with enough elongation should be selected since for $\varepsilon_{fS} < \varepsilon_{fG}$ the stress-strain curve of SCG stops at $\varepsilon = \varepsilon_{fS}$ and the stress-strain curve of SCG for $\varepsilon_{fS} < \varepsilon \leq \varepsilon_{fG}$ is missing. This undesired situation is schematically plotted in Fig. S3b.

Another possible scenario of failure is the case that the SCG suddenly fails at its ultimate strength, $\sigma_{uG}$, where $\varepsilon = \varepsilon_{uG}$ and its reinforcing role vanishes at once which means the whole load carried by the bilayer is transferred to the substrate. The substrate will fail if this load is higher than its maximum load-bearing capacity, $v_S\sigma_{us}$. The stress in bilayer at the failure strain of the SCG is:

$$\sigma_B(\varepsilon_{uG}) = (1 - v_S)\sigma_{uG} + v_S\sigma_s(\varepsilon_{uG}) \tag{S8}$$

The failure of the substrate needs $\sigma_B(\varepsilon_{uG})$ be greater than the ultimate load capacity of the substrate:

$$(1 - v_S)\sigma_{uG} + v_S\sigma_s(\varepsilon_{uG}) > v_S\sigma_{us} \tag{S9}$$

Solving Eq. S9 for $v_S$ yields a critical volume fraction, $v_{cS}$, as:

$$v_{cS} = \frac{\sigma_{uG}}{[\sigma_{us} - \sigma_s(\varepsilon_{uG})] + \sigma_{uG}} \tag{S10}$$

For $v_s < v_{cS}$, the failure of the SCG results in the failure of the substrate and the bilayer, however, for the fractions higher than the critical value, the substrate can bear the applied load alone to its ultimate strength, $\sigma_{us}$. Fig. S3c shows schematically the scenario of sudden failure of the SCG. The critical fraction corresponds to a critical thickness of the substrate, $t_{cS}$ as:

$$t_{cS} = \frac{v_{cS}}{1 - v_{cS}} t_G \tag{S11}$$

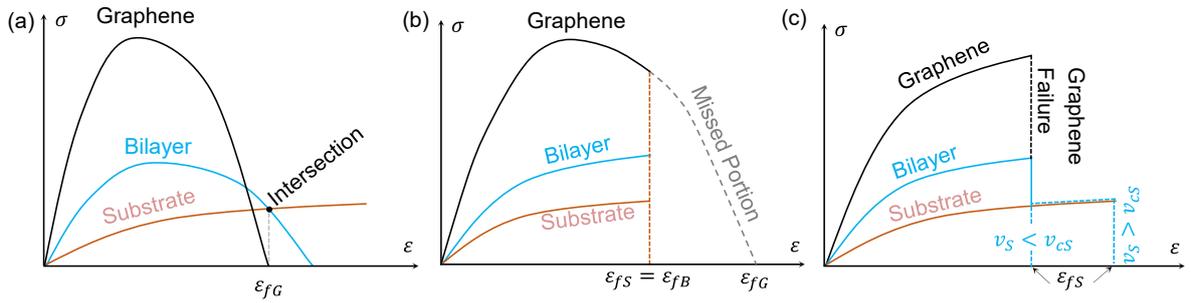

**Fig. S3:** Schematic of different failure scenarios of the SCG-substrate bilayer.

*3.2. The constitutive law of the macroscale SCG*

Assuming that the elongation of the substrate is greater than that of the SCG, one can extract the full stress-strain curve of the SCG by employing Eq. S11. The schematic of the extracted stress-strain curve of the SCG is shown in Fig. S4. By increasing the applied stain, prior to a proportional limit at $\varepsilon = \varepsilon_{lG}$, the constitutive law is assumed to be linear. Then, the nonlinear softening behavior becomes visible and continues to the breakage at $\varepsilon = \varepsilon_{uG}$ where the ultimate strength of SCG, $\sigma_{uG}$, is achieved. At this point the independent unidirectional load bearing of the SCG is impossible since $d\bar{\sigma}_G(\varepsilon)/d\varepsilon = 0$. The SCG on the substrate may (Fig. S4a) or may not (Fig. S4b) (on average) be able to keep its ultimate load bearing until a secondary ultimate strain, $\varepsilon = \varepsilon'_{uG}$. After that, the average load carried by the SCG drops gradually to be completely vanished at its failure strain, $\varepsilon = \varepsilon_{fG}$. Note that the post-breakage ($\varepsilon_{uG} \leq \varepsilon \leq \varepsilon_{fG}$) behavior of SCG is the reinforcing effect of broken SCG on the substrate which may not be observed for the case of an isolated macroscale SLG.

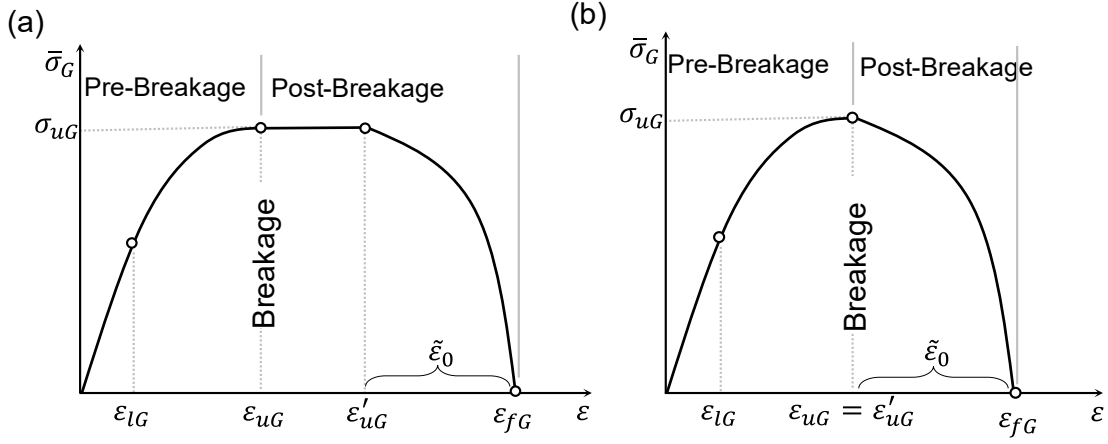

**Fig. S4: The schematic of back calculated stress-strain curve of an SCG.** (a) Constant load carrying after breakage. (b) No constant load standing after breakage.

At this stage, a constitutive law of SCG, which mathematically describes the stress-strain relationship in both pre- and post-breakage zones, is introduced.

*The pre-breakage zone* ($0 \leq \varepsilon \leq \varepsilon_{uG}$): The observed softening behavior is imposed to the unidirectional constitutive law by relating the tangent slope of the stress-strain curve to the applied strain through a power-law equation as:

$$d\bar{\sigma}_G(\varepsilon)/d\varepsilon = E_G(1 - (\hat{\varepsilon})^n) \tag{S12}$$

where the Young's modulus $E_G$ is the initial slope at $\varepsilon = 0$ and $\hat{\varepsilon} = \varepsilon/\varepsilon_u$ is the dimensionless strain in the pre-breakage zone, $0 \leq \hat{\varepsilon} \leq 1$. Note that Eq. S12 inherently satisfies a zero slope at $\varepsilon = \varepsilon_{uG}$. One can obtain the corresponding nonlinear constitutive law by integrating Eq. S12 as:

$$\bar{\sigma}_G(\hat{\varepsilon}) = E_G \varepsilon \left(1 - \frac{1}{n+1}(\hat{\varepsilon})^n\right) \tag{S13}$$

Considering that at the breakage point, $\hat{\varepsilon} = 1$, the stress is $\bar{\sigma}_G = \sigma_{uG}$, the Young's modulus is obtained as:

$$E_G = \frac{\sigma_{uG}}{\varepsilon_{uG}}\left(\frac{n+1}{n}\right) \tag{S14}$$

The proportional limit, $\varepsilon_l$, can be defined as the strain where the dimensionless deviation in the slope of stress-strain curve with respect to its initial value, $E_G$, remains under a threshold of $\delta\hat{E} = (slope - E_G)/E_G$. From Eq. S12:

$$\frac{d\bar{\sigma}_G(\varepsilon)/d\varepsilon}{E_G} = (1 - (\hat{\varepsilon})^n) \leq 1 - \delta\hat{E} \quad \text{for } 0 \leq \varepsilon \leq \varepsilon_{lG} \tag{S15}$$

Setting $\varepsilon = \varepsilon_l$ and equating the left and the right terms in Eq. S15, gives the proportional limit:

$$\frac{\varepsilon_{lG}}{\varepsilon_{uG}} = \hat{\varepsilon}_l = (\delta\hat{E}_G)^{\frac{1}{n}} \tag{S16}$$

For engineering applications, a deviation of $\delta\hat{E} = 0.04$ is acceptable for estimating the proportional limit.

Fig. S5 parametrically represents our power-law model for the pre-breakage zone. As seen, the power $n$, governs the shape of the stress-strain curve between the origin and the breakage point. For $n = 1$, a linear reduction in the slope happens (linear stiffness softening), the dimensionless proportional limit is as small as the assumed deviation, $\hat{\varepsilon}_l = \delta E_G$ and the stress-strain relationship obeys a parabolic curve. By increasing $n$, the reduction in the tangent stiffness (slope) lessens since the proportional limit is extended. In other words, the rate of softening in the stiffness from its initial value, $E_G$, at $\hat{\varepsilon} = 0$ to zero at $\hat{\varepsilon} = 1$, is delayed by increasing the power, $n$.

*The post-breakage zone* ($\varepsilon_{uG} \leq \varepsilon \leq \varepsilon_{fG}$): This zone represents the enhancing effect of the SCG on the substrate after its ultimate stress. If it keeps its load bearing, as shown in Fig. S4a, a horizontal portion is simply defined in the constitutive law as:

$$\bar{\sigma}_G(\varepsilon) = \sigma_{uG} \quad \text{for} \quad \varepsilon_{uG} \leq \varepsilon \leq \varepsilon'_{uG} \tag{S17}$$

otherwise, the horizontal part is omitted and $\varepsilon'_{uG} = \varepsilon_{uG}$.

For $\varepsilon'_{uG} \leq \varepsilon \leq \varepsilon_{fG}$, the average load carried by the SCG drops and finally vanished at $\varepsilon = \varepsilon_{fG}$. A dimensionless post-breakage strain, $0 \leq \tilde{\varepsilon} \leq 1$, is defined as:

$$\tilde{\varepsilon} = \frac{\varepsilon - \varepsilon'_{uG}}{\tilde{\varepsilon}_0} \quad \text{where} \quad \tilde{\varepsilon}_0 = \varepsilon_{fG} - \varepsilon'_{uG} \tag{S18}$$

The corresponding stress-strain relationship is suggested to be a power-law in the form of:

$$\bar{\sigma}_G(\tilde{\varepsilon}) = (1 - \tilde{\varepsilon})^m \sigma_{uG} \tag{S19}$$

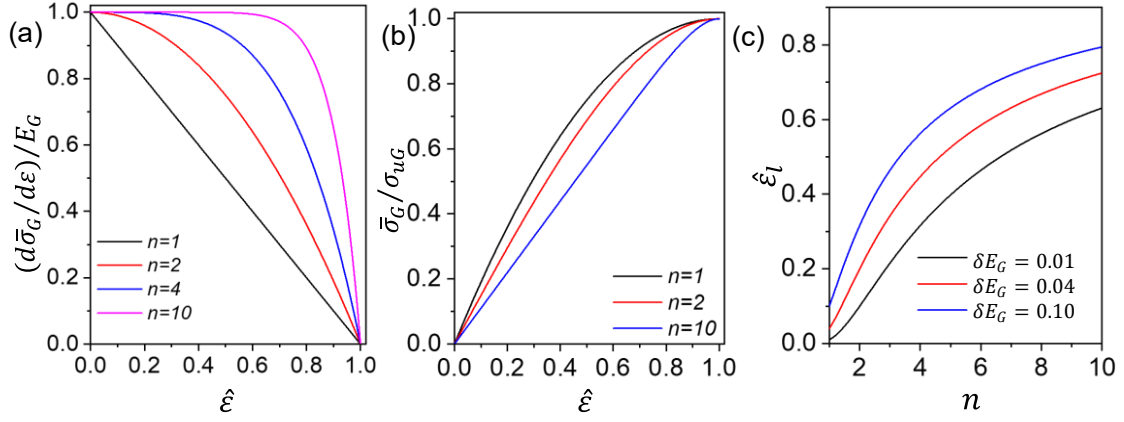

**Fig. S5: The parametric representation of this model for the pre-breakage zone.** (a) Variation of the tangent to initial slope ratio, $(d\bar{\sigma}_G/d\varepsilon)/E_G$, defined in Eq. S12, versus the dimensionless strain, $\hat{\varepsilon}$, for different powers, $n$. (b) Variation of the averaged stress of SCG to its ultimate strength ratio, $\bar{\sigma}_G/\sigma_{uG}$, defined in Eqs. S13 and S14, versus the dimensionless strain, $\hat{\varepsilon}$, for different powers, $n$. (c) Variation of the dimensionless proportional limit, $\hat{\varepsilon}_l$, versus the power $n$ for different deviations in the initial slope, $\delta E_G$.

The power (exponent) $m$ controls the reduction in the average stress carried by SCG from a maximum of $\bar{\sigma}_G = \sigma_{uG}$ at $\tilde{\varepsilon} = 0$ to $\bar{\sigma}_G = 0$ at $\tilde{\varepsilon} = 1$. Note that the term $(1 - \tilde{\varepsilon})$ is the remained portion of the post-breakage strain. For $m = 1$ a linear decreasing trend with a constant negative slope of $-\tilde{E}_G = -\sigma_{uG}/\tilde{\varepsilon}_0$ happens while $m < 1$ results in a nonlinear decrease with a downward concavity. The slope of the curve is obtained by taking the derivative of Eq. S19:

$$d\bar{\sigma}_G(\tilde{\varepsilon})/d\tilde{\varepsilon} = -\frac{m}{(1-\tilde{\varepsilon})^{(1-m)}}\tilde{E}_G \text{ where } \tilde{E}_G = \sigma_{uG}/\tilde{\varepsilon}_0. \tag{S20}$$

The starting slope is $-m\tilde{E}_G$ at $\tilde{\varepsilon} = 0$ and it approaches $-\infty$ at $\tilde{\varepsilon} = 1$ which corresponds to the complete failure of the SCG on the substrate. The effect of the power, $m$, on the reinforcing role is demonstrated in Fig. S6. It is seen that by approaching $m$ from one to zero, the amount of the starting slope $(-m\tilde{E}_G)$ decreases. It means that the reinforcing effect of the SCG can be kept further when $m \to 0$, although it results in a sharper removal of the reinforcing effect when the applied strain approaches to the failure strain of SCG, at $\tilde{\varepsilon} = 1$.

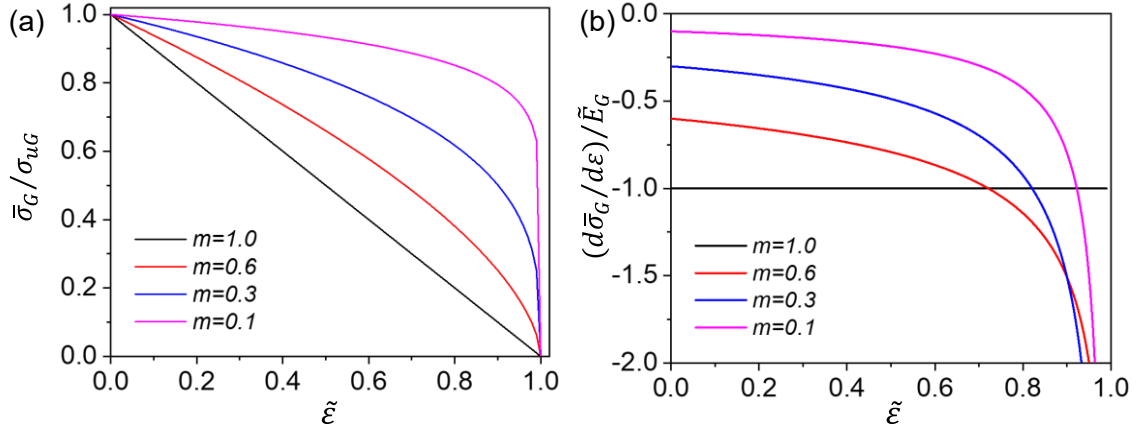

**Fig. S6: The parametric representation of the developed model for the post-breakage zone and $\varepsilon \geq \varepsilon'_{uG}$.** (a) Variation of the averaged stress carried by the SCG to its ultimate strength ratio, $\bar{\sigma}_G/\sigma_{uG}$, defined in Eq. S19, versus the dimensionless strain, $\tilde{\varepsilon}$, for different powers, $m$. (b) Variation of the tangent slope, $(d\bar{\sigma}_G/d\tilde{\varepsilon})$, to the assumed linear slope in the post-breakage zone, $\tilde{E}_G$, defined in Eq. S20, versus the dimensionless strain, $\tilde{\varepsilon}$, for different powers, $m$.

*3.3. The damage evolution variable of the macroscale SCG*

This subsection addresses a damage evolution model to establish a physical description for the softening behavior observed at the pre-breakage and the decrease in reinforcing effect in the post-breakage zone. The term *damage* encompasses a wide range of situations which causes a stiffness reduction and finally results in failure. Damage evolution models are often constructed by introducing a damage variable in the constitutive law of the material rooted in mechanistic processes within the material, such as the initiation and growth of voids, cavities, microcracks, and other nano-sized, microscopic, or mesoscopic defects. The general form of the modified unidirectional constitutive law of macroscale SCG including damage evolution can be defined as:

$$\bar{\sigma}_G(\varepsilon) = (1 - D(\varepsilon))E_G \varepsilon \qquad (S21)$$

where $D(\varepsilon)$ is the damage variable which may be a function of applied strain, generally. When $D = 0$ no damage is introduced and when $D = 1$ the material is completely failed. Comparing Eq. S21 to the developed stress-strain relationships in Eqs. S13 and S14, yields the corresponding damage variables, $D_1(\hat{\varepsilon})$, and $D_2(\tilde{\varepsilon})$ for pre- and post-breakage zones, respectively.

$$D_1(\hat{\varepsilon}) = \frac{1}{n+1}(\hat{\varepsilon})^n \qquad \text{for} \qquad 0 \leq \hat{\varepsilon} \leq 1 \qquad (S22)$$

$$D_2(\tilde{\varepsilon}) = 1 - \left(\frac{\sigma_{uG}}{E_G}\right)\left(\frac{\varepsilon'_{uG}}{\varepsilon_{uG}}\right)\frac{(1-\tilde{\varepsilon})^m}{\varepsilon} \qquad \text{for} \qquad 0 \leq \tilde{\varepsilon} \leq 1 \qquad (S23)$$

The two damage variables are equal to $D_u$ at their intersection at the breakage:

$$D_1(1) = D_2(0) = D_u = \frac{1}{n+1} \tag{S24}$$

Eq. S24 guarantees the continuity condition of the stress-strain law at the breakage point. For the case of keeping the ultimate stress in the strain range of $\varepsilon_{uG} \leq \varepsilon \leq \varepsilon'_{uG}$, as demonstrated in Fig. S4a, the damage variable is constant and equal to $D_u$. The damage parameter is $D_2 = 1$, at the failure strain, $\tilde{\varepsilon} = 1$.

### 3.4. The toughness modulus of SCG

Given the constitutive law, the toughness modulus of SCG can be determined by calculating the area under the stress-strain diagram as schematically shown in Fig. S7. In the pre-breakage area, denoted as $U_1$, the area under the stress-strain diagram is affected by the evolution of damage and the gradual deviation from linear behavior. This area can be divided into two components: stored elastic energy density, $U_1^{(e)}$, and dissipated energy density, $U_1^{(d)}$. By integrating Eq. S13 within the pre-breakage range, $0 \leq \hat{\varepsilon} \leq 1$, these two terms of strain energy density can be calculated as follows,

$$U_1^{(e)} = \frac{1}{2} \sigma_{uG} \varepsilon_{uG} \tag{S25}$$

$$U_1^{(d)} = \frac{1}{(n+1)(n+2)} \sigma_{uG} \varepsilon_{uG} \tag{S26}$$

Note that as $n$ increases, the damage evolution diminishes, see Eq. S24, leading to a reduction in the dissipative term in Eq. S26. The area under the curve in the post-breakage range is determined as follows by integrating Eq. S14 (and Eq. S17 in case a plastic plateau is observed, $\varepsilon_{uG} \neq \varepsilon'_{uG}$), indicating the strain energy density associated with the effective axial stress carried by the fragments of broken SCG.

$$U_2 = \sigma_{uG}\left((\varepsilon'_{uG} - \varepsilon_{uG}) + \frac{\varepsilon_{fG} - \varepsilon'_{uG}}{m+1}\right) \tag{S27}$$

Fig. S8 plots the variation of $U_1^{(d)}$ and $U_2$ versus $n$ and $m$. The toughness modulus of macroscale SCG, $T_G$, is calculated as the total area under the curve, as,

$$\begin{aligned} T_G = U_{tot} &= U_1^{(e)} + U_1^{(d)} + U_2 \\ &= \sigma_{uG}\varepsilon_{uG}\left(\frac{1}{2} + \frac{1}{(n+1)(n+2)}\right) + \sigma_{uG}\left((\varepsilon'_{uG} - \varepsilon_{uG}) + \frac{\varepsilon_{fG} - \varepsilon'_{uG}}{m+1}\right) \end{aligned} \tag{S28}$$

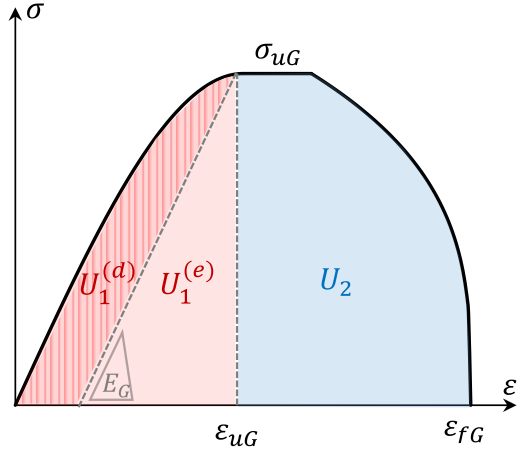

**Fig. S7: The schematic visualization of strain energy density terms.** $U_1$ is the area under the stress-strain curve in pre-breakage zone, divided into the stored elastic, $U_1^{(e)}$, and the dissipated energy density, $U_1^{(d)}$, while $U_2$, as the area under the stress-strain curve in post-breakage zones, reflects the strain energy density corresponds to the effective axial stress carried by the fragments of broken SCG. The toughness modulus is defined as the entire area under the curve.

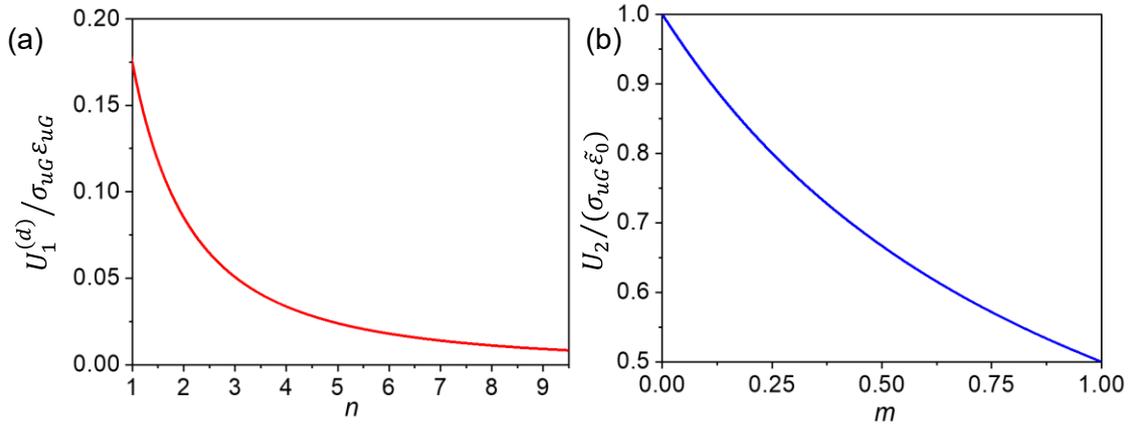

**Fig. S8: The variation of strain energy density terms.** (a) Variation of dissipated strain energy density in the pre-breakage zone, $U_1^{(d)}$, versus the fitting power, $n$. (b) Variation of dissipated strain energy density in the post-breakage zone versus the fitting power, $m$.

## 3.5. The calibrated damage evolution model for centimeter-scale SCG

Here, using the average mechanical properties of chiral, zigzag, and armchair centimeter-scaled SCGs, listed in Tables S1, the proposed damage evolution model is calibrated. This is done by calculating the exponent $n$ for every configuration as (see Eq. S14),

$$n = \frac{\sigma_{uG}}{E_G \varepsilon_{uG} - \sigma_{uG}} \tag{S29}$$

The calculated values of $n$ from Eq. S39 and the calibrated $m$ values on the post-breakage curves are provided in Table S1. One can evaluate the terms $U_1$ and $U_2$ in Eq. S28 to calculate the toughness of SCGS in the pre- and post-breakage zone, as presented in Table S1 for chiral, zigzag, and armchair SCGs. The damage evolution parameters listed in Table S1 yields the closed-form full stress-strain constitutive law of SCG via Eq. S21 as,

$$\sigma_G(\varepsilon) = 748.9\varepsilon(1 - D(\varepsilon))[\text{GPa}] \quad \text{for chiral,} \tag{S30a}$$

$$\sigma_G(\varepsilon) = 1109\varepsilon(1 - D(\varepsilon))[\text{GPa}] \quad \text{for zigzag,} \tag{S30b}$$

$$\sigma_G(\varepsilon) = 1009\varepsilon(1 - D(\varepsilon))[\text{GPa}] \quad \text{for armchair,} \tag{S30c}$$

where the corresponding damage evolution variable in the pre-breakage zone is calculated by Eq. S22 as,

$$D(\varepsilon) = \frac{1}{3.273}\left(\frac{\varepsilon}{0.0453}\right)^{2.273} \quad \text{for chiral, } 0 \leq \varepsilon \leq 0.0453 \tag{S31a}$$

$$D(\varepsilon) = \frac{1}{1.698}\left(\frac{\varepsilon}{0.0601}\right)^{0.698} \quad \text{for zigzag, } 0 \leq \varepsilon \leq 0.0601 \tag{S31b}$$

$$D(\varepsilon) = \frac{1}{2.187}\left(\frac{\varepsilon}{0.0369}\right)^{1.187} \quad \text{for armchair, } 0 \leq \varepsilon \leq 0.0369 \tag{S31c}$$

and the corresponding damage variable in the post-breakage zone are calculated from Eq. S23 as,

$$D(\varepsilon) = 1 - \frac{23.57}{748.9}\frac{\left(1 - \frac{\varepsilon - 0.0453}{0.0769 - 0.0453}\right)^{0.39}}{\varepsilon} \quad \text{for chiral, } 0.0453 \leq \varepsilon \leq 0.0769 \tag{S32a}$$

$$D(\varepsilon) = 1 - \frac{27.40}{1109}\frac{\left(1 - \frac{\varepsilon - 0.0601}{0.0779 - 0.0601}\right)^{0.42}}{\varepsilon} \quad \text{for zigzag, } 0.0601 \leq \varepsilon \leq 0.0779 \tag{S32b}$$

$$D(\varepsilon) = 1 - \frac{20.21}{1009}\frac{\left(1 - \frac{\varepsilon - 0.0369}{0.1010 - 0.0369}\right)^{0.67}}{\varepsilon} \quad \text{for armchair, } 0.0369 \leq \varepsilon \leq 0.1010 \tag{S32c}$$

*3.6. The stress-strain of a stack made of folded SCG*

Here, based on the developed analytical model, a closed-form expression for predicting the stress-strain relationship of a stack made by folding SCG-PC samples is provided. This stack is composed of $N$ layers of PC and $N$ layers of SCG among them. While two SCG layers at the top and the bottom layers experience the edge damage due to cutting, the remaining layers possess rounded edges which are free from cutting defects, as schematically shown in Fig. S9a. First, we define the stress-strain behavior of all three components in the stack, i.e., the PC layers, the cut SCG, and the folded SCG. Then, we use them to estimate the stress-strain curve of the stack. The stress-strain equation governing PC layers can be derived using the current development damage model, similar to what we did for SCG. By utilizing the experimental tensile test results on 200-nm thick PC samples, which are provided in Table S4, we can determine the stress sustained by PC as a function of applied strain as,

$$\sigma_s(\varepsilon) = 2010\varepsilon \left(1 - \frac{1}{3.57}\left(\frac{\varepsilon}{0.048}\right)^{2.57}\right) \text{[MPa] for } 0 \leq \varepsilon \leq 0.048, \quad \text{(S33)}$$

while it keeps the maximum stress of 69.44 MPa after the yielding at $\varepsilon = 0.048$ until the failure strain, see Fig. S9b. The stress carried by chiral SCGs with damaged edges from cutting is calculated using Eq. S30a, yielding a strength of 23.57 GPa at 4.5% strain. In contrast, for folded SCGs with rounded edges, in an ideal scenario the damage is entirely eliminated. It can be considered in the constitutive law by setting the damage variable to zero in Eq. S30a, leading to a strength enhancement, reaching as high as 33.7 GPa at 4.5% strain. Fig. S9c plots the stress-strain curves for chiral SCGs with cut edges and folded edges. The maximum stress carried by an ideal defect-free SCG with a Young's modulus of 1 TPa and a third-order modulus of -2 TPa ($\sigma = \varepsilon - 2\varepsilon^2$ TPa) is also plotted as the theoretical upper limit.

As a parallel layered system, the stress carried by the stack can be calculated as the sum of stress carried by the components:

$$\sigma_{stack}(\varepsilon) = v_{G1}\sigma_{G1}(\varepsilon) + v_{G2}\sigma_{G2}(\varepsilon) + v_s\sigma_s(\varepsilon) \quad \text{(S34)}$$

where subscripts 1 and 2 indicate the SCG with a cut edge and folded edges, respectively. In Eq. S34, $v_{G1} = 2(t_G/t_T)$, $v_{G2} = (N-2)(t_G/t_T)$, and $v_S = N(t_S/t_T)$, are the corresponding volume fractions of the components, where $t_G$, and $t_S$ are the thickness of the SCG and a single substrate layer, respectively, and $t_T = N(t_G + t_S)$ is the total thickness of the stack.

For a SCG-PC sample with a nominal thickness of 200 nm, the volume fraction of SCG is consistently 0.17% across stacks of varying layer numbers, assuming no voids and $t_G = 0.34$ nm. However, as the number of layers, $N$, increases, the portion of SCGs with significant edge defects decreases. For example, with $N=4$, 50% of the SCGs (2 out of 4 layers) have cutting-induced edge defects, while with $N=100$, this reduces to 2%. The predicted stress-strain curves of the stacks made by folding 200 nm chiral SCG-PC samples, obtained from Eq. S34, is presented in Fig. S9d for $N= 4$, and 100, and is compared to the bare PC, the original SCG-PC before folding, and an ideal SCG-PC sample of the same thickness including an ideal defect-free SCG. As seen, increasing $N$ leads to a higher portion of rounded edges, thereby reducing the overall damage in the stack and enhancing its strength. Consequently, at the failure strain of chiral SCG, 4.53%, the stack with 100 layers sustains a stress of 125.15 MPa, representing an 80% and 19% increase in strength compared to PC and bilayer SCG-PC, respectively.

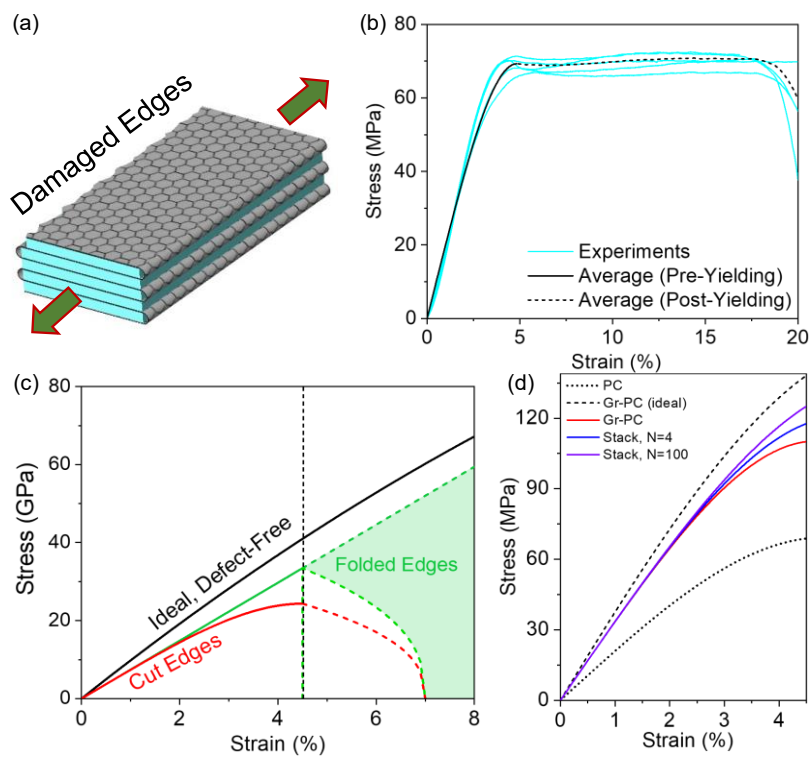

**Fig. S9: The stress-strain diagrams of the components of a SCG-PC stack.** (a) The schematic of an $N$-layer stack made by folding SCG-PC samples. The edges under cutting experience significant damage, while the folded edges are assumed to be free of cutting-edge damage. (b) The average stress-strain diagram of 200 nm PC layer, presented in Eq. S33, compared to the experimentally measured ones. (c) Comparison between the stress-strain diagram of SCG with edges damaged by cutting, evaluated from Eq. S30a and that of folded SCG with rounded edge estimated by setting the damage variable, $D$, to zero in Eq. S30a. The ideal defect-free, infinite SCG with a Young's modulus of 1 TPa and a third-order modulus of -2 TPa ($\sigma = \varepsilon - 2\varepsilon^2$ TPa) is plotted as the upper limit. (d) The predicted stress-strain curves of the stacks ($N= 4$, and 100) made by folding 200 nm SCG-PC samples and comparison to the PC, the original SCG-PC before folding, and an ideal SCG-PC sample of the same thickness including an ideal defect-free SCG.

*3.7. Microcrack-macrocrack interaction hypothesis as a toughening mechanism in Macro-scale SCGs*

In brittle materials like concrete, rocks, and ceramics, resistance to macrocrack propagation can be linked to the shielding effect created by microcracks surrounding the macrocrack tip. Building on this concept, we hypothesize that microcracks may similarly act as a shield against macrocrack growth in centimeter-scale SCG on an ultra-thin PC substrate. The microcrack density eventually reaches a saturated value, $\omega = \omega_s$. The effective elastic modulus transitions smoothly from its initial value, $E_G$, in the undamaged state, to $E_{sG}$ at microcrack saturation, as illustrated in the schematic stress-strain curve shown in Fig. S10a(*8*). When the macrocrack tip is surrounded by this localized zone saturated with microcracks, the stress intensity factor is reduced due to this softening, known as microcrack toughening. Assuming isotropy, the Poisson's ratio of 1/3, and a dilute self-consistent model for relating the defect density to softening, it is shown(*8, 41-43*) that the shielding factor is roughly proportional to the softening ratio:

$$\frac{\sigma_{tip}(\omega_s)}{\sigma_{tip}(\omega=0)} = \frac{K_{tip}(\omega_s)}{K_{tip}(\omega=0)} \approx \frac{E_{sG}}{E_G} \tag{S35}$$

Fig. S10b illustrates this behavior in a typical stress-strain curve for our SCGs, derived from the chiral SCG-PC sample #4 and PC sample #1 using Eq. S6. The similarity to Fig. S10a supports the hypothesis of a microcrack toughening mechanism in SCG damage evolution during macrocrack growth.

Although more detailed studies are required for an accurate quantitative evaluation, to roughly estimate the shielding effect, we analyzed 25 stress-strain diagrams for chiral SCG, combining data from five 200 nm SCG-PC samples (Table S5) and five 200 nm PC samples (Table S4). The stiffness in the microcrack saturation region, $E_{sG}$, was estimated by measuring the slope of the stress-strain curves within the strain range of 0.025 to 0.035. On average, the ratio of this saturated stiffness to the initial modulus, $E_G$, was 0.58 ± 0.12. Applying this value to Eq. S35 suggests a 58% reduction in the stress intensity factor.

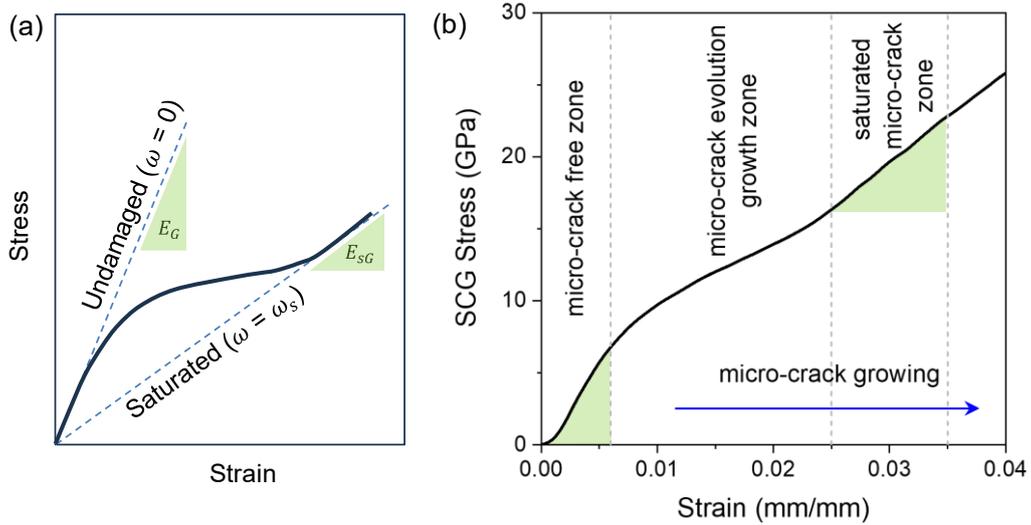

**Fig. S10: The Microcrack-Macrocrack Interaction.** (a) A schematic stress-strain curve in case of microcrack saturation(*40*). (b) The evidence of microcrack initiation and saturation and its softening effect in a sample back-calculated stress-strain curve of SCGs (extracted from SCG-PC sample #4 and PC sample #1). Microcrack saturation can act as a shield by softening the area around the macrocrack tip, reducing the stress intensity factor.

## 4. Role of Polymer Thickness in Calculating SCG Mechanics

To gain a deeper insight into SCG mechanics, we conducted the tensile test measurements on additional SCG-PC and bare PC samples with varying PC thicknesses (80 nm, 150 nm, and 300 nm, see Fig. S17 and S18 for stress-strain curves). By comparing the stress-strain responses of bare PC and SCG-PC samples, we extracted the mechanical properties of the SCG. Detailed calculations for all thicknesses are provided in Table S7-S9. The calculated results of $E$ and $\sigma$ of SCG for all four different PC thicknesses (including 200 nm) are compared in Fig. S11a. The PC thickness significantly impacts the extracted intrinsic mechanical properties ($E$ and $\sigma$) of SCG. For example, the 80 nm SCG-PC samples exhibited an average Young's modulus and strength approximately half that of the 200 nm samples (374.5 GPa and 10.66 GPa, respectively). The 150 nm samples showed 86% of $E$ and $\sigma$ obtained from the 200 nm samples. This suggests that a thicker PC substrate enhances the measured mechanical properties of SCG-PC samples, leading to extracted SCG values closer to the theoretical limits(*6, 29, 34*). One can ask: Perhaps degradation of SCG quality during the transfer process could explain the inferior mechanical properties observed for lower PC thickness? But as described next, the answer is "no".

We investigated the quality of SCG in SCG-PC samples across all four PC thicknesses (80, 150, 200, and 300 nm). After transferring these samples onto Si/SiO$_2$ wafers, we performed Raman spectroscopic mapping over a large area (9 mm x 2.2 mm) to assess SCG quality (Fig. S30-S33). The resulting Raman maps revealed that the D peak (~1350 cm$^{-1}$), indicative of structural defects, was confined exclusively to the edges of the SCG-PC samples, while the majority of the area exhibited no D peak. This observation suggests that SCG quality remained largely consistent across SCG-PC samples with different PC thicknesses, and that the thickness-dependent intrinsic mechanical properties of the PC layer played a crucial role in determining the mechanical response of SCG.

Analysis of stress-strain curves (Fig. S11b) showed that SCG exhibits a consistent failure strain (4.45-4.62%) across various PC thicknesses, indicating uniform SCG quality that is independent of PC thickness. However, crack initiation strain in SCG increased with PC thickness (3.0% for 80 and 150 nm, 3.75% for 200 nm, 5.25% for 300nm, Fig. S11b). Optical images (Fig. S20, Fig. S27-S29) suggest that the variation in mechanical response depends on intrinsic PC mechanical properties. For 200 nm PC films, $E$ and $\sigma$ are notably higher (2.01 GPa and 69.44 MPa, approaching bulk values) compared to 80 nm (0.77 GPa and 20.35 MPa) and 150 nm (1.07 GPa and 28.56 MPa) PC films, (Fig. S11c-d). This variation in mechanical responses with PC film thickness could be attributed to 1) the degree of entanglement in the polymer, 2) higher chain mobility at the surface, 3) surface roughness or non-uniform thickness, and 4) the degree of deformation (shear deformation or 2D/3D crazing)(*44-49*). The surface roughness of the SCG-PC samples with varying PC thickness is compared in Fig. S34. The roughness dropped from 1.97 nm to 0.34 nm for an increase in thickness from 150 nm to 200 nm.

We thus found a PC film thickness (200 nm) that maximized the performance of SCG-PC composites, for the 4 thicknesses we tested. We found that while PC films with thicknesses of 200 nm and 300 nm exhibit similar $E$ and $\sigma$ to bulk PC ($E$ = 2.3-2.4 GPa and $\sigma$ = 55-65 MPa)(*50*), thinner films (80 nm and 150 nm) have lower $E$ and $\sigma$ values and rougher surfaces. This roughness reduces the contact area between the PC and SCG, hindering effective stress transfer and resulting in lower mechanical performance for the SCG-PC composite.

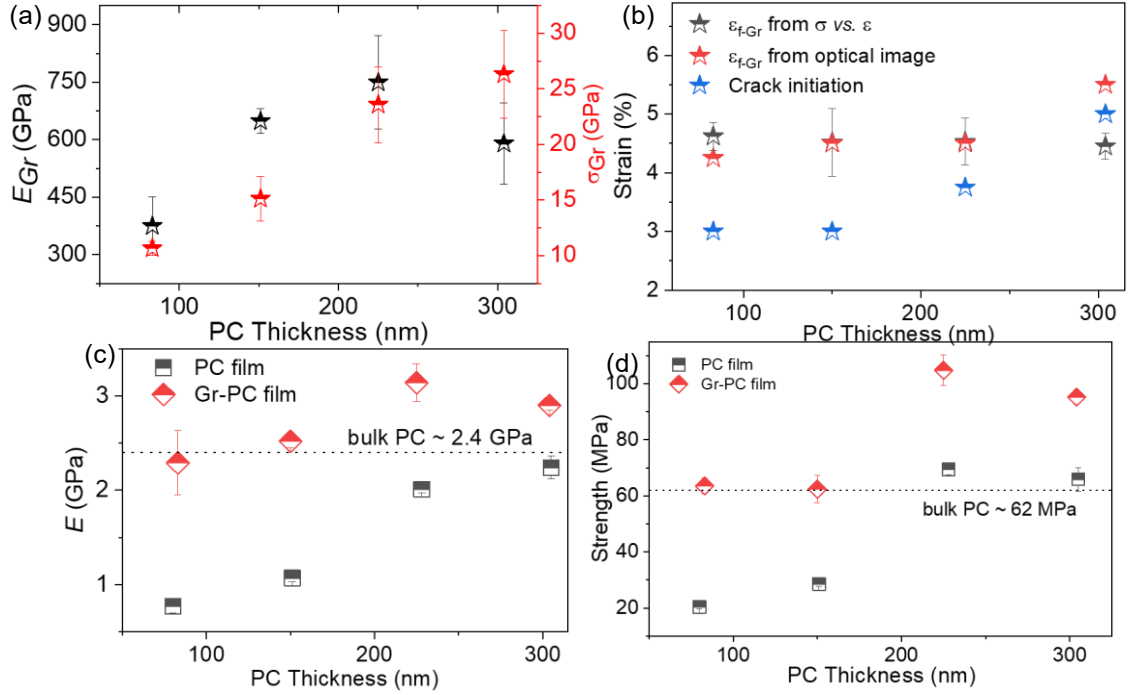

**Fig. S11: Origin of graphene mechanical properties with variation of PC thickness.** (a) Young's modulus and tensile strength of SCG extracted by comparing stress-strain plots of SCG-PC and PC films with varying PC thickness. (b) Comparison of different strain values for SCG, SCG-PC, and PC films for different PC thicknesses. (c) Young's modulus and (d) tensile strength of PC and SCG-PC films. The edge orientation of the SCG samples for all the thicknesses are chiral.

Despite both 200 nm and 300 nm thick PC films having a similar $E$ and $\sigma$ to bulk PC, the 300 nm PC film led to a lower $E$ (589.9 ± 105.7 GPa) and a higher $\sigma$ (26.31 ± 3.93 GPa) for the SCG compared to the 200 nm PC film. This seemingly contradictory behavior arises from the interplay between stress transfer and stress relaxation. Stress relaxation tests (Fig. S40) reveal significant relaxation for the 300 nm PC film, particularly at 4% strain. Thicker PC films allow greater mobility of polymer chains, enabling them to rearrange and dissipate stress more rapidly(*50*). While rapid stress relaxation in the 300 nm PC film allows it to absorb more strain, it also reduces stress transfer to the SCG, limiting its contribution to the composite's overall stress. Essentially, the PC film "softens" under stress, leaving less strain for the SCG. Conversely, the 200 nm PC film achieves a balance between effective stress transfer to the SCG and moderate stress relaxation, maximizing the composite's stiffness. Therefore, the 200 nm PC layer provides the optimal balance, effectively transferring stress to the graphene to enhance the composite's strength without excessive relaxation, which would weaken the overall system by reducing its stress.

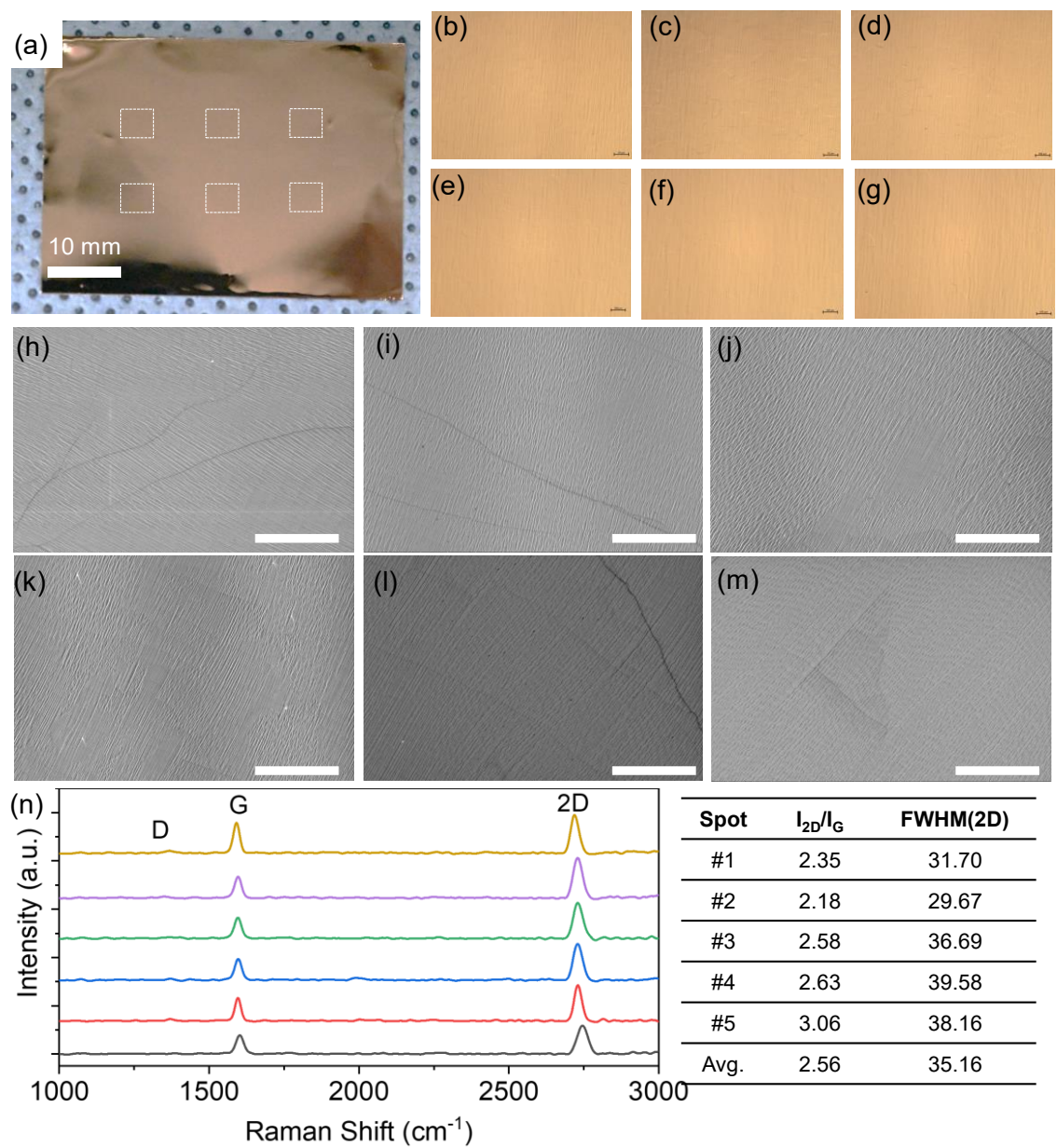

**Fig. S12: Characterization of as-grown single crystal monolayer graphene (SCG) on 50 μm Cu(111) foil.** (a) Optical image displaying a typical 5 cm x 4 cm Cu(111) foil with single crystal single-layer graphene. (b-g) Optical micrographs of graphene/Cu(111) foil corresponding to six different marked regions in (a). Scale 200 μm. (h-m) Scanning electron microscopy (SEM) images of the graphene/Cu(111) foil captured from six different regions, illustrating uniform single-layer graphene with few parallel folds. Scale 10 μm. (n) Raman spectra of the single-layer graphene on Cu(111) shows high quality graphene with an $I_{2D}/I_G$ of 2.56 and $FWHM(2D)$ of 35.2 cm$^{-1}$. No D peak (~1350 cm$^{-1}$) confirms the quality of graphene. The table at bottom right shows the calculated $I_{2D}/I_G$ and $FWHM(2D)$ from six different spots.

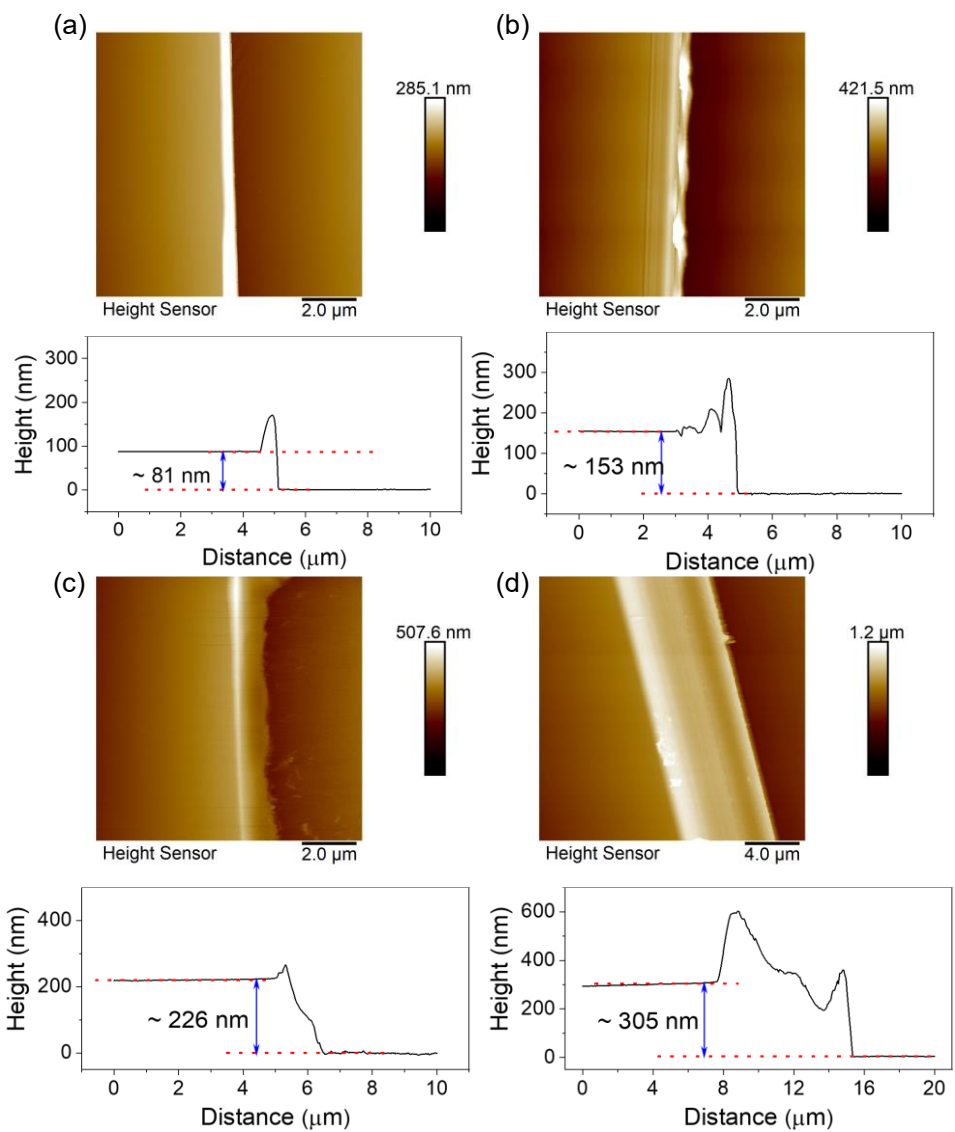

**Fig. S13: Determination of PC Film Thickness.** Atomic Force Microscopy (AFM) height images and respective height profiles illustrating the thickness of polycarbonate (PC) films with varying concentrations (0.5 - 5 wt%) and coating speeds. The film thicknesses for (a) 80 nm, (b) 150 nm, (c) 200 nm, and (d) 300 nm were controlled during fabrication.

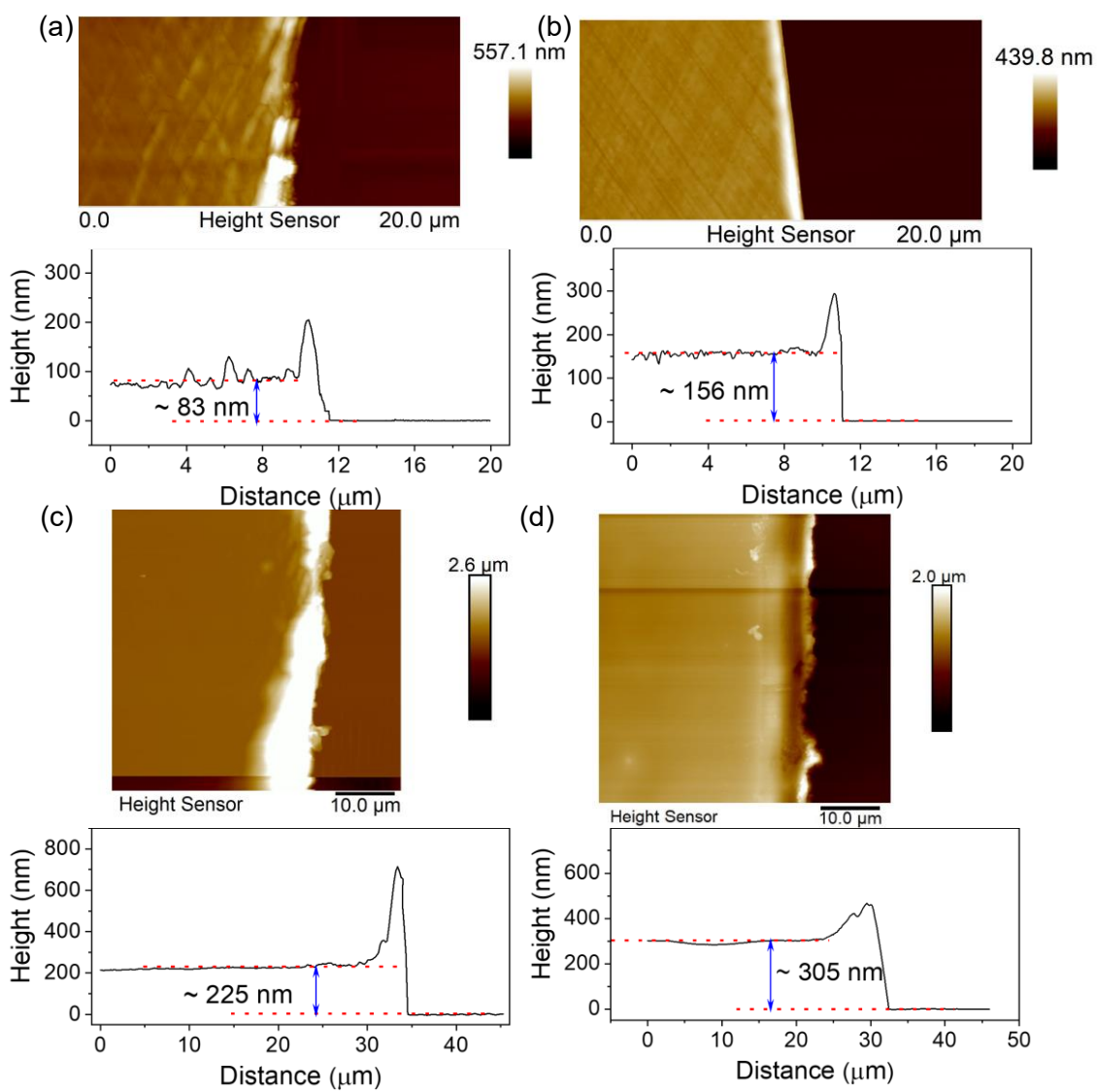

**Fig. S14: Determining SCG-PC film thickness.** Atomic Force Microscopy (AFM) height image and corresponding height profile depicting the thickness of SCG-PC films for (a) 80 nm and (b) 150 nm, (c) 200 nm, and (d) 300 nm PC films.

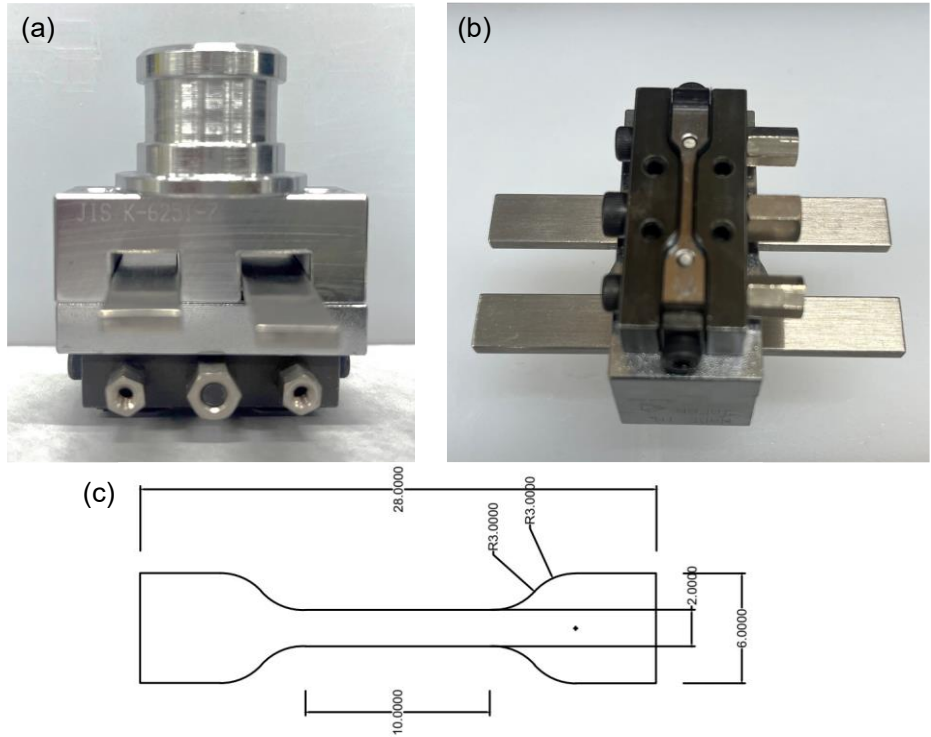

**Fig. S15: Mechanical cutter (Model No. SDMP-1000; Dumbell Co. Ltd.).** (a) Side and (b) Top view of the cutter. (c) Dimensions of dogbone cutter in millimeter scale.

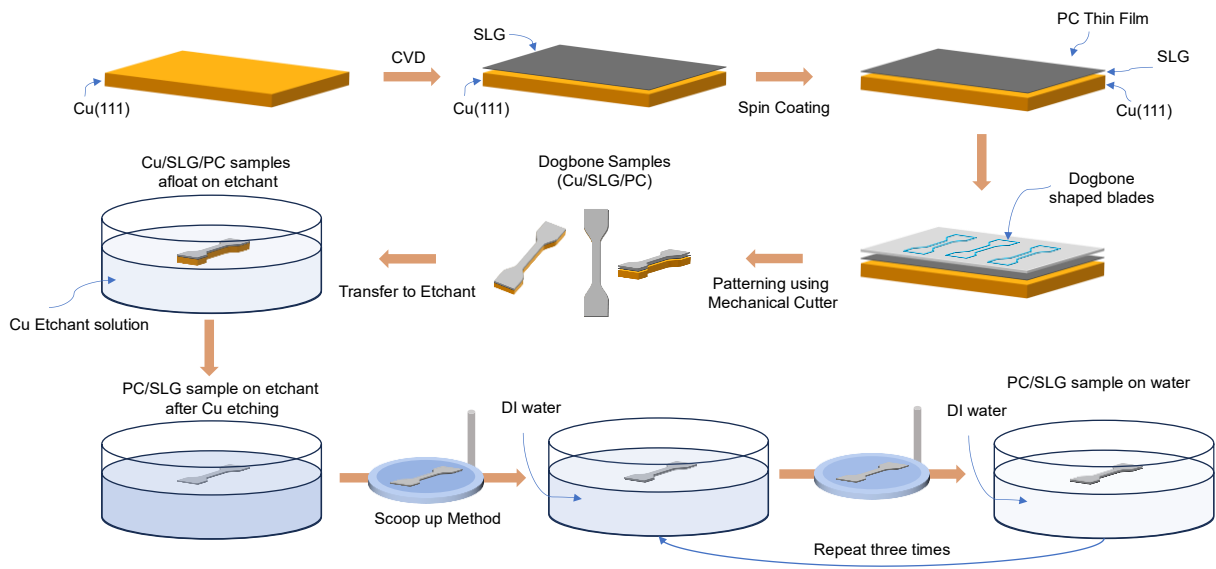

**Fig. S16:** Schematic representation of the process flow for preparation of "dogbone" samples for tensile test measurement in FOW system.

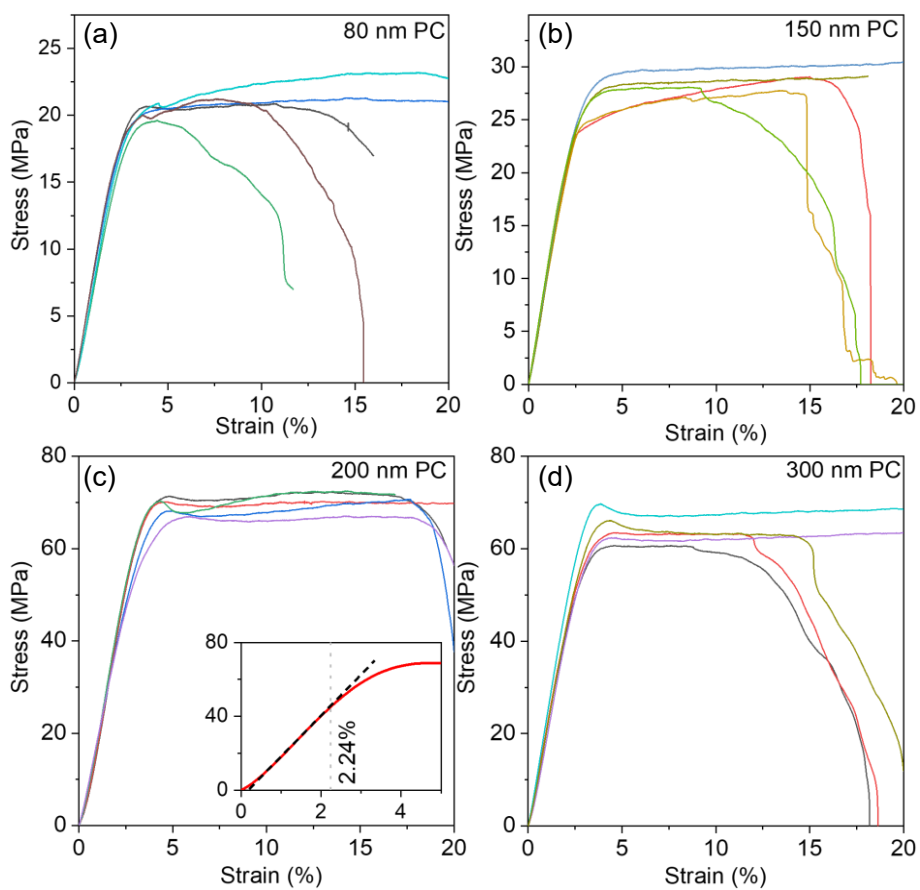

**Fig. S17: Stress-Strain Curves of PC films.** Stress-strain profiles for PC films of various thicknesses (a) 80 nm, (b) 150 nm, (c) 200 nm, and (d) 300 nm measured in the FOW system. Inset of (c) denotes the strain at which linear to non-linear transition occurs at 2.24% strain for 200 nm PC.

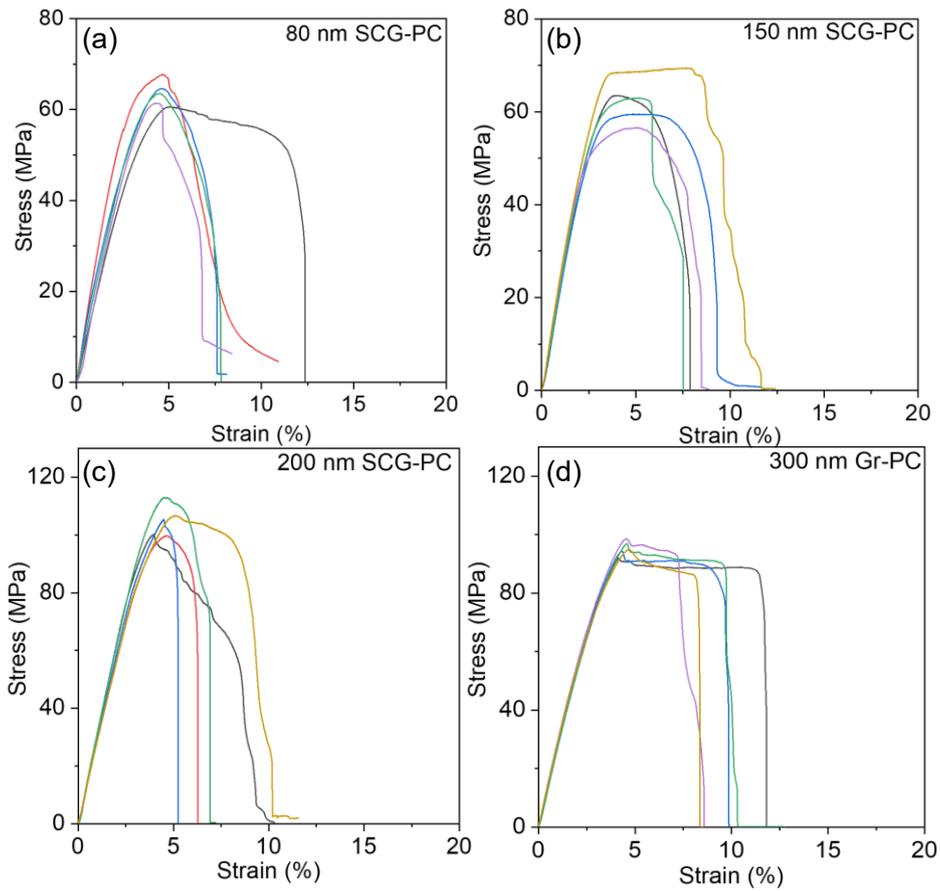

**Fig. S18: Stress-Strain Curves of SCG-PC films.** Stress-strain profiles for SCG-PC films of various thicknesses (a) 80 nm, (b) 150 nm, (c) 200 nm, and (d) 300 nm measured in the FOW system.

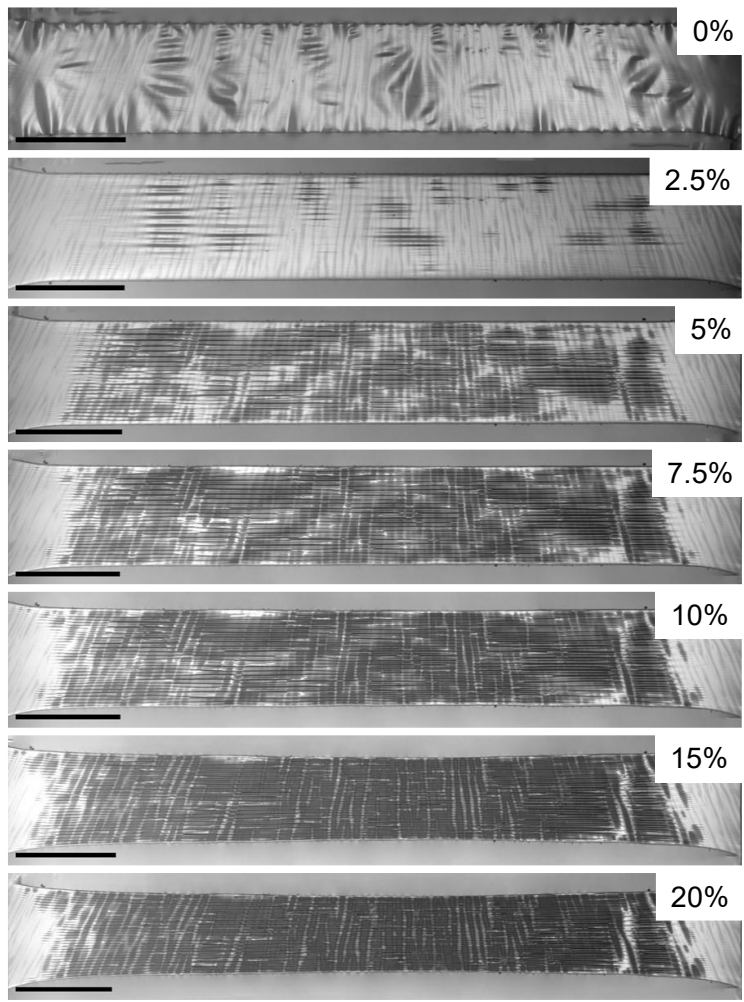

**Fig. S19: Evolution of 200 nm PC Films under Tensile Loading.** Monochromatic optical images showcasing the progressive changes in 200 nm PC films subjected to uniform tensile loading mentioned in inset. Scale bar ~ 2 mm.

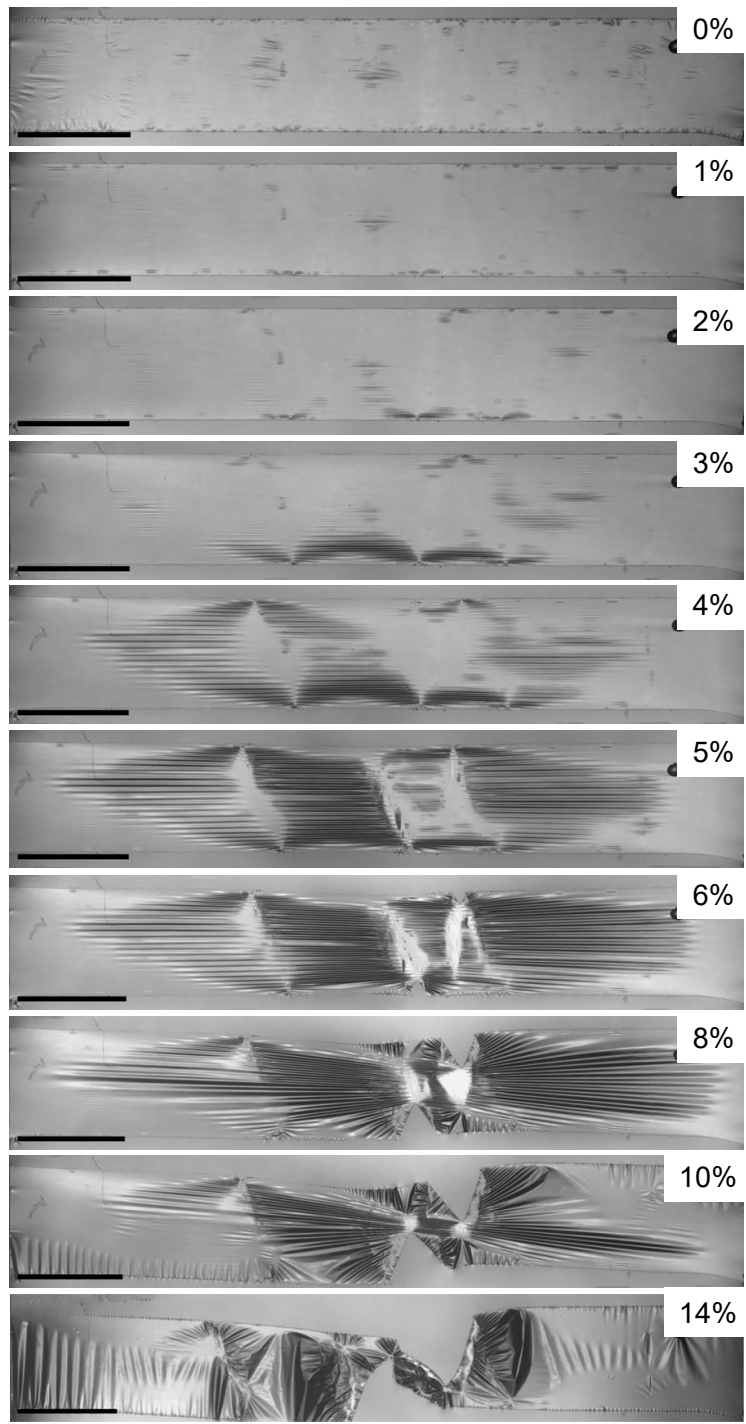

**Fig. S20: Evolution of 200 nm SCG-PC Films under Tensile Loading.** Monochromatic optical images showcasing the progressive changes in 200 nm SCG-PC films subjected to uniform tensile loading mentioned in inset. Scale bar ~ 2 mm.

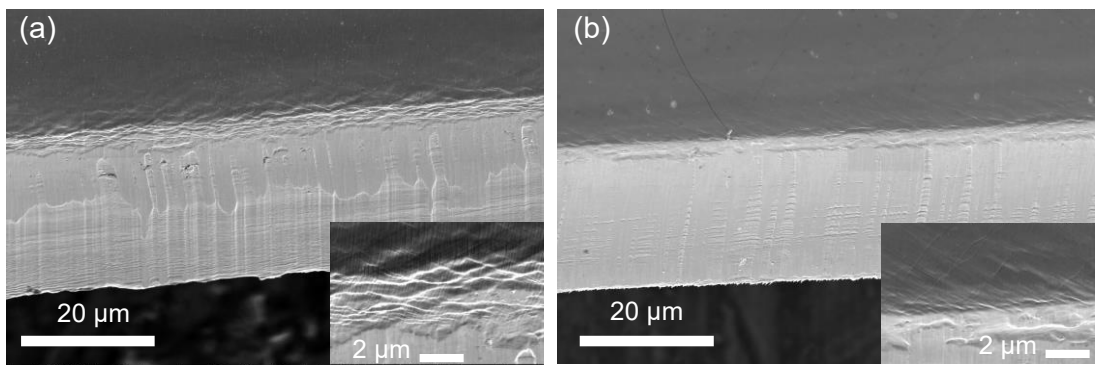

**Fig. S21: SEM images of edge defects formed SCG/Cu(111) Edges.** SEM images of SCG/Cu(111) edges after dogbone shape fabrication in (a) 80 nm and (b) 200 nm SCG-PC films, reveal defects in graphene, particularly in thinner PC layers.

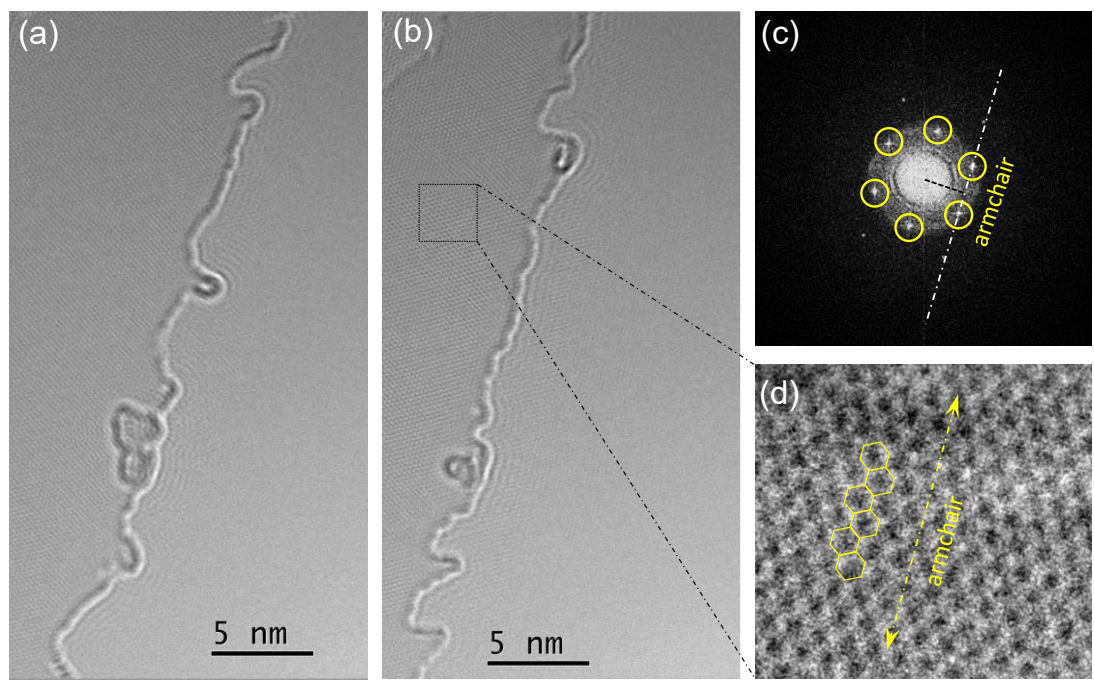

**Fig. S22: Transmission Electron Microscopy (TEM) Micrographs of SCG Edges.** (a-b) TEM images of SCG edges, where the SCG samples were cut parallel to the Cu(111) rolling marks. Nanoscale edge defects are at the SCG edges. (c) Diffraction pattern of the SCG sample. (d) High resolution TEM image shows the atomic arrangement and armchair orientation of SCG along the edge.

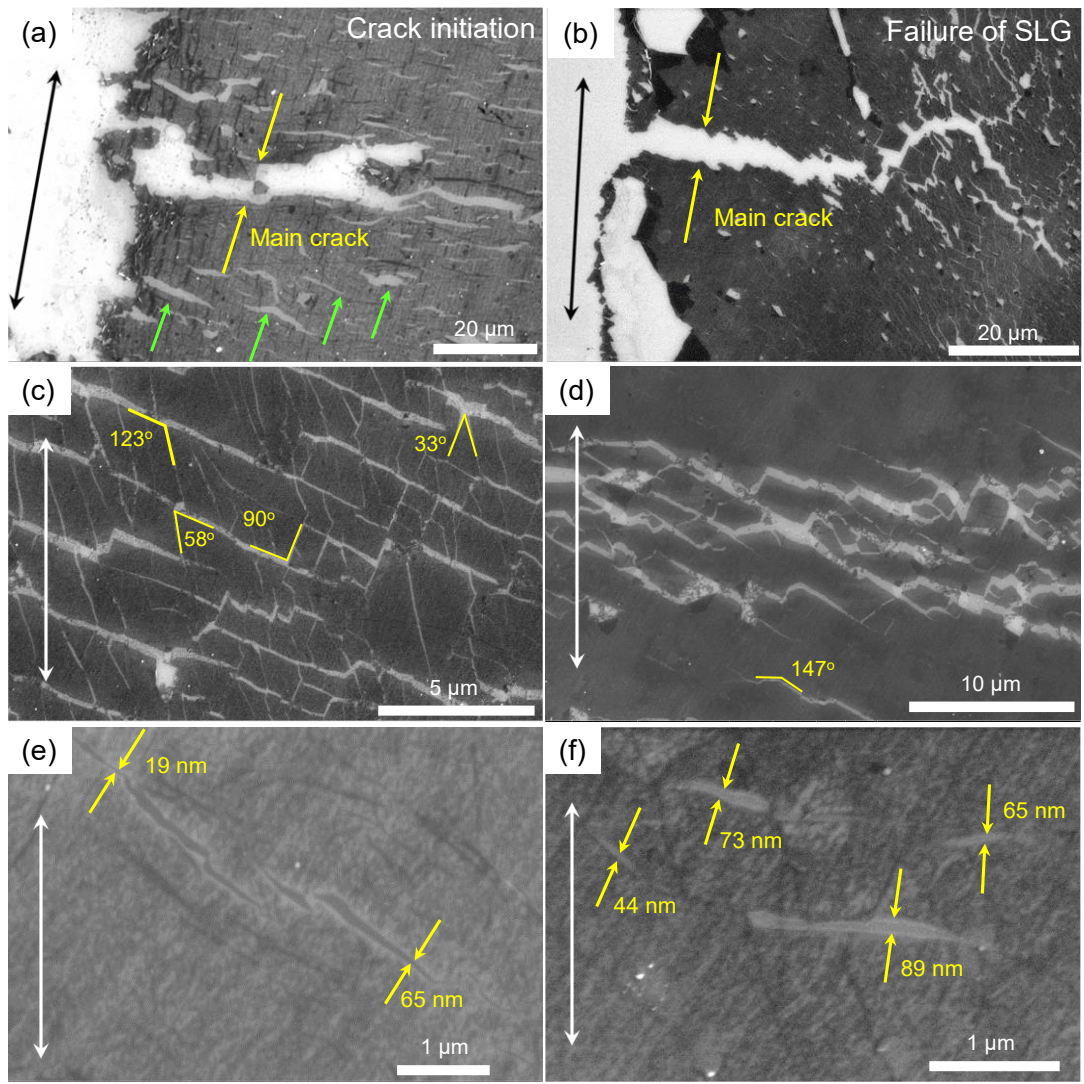

**Fig. S23: SEM Micrographs of cracks in SCG.** SEM image of SCG at (a) crack initiation point and (b) after failure of SCG. The black arrow defines the tensile loading direction. The tensile loading was stopped at 2% (crack initiation) and 4.5% (failure of SCG), and SCG-PC sample was transferred on SiO$_2$/Si substrate followed by removal of PC for SEM imaging. (c-d) The relative angle of crack edges shows the crack propagation is along the armchair or zigzag direction of SCG. (e-f) Different shapes and width of crack in SCG. The black/white arrow defines the tensile loading or elongation direction.

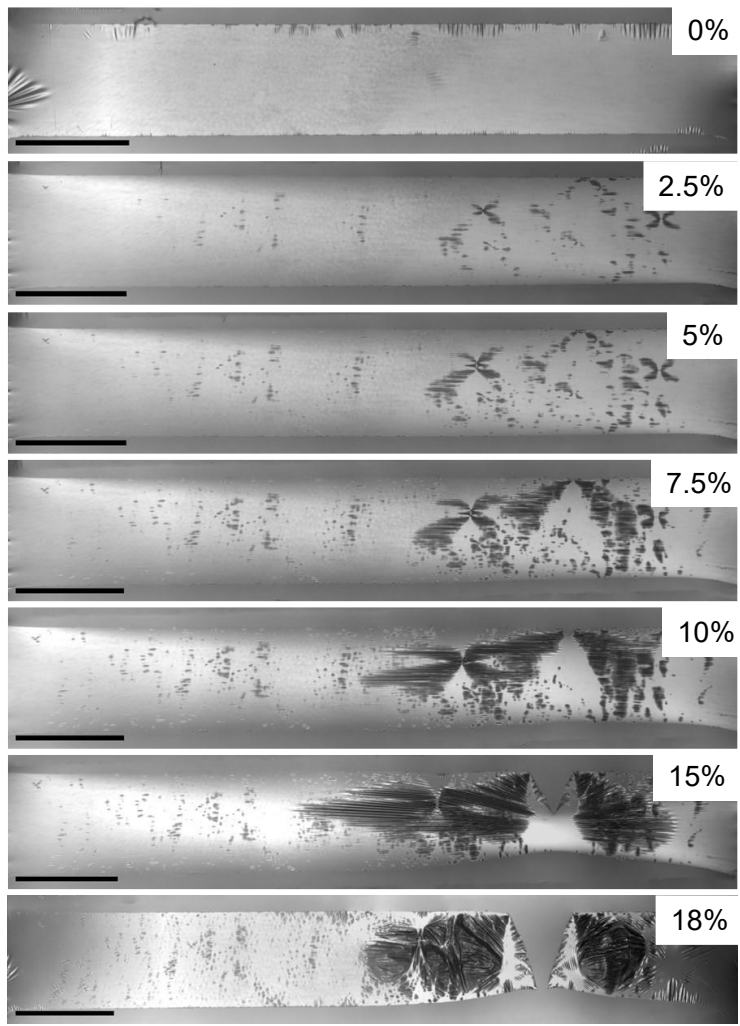

**Fig. S24: Evolution of 80 nm PC Films under Tensile Loading.** Monochromatic optical images showcasing the progressive changes in 80 nm PC films subjected to uniform tensile loading to the strain shown in the inset. Scale bar ~ 2 mm.

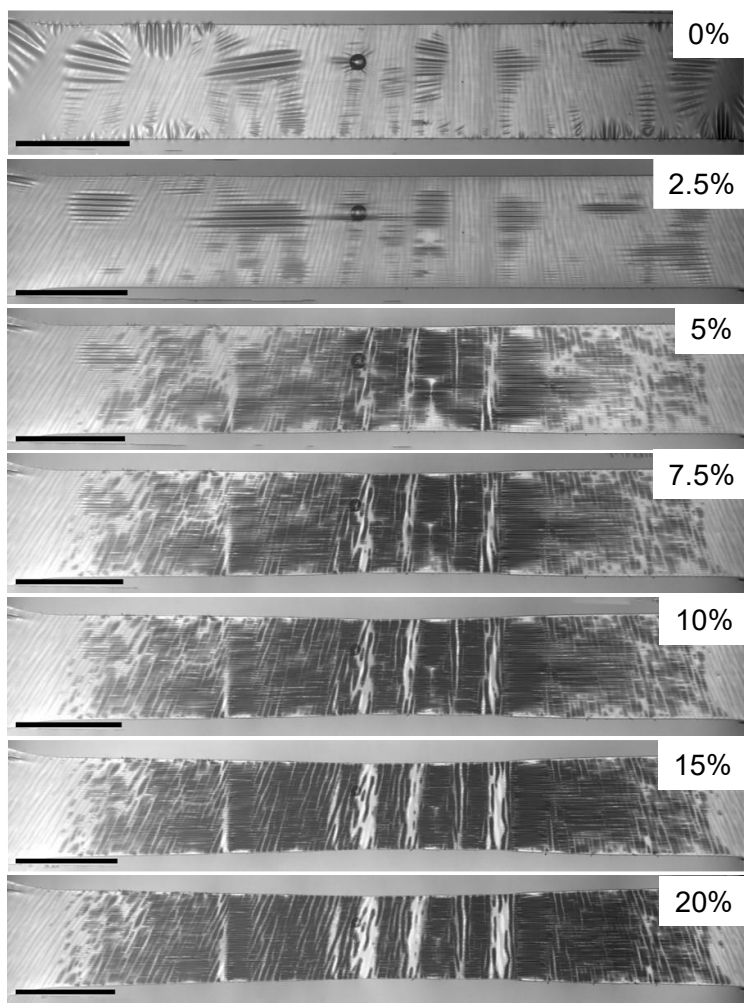

**Fig. S25: Evolution of 150 nm PC Films under Tensile Loading.** Monochromatic optical images showcasing the progressive changes in 150 nm PC films subjected to uniform tensile loading to the strain shown in the inset. Scale bar ~ 2 mm.

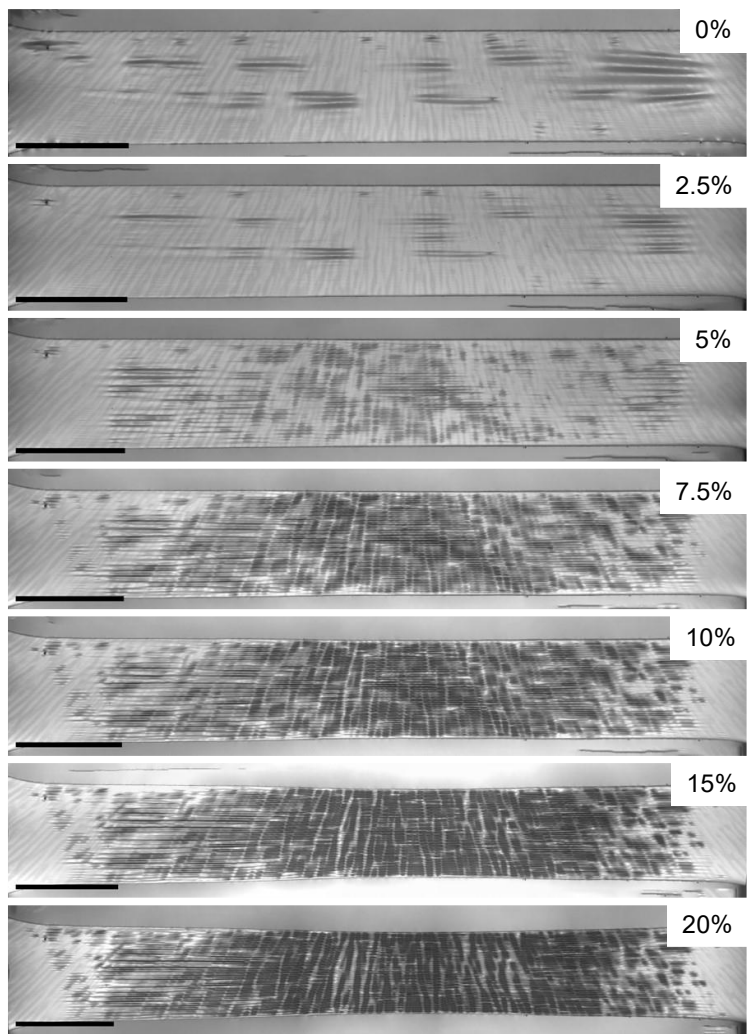

**Fig. S26: Evolution of 300 nm PC Films under Tensile Loading.** Monochromatic optical images showcasing the progressive changes in 300 nm PC films subjected to uniform tensile loading to the strain shown in the inset. Scale bar ~ 2 mm.

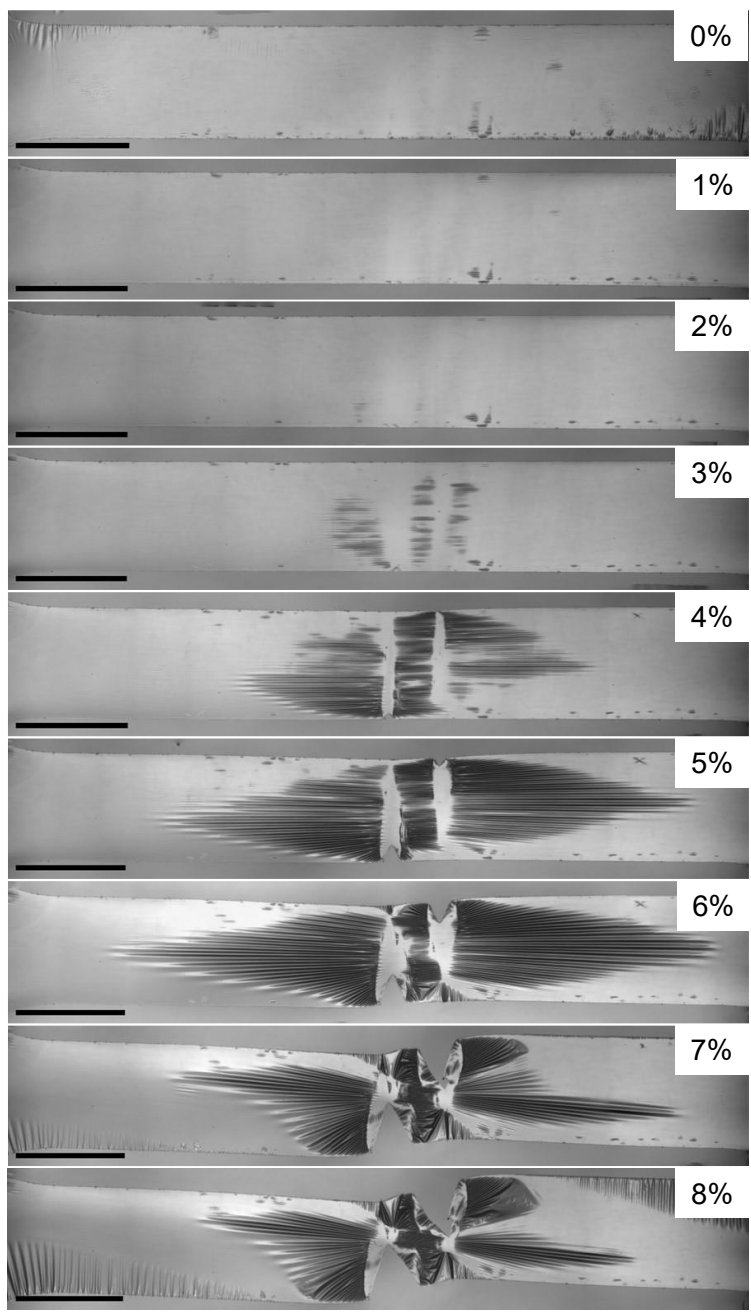

**Fig. S27: Evolution of 80 nm SCG-PC Films under Tensile Loading.** Monochromatic optical images showcasing the progressive changes in 80 nm SCG-PC films subjected to uniform tensile loading to the strain shown in the inset. Scale bar ~ 2 mm.

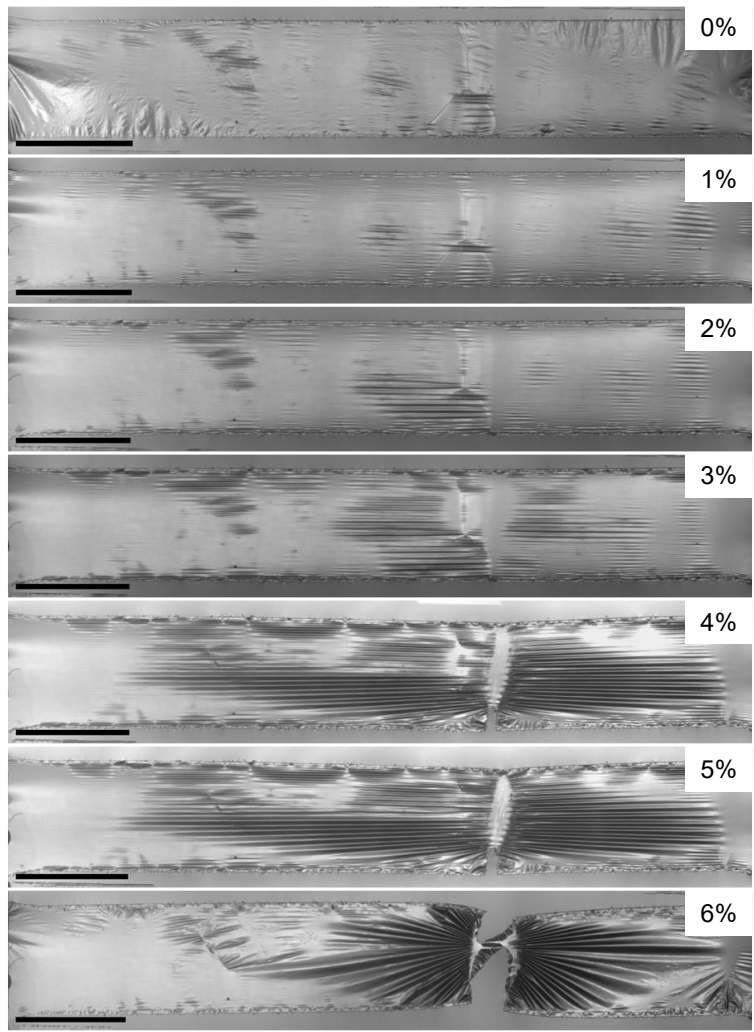

**Fig. S28: Evolution of 150 nm SCG-PC Films under Tensile Loading.** Monochromatic optical images showcasing the progressive changes in 150 nm SCG-PC films subjected to uniform tensile loading to the strain shown in the inset. Scale bar ~ 2 mm.

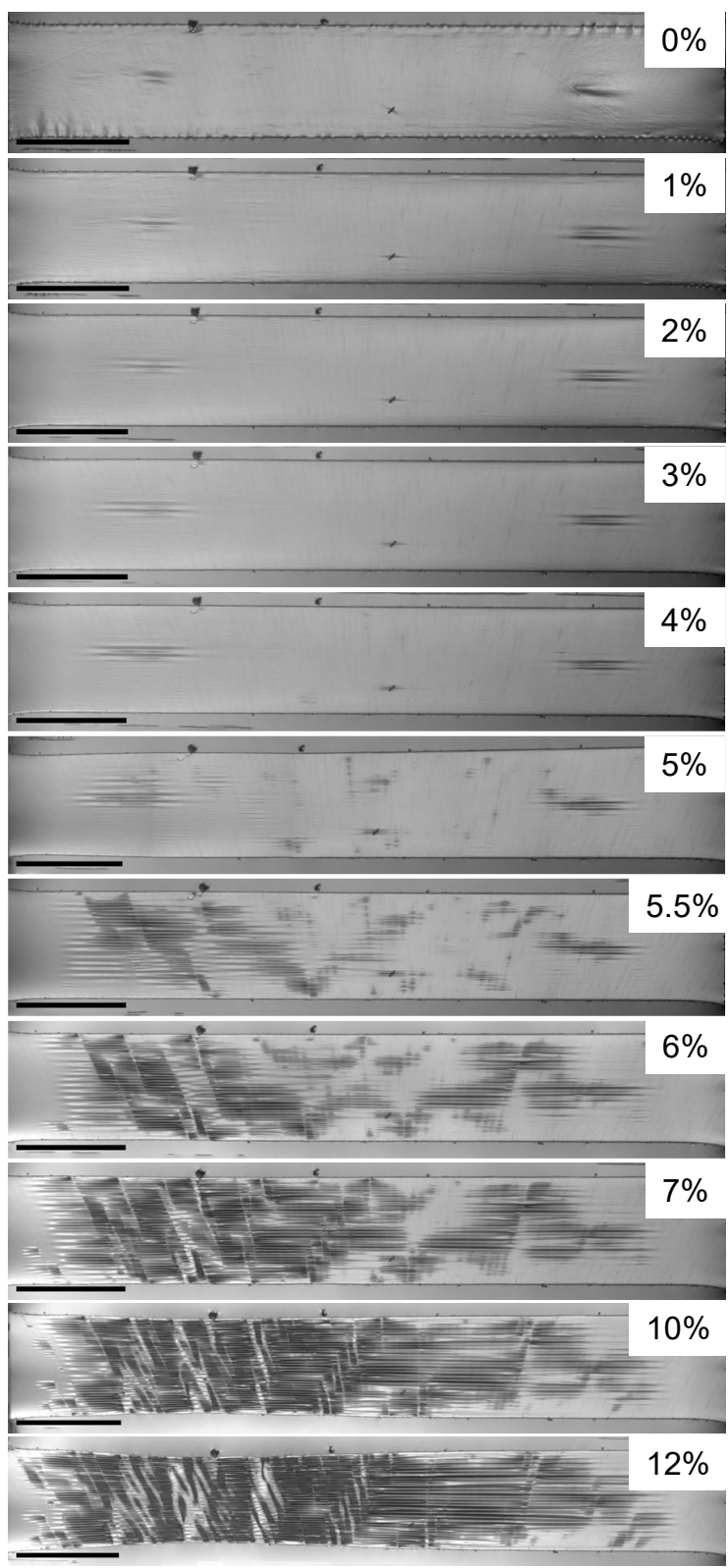

**Fig. S29: Evolution of 300 nm SCG-PC Films under Tensile Loading.** Monochromatic optical images showcasing the progressive changes in 300 nm SCG-PC films subjected to uniform tensile loading to the strain shown in the inset. Scale bar ~ 2 mm.

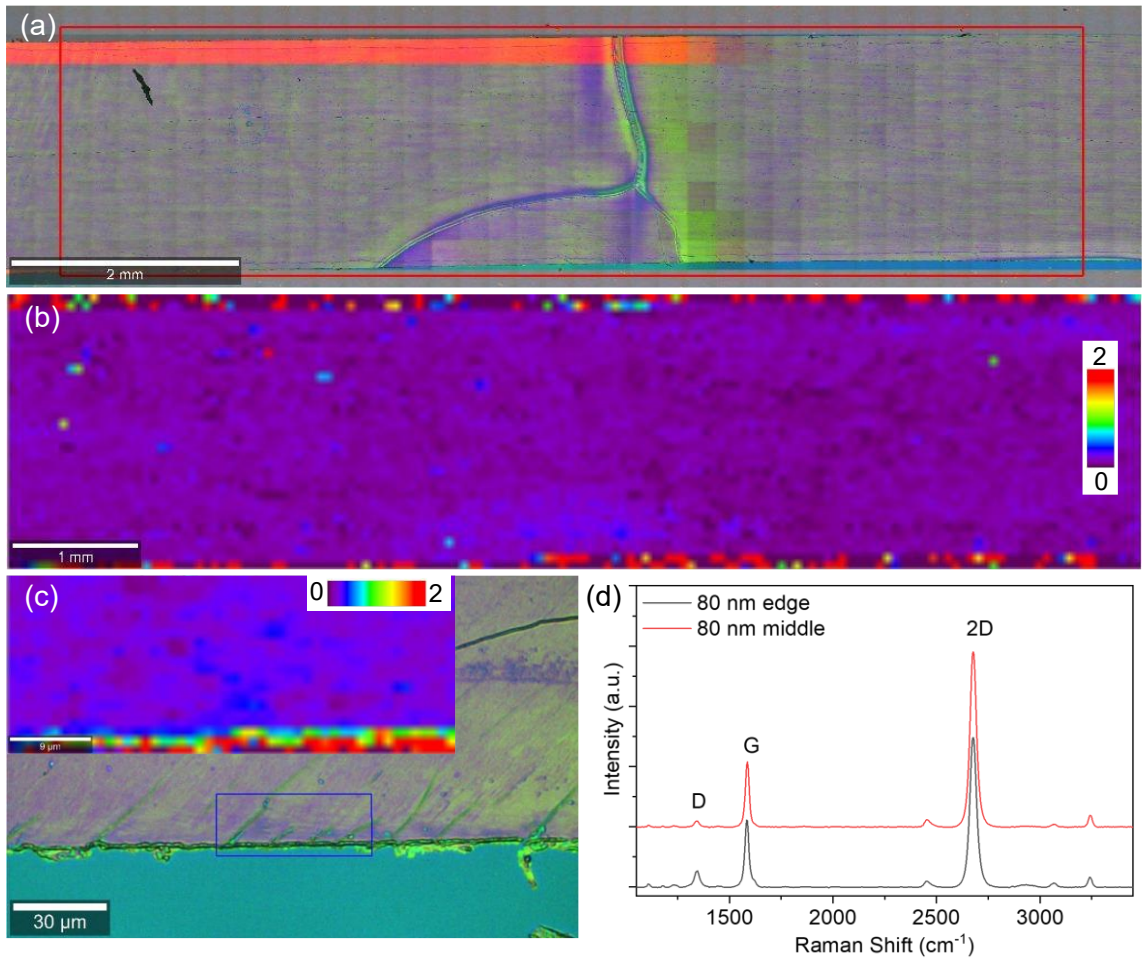

**Fig. S30: Large scale Raman mapping of 80 nm SCG-PC samples.** (a) Optical image and (b) $I_D/I_G$ map of the marked region in (a). (c) $I_D/I_G$ map at the edges of 80 nm SCG-PC samples shows the presence of the D peak at the edge region originated due to the undamaged or damaged edge in SCG. (d) comparison of Raman spectra at the edges and middle region of the SCG-PC sample.

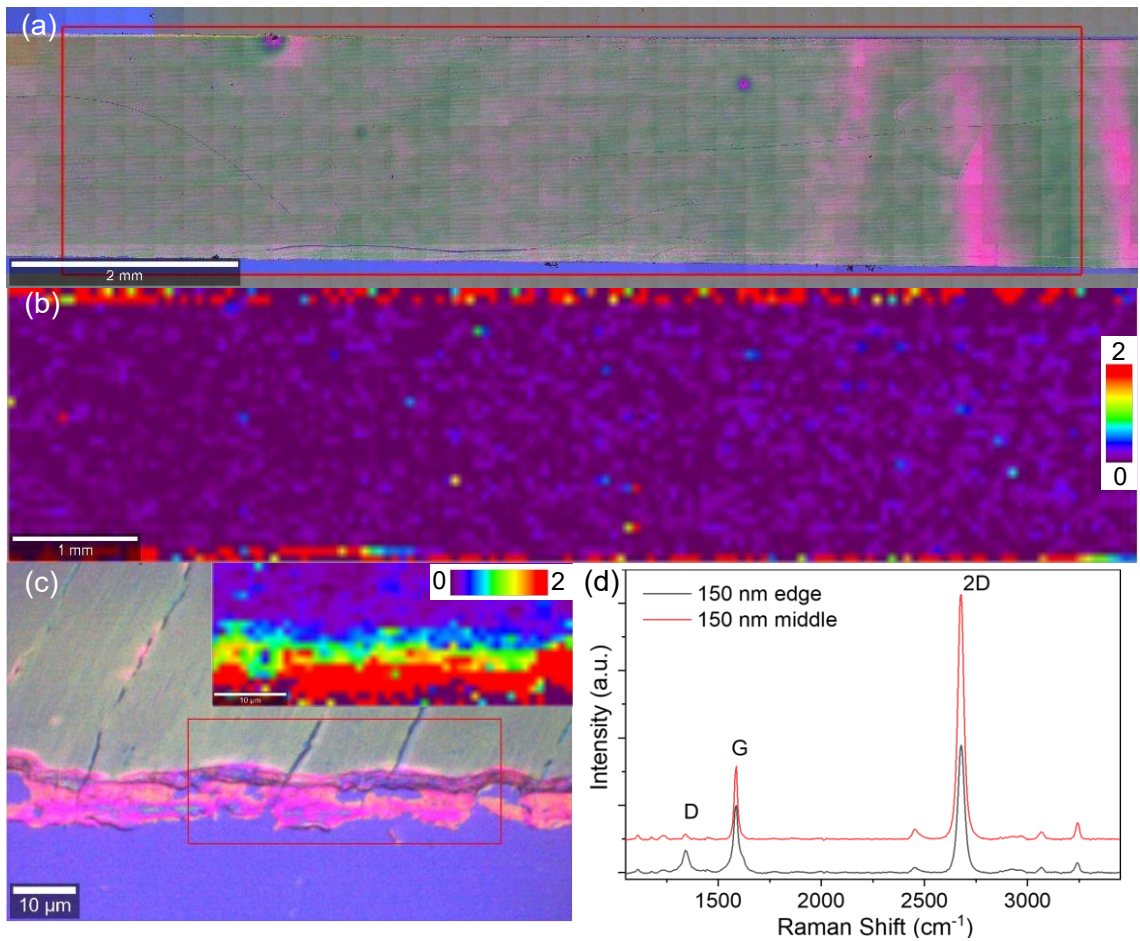

**Fig. S31: Large scale Raman mapping of 150 nm SCG-PC samples.** (a) Optical image and (b) $I_D/I_G$ map of the marked region in (a). (c) $I_D/I_G$ map at the edges of 150 nm SCG-PC samples shows the presence of D peak at the edge region originated due to the undamaged or damaged edge in SCG. (d) comparison of Raman spectra at the edges and middle region of the SCG-PC sample.

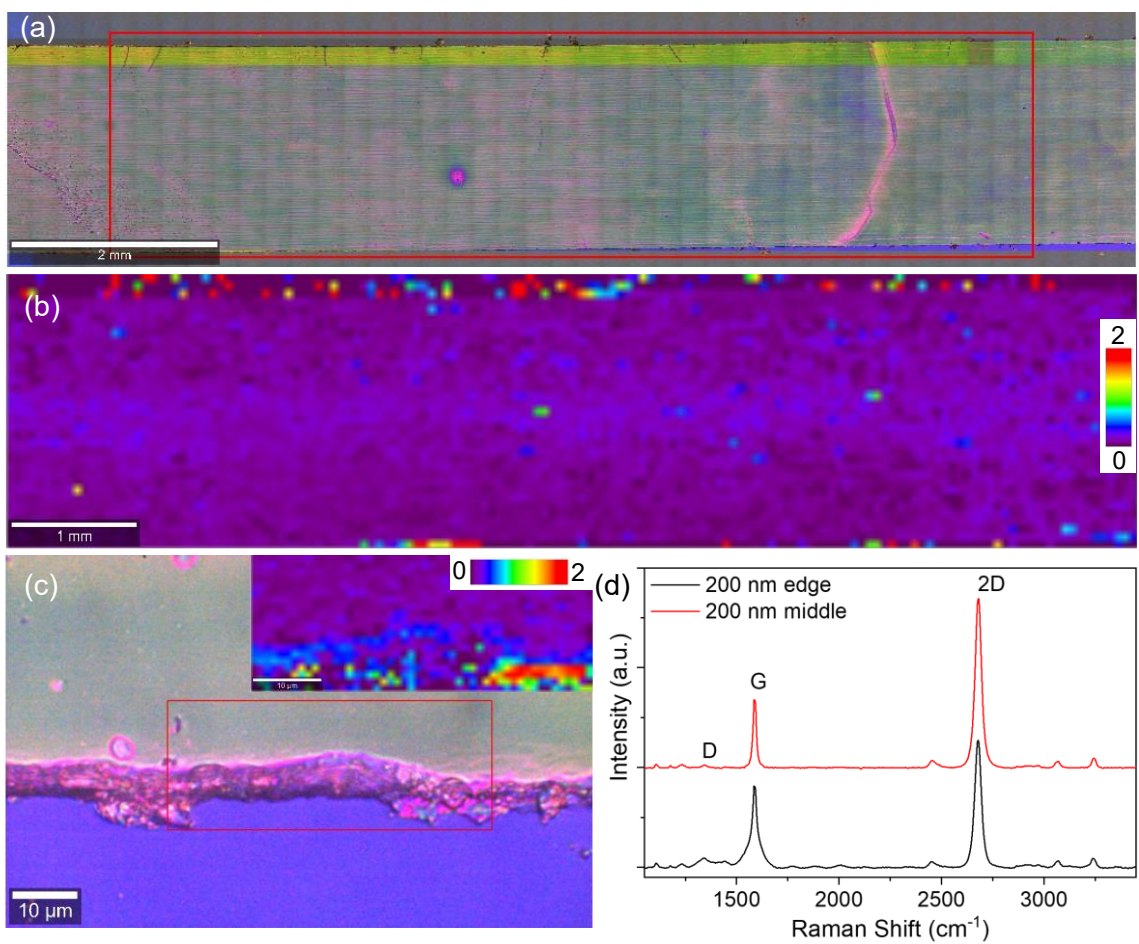

**Fig. S32: Large scale Raman mapping of 200 nm SCG-PC samples.** (a) Optical image and (b) $I_D/I_G$ map of the marked region in (a). (c) $I_D/I_G$ map at the edges of 200 nm SCG-PC samples shows the presence of D peak at the edge region originated due to the undamaged or damaged edge in SCG. (d) comparison of Raman spectra at the edges and middle region of the SCG-PC sample.

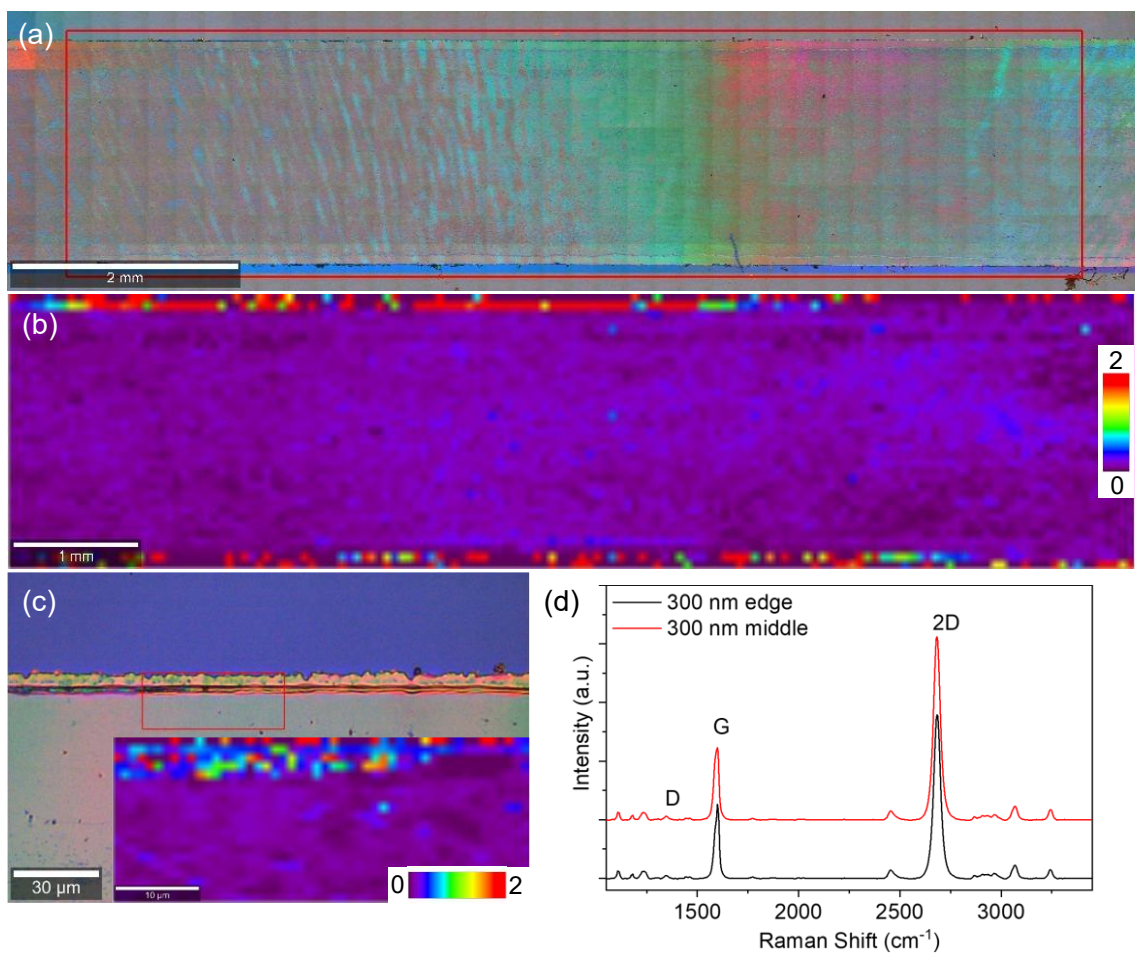

**Fig. S33: Large scale Raman mapping of 300 nm SCG-PC samples.** (a) Optical image and (b) $I_D/I_G$ map of the marked region in (a). (c) $I_D/I_G$ map at the edges of 300 nm SCG-PC samples shows the presence of D peak at the edge region originated due to the pure or damaged edge in SCG. (d) comparison of Raman spectra at the edges and middle region of the SCG-PC sample.

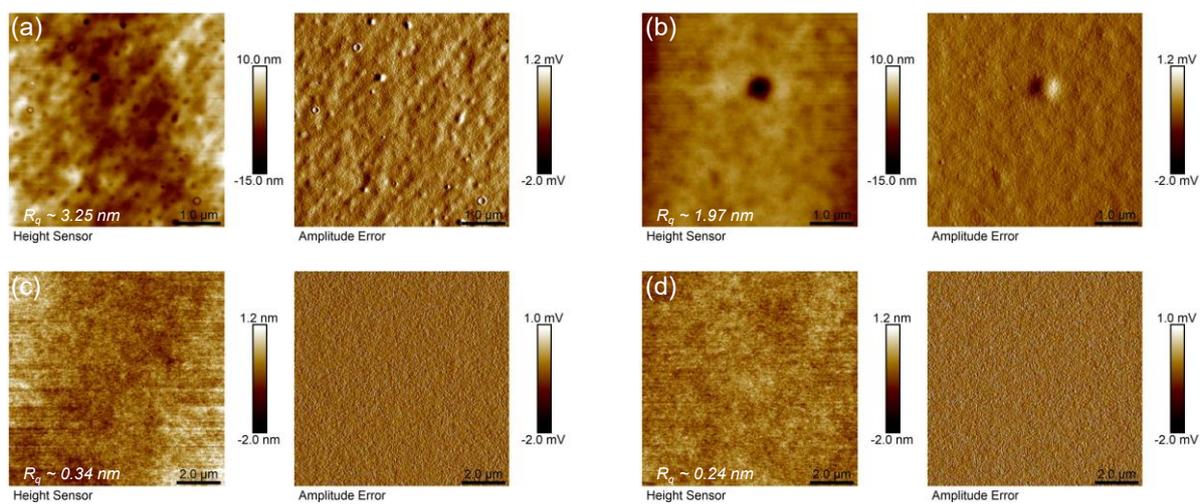

**Fig. S34:** AFM height and amplitude error image for SCG-PC samples with (a) 80 nm, (b) 150 nm, (c) 200 nm, and (d) 300 nm PC thickness.

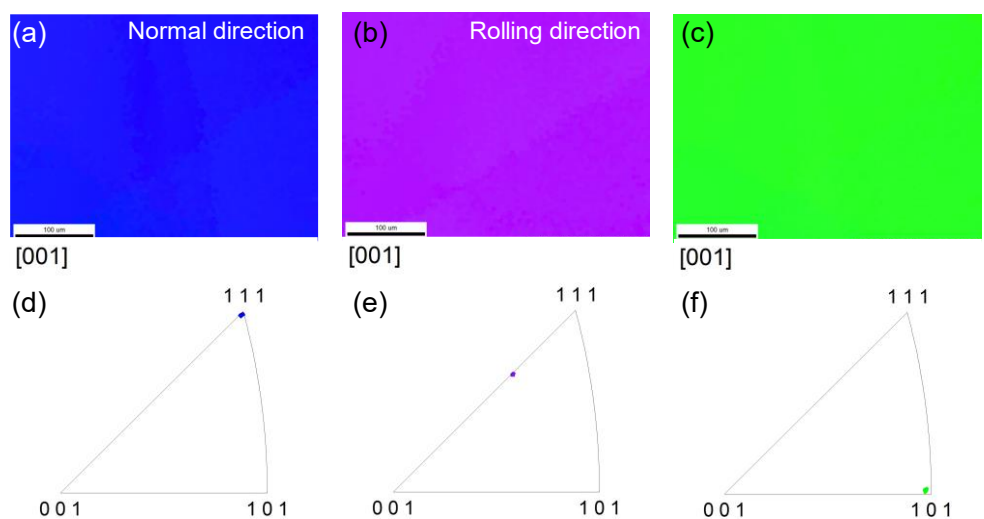

**Fig. S35: EBSD of Cu(111) foil.** (a-c) EBSD map data of the Cu(111) foils along z, y and x directions representing the (111), (112) and (110) direction of single crystal Cu(111) foil. (d-f) Inverse Pole Figure (IPF) plot of the Cu(111) foil along the z, y and x directions. The y and x directions denote the parallel and perpendicular directions with respect to the rolling marks.

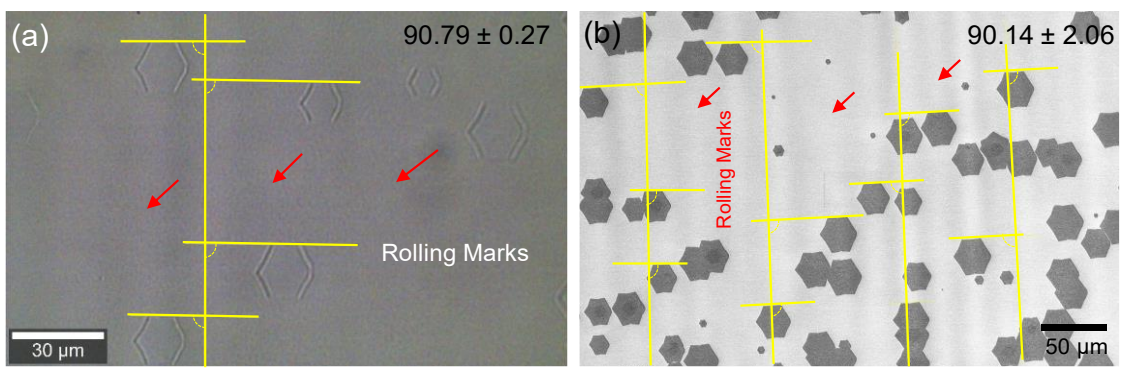

**Fig. S36:** (a) Optical image and (b) SEM micrograph of hexagonal SCG on Cu(111) showing the direction of the edges along the perpendicular direction of Cu rolling marks. The red arrows denote the rolling marks of Cu(111). The vertical lines (yellow) represent the Cu rolling marks and the horizontal lines (yellow) represent the graphene edges.

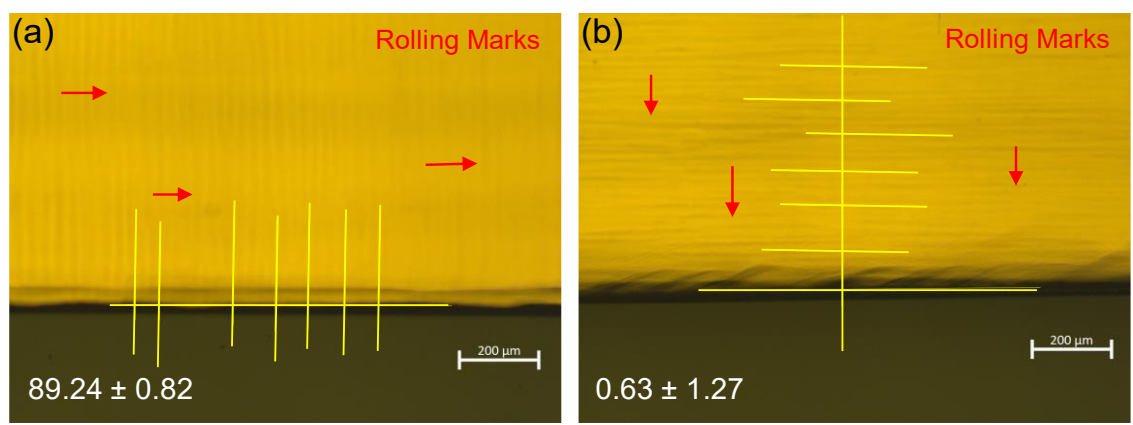

**Fig. S37:** Orientation of dogbone shaped SCG edges (after using mechanical cutter) with the (a) perpendicular and (b) parallel to the rolling marks of Cu(111). The red arrows indicated the Cu(111) rolling marks.

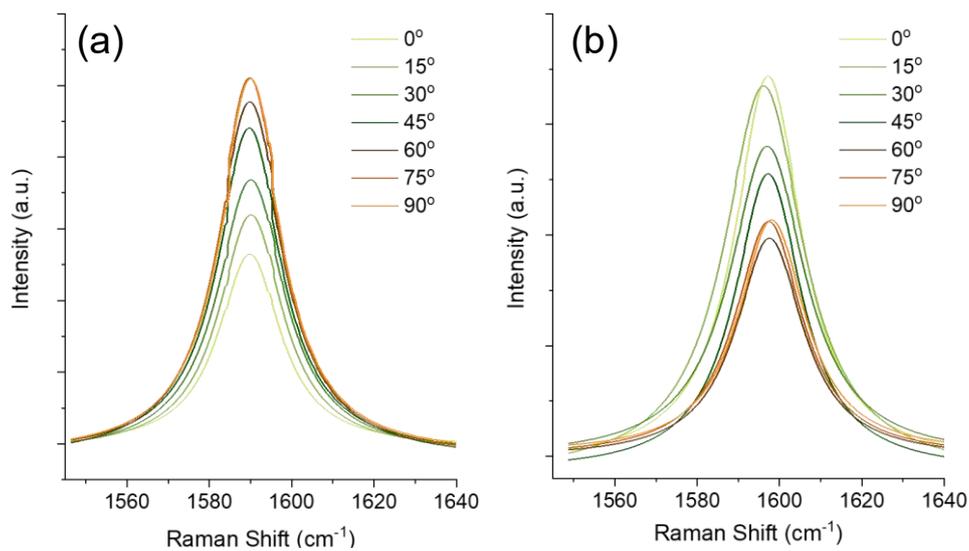

**Fig. S38:** Polarized Raman spectra of SLG at the (a) zigzag edge (perpendicular to rolling marks) and (b) armchair edge (parallel to the rolling marks) of Cu(111).

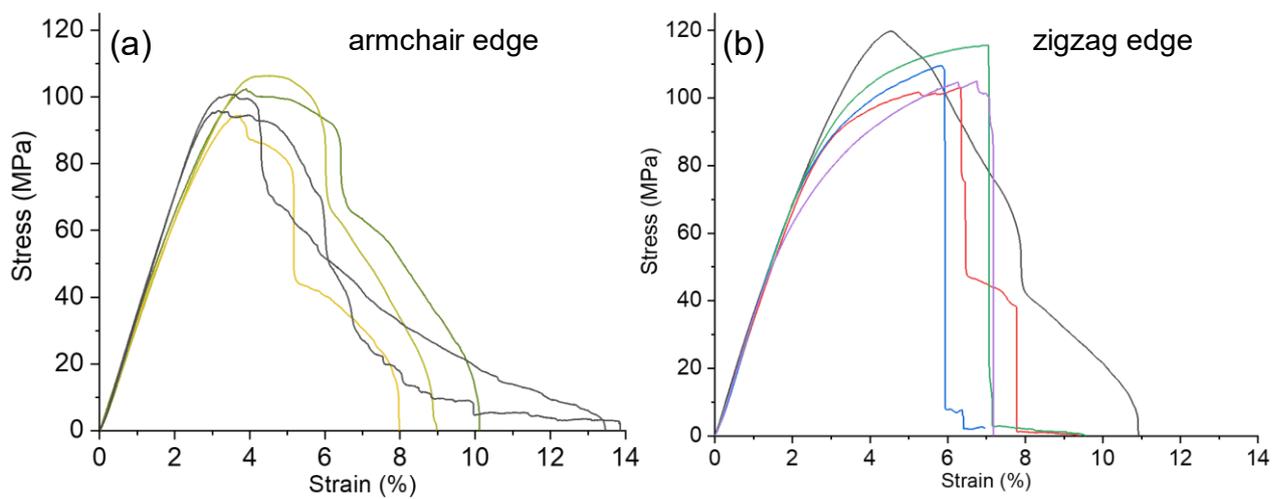

**Fig. S39:** Complete stress strain curve of (a) armchair and (b) zigzag 200 nm SCG-PC samples.

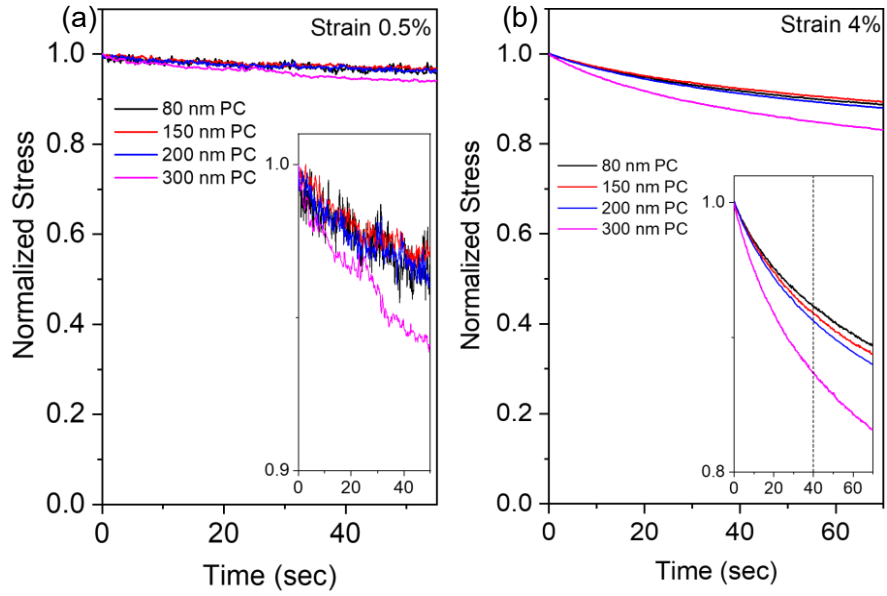

**Fig. S40:** Stress relaxation behavior of 80, 150, 200 and 300 nm PC film at (a) 0.5% and (b) 4% strain. The distinct strain relaxation of 300 nm is observed at 0.5% strain also. The inset shows the enlarged plot. A vertical dotted line was drawn at 40 sec in the inset of (b) to understand the stress relaxation. For 4% strain, at 40 sec, 300 nm PC film shows 13% reduction in stress, whereas the 80, 150, and 200 nm PC films show stress reduction of 7.6, 8.2, and 8.8%, respectively.

**Table S1:** The damage evolution model parameters for chiral, zigzag, and armchair SCGs.

|  | $E_G$ [GPa] | $\sigma_{uG}$ [GPa] | $\varepsilon_{uG}$ | $\varepsilon_{uf}$ | $n$ | $m$ | $U_1$ [GJ m$^{-3}$] | $U_2$ [GJ m$^{-3}$] |
|---|---|---|---|---|---|---|---|---|
| Chiral | 748.9 | 23.56 | 0.0453 | 0.0769 | 2.273 | 0.39 | 0.610 | 0.536 |
| Zigzag | 1109 | 27.40 | 0.0601 | 0.0779 | 0.698 | 0.42 | 1.183 | 0.343 |
| Armchair | 1009 | 20.21 | 0.0369 | 0.1010 | 1.187 | 0.67 | 0.480 | 0.776 |

**Table S2:** Thickness summary of PC films.

| Sample Name | Sample number | Thickness (nm) |
|---|---|---|
| **80 nm PC** | #1 | 80.3 |
| | #2 | 77.9 |
| | #3 | 81.1 |
| | #4 | 82.4 |
| | #5 | 80.2 |
| **Average** | | 80.4 |
| **150 nm PC** | #1 | 150.5 |
| | #2 | 153.7 |
| | #3 | 155.1 |
| | #4 | 152.6 |
| | #5 | 149.9 |
| **Average** | | 151.6 |
| **200 nm PC** | #1 | 229.4 |
| | #2 | 231.2 |
| | #3 | 227.1 |
| | #4 | 234.6 |
| | #5 | 229.9 |
| **Average** | | 228.8 |
| **300 nm PC** | #1 | 304.7 |
| | #2 | 302.5 |
| | #3 | 301.6 |
| | #4 | 307.3 |
| | #5 | 306.9 |
| **Average** | | 304.6 |

**Table S3:** Thickness summary of SCG-PC films.

| Sample Name | Sample number | Thickness (nm) |
|---|---|---|
| **80 nm SCG-PC** | #1 | 86.1 |
| | #2 | 84.7 |
| | #3 | 81.4 |
| | #4 | 84.5 |
| | #5 | 80.1 |
| **Average** | | 83.4 |
| **150 nm SCG-PC** | #1 | 151.1 |
| | #2 | 147.6 |
| | #3 | 156.3 |
| | #4 | 150.7 |
| | #5 | 150.2 |
| **Average** | | 151.2 |
| **200 nm SCG-PC** | #1 | 228.6 |
| | #2 | 219.9 |
| | #3 | 228.9 |
| | #4 | 221.5 |
| | #5 | 227.1 |
| **Average** | | 225.2 |
| **300 nm SCG-PC** | #1 | 304.1 |
| | #2 | 302.4 |
| | #3 | 301.5 |
| | #4 | 300.6 |
| | #5 | 307.3 |
| **Average** | | 303.3 |

**Table S4:** Mechanical properties of PC films of different thicknesses.

| Sample | Sample number | $E_{PC}$ (GPa) | $\sigma_{PC}$ (MPa) | Strain at max stress (%) | Failure Strain (%) |
|---|---|---|---|---|---|
| **80 nm PC** | #1 | 0.80 | 20.67 | 3.85 | 15.94 |
| | #2 | 0.82 | 20.60 | 5.26 | 23.82 |
| | #3 | 0.68 | 19.57 | 4.45 | 11.64 |
| | #4 | 0.72 | 20.87 | 4.52 | 25.03 |
| | #5 | 0.83 | 20.02 | 3.70 | 15.46 |
| **80 nm PC average** | | **0.77 ± 0.07** | **20.35 ± 0.54** | **4.36 ± 0.62** | **18.38 ± 5.78** |
| **150 nm PC** | #1 | 1.06 | 29.02 | 14.81 | 18.24 |
| | #2 | 1.02 | 27.76 | 8.20 | 16.79 |
| | #3 | 1.04 | 28.45 | 5.49 | 18.03 |
| | #4 | 1.09 | 29.71 | 6.08 | 26.17 |
| | #5 | 1.12 | 27.88 | 4.44 | 17.43 |
| **150 nm PC average** | | **1.07 ± 0.04** | **28.56 ± 0.81** | **7.80 ± 4.14** | **19.33 ± 3.86** |
| **200 nm PC** | #1 | 1.95 | 71.41 | 4.77 | 22.43 |
| | #2 | 2.03 | 70.22 | 4.44 | 24.89 |
| | #3 | 2.06 | 68.25 | 4.76 | 20.43 |
| | #4 | 2.02 | 70.34 | 4.37 | 16.84 |
| | #5 | 1.99 | 66.96 | 5.88 | 21.25 |
| **200 nm PC average** | | **2.01 ± 0.04** | **69.44 ± 1.79** | **4.84 ± 0.61** | **21.17 ± 2.95** |
| **300 nm PC** | #1 | 2.16 | 60.45 | 4.44 | 18.17 |
| | #2 | 2.19 | 63.30 | 4.41 | 18.62 |
| | #3 | 2.40 | 69.59 | 4.30 | 26.70 |
| | #4 | 2.12 | 65.96 | 4.55 | 33.86 |
| | #5 | 2.33 | 70.10 | 3.79 | 20.32 |
| **300 nm PC average** | | **2.24 ± 0.12** | **65.88 ± 4.11** | **4.30 ± 0.30** | **23.53 ± 6.71** |

**Table S5:** Mechanical properties of SCG-PC films of different thicknesses.

| Sample | Sample number | $E_{SCG-PC}$ (GPa) | $\sigma_{SCG-PC}$ (MPa) | Strain at max stress (%) | Failure Strain (%) |
|---|---|---|---|---|---|
| **80 nm SCG-PC** | #1 | 1.83 | 60.54 | 5.02 | 12.38 |
| | #2 | 2.59 | 67.68 | 4.61 | 8.63 |
| | #3 | 2.64 | 64.62 | 4.59 | 7.62 |
| | #4 | 2.14 | 63.54 | 4.46 | 7.82 |
| | #5 | 2.24 | 61.42 | 4.41 | 6.81 |
| **80 nm SCG-PC average** | | **2.29 ± 0.34** | **63.56 ± 2.82** | **4.62 ± 0.24** | **8.65 ± 2.18** |
| **150 nm SCG-PC** | #1 | 2.53 | 63.53 | 3.90 | 7.91 |
| | #2 | 2.44 | 59.52 | 4.66 | 9.33 |
| | #3 | 2.46 | 62.96 | 5.11 | 7.51 |
| | #4 | 2.55 | 56.62 | 5.02 | 8.49 |
| | #5 | 2.61 | 69.40 | 3.92 | 10.76 |
| **150 nm SCG-PC average** | | **2.52 ± 0.07** | **62.41 ± 4.80** | **4.52 ± 0.58** | **8.80 ± 1.29** |
| **200 nm SCG-PC** | #1 | 3.24 | 100.1 | 3.96 | 9.84 |
| | #2 | 2.95 | 99.76 | 4.66 | 6.25 |
| | #3 | 3.13 | 105.1 | 4.45 | 5.25 |
| | #4 | 3.41 | 112.8 | 4.50 | 6.93 |
| | #5 | 2.95 | 106.5 | 5.08 | 10.20 |
| **200 nm SCG-PC average** | | **3.14 ± 0.20** | **104.9 ± 5.35** | **4.53 ± 0.40** | **7.69 ± 2.21** |
| **300 nm SCG-PC** | #1 | 2.85 | 91.86 | 4.14 | 11.82 |
| | #2 | 2.90 | 94.43 | 4.33 | 9.86 |
| | #3 | 2.86 | 96.97 | 4.60 | 10.34 |
| | #4 | 2.98 | 98.44 | 4.51 | 8.58 |
| | #5 | 2.90 | 94.61 | 4.68 | 8.35 |
| **300 nm SCG-PC average** | | **2.90 ± 0.05** | **95.26 ± 2.54** | **4.45 ± 0.22** | **9.79 ± 1.41** |

Table S6: Mechanical properties of SCG extracted by comparing 200 nm SCG-PC and PC films.

| 200 nm SCG-PC | 200 nm PC | $E$ (GPa) | $\sigma$ (GPa) |
|---|---|---|---|
| #1 | #1 | 857.7 | 19.10 |
| | #2 | 804.7 | 19.89 |
| | #3 | 784.8 | 21.20 |
| | #4 | 811.3 | 19.81 |
| | #5 | 831.2 | 22.05 |
| **Average #1** | | **817.9 ± 27.7** | **20.41 ± 1.19** |
| #2 | #1 | 665.3 | 18.88 |
| | #2 | 612.3 | 19.67 |
| | #3 | 592.4 | 20.97 |
| | #4 | 618.9 | 19.59 |
| | #5 | 638.8 | 21.82 |
| **Average #2** | | **625.5 ± 27.7** | **20.19 ± 1.19** |
| #3 | #1 | 784.7 | 22.42 |
| | #2 | 731.7 | 23.21 |
| | #3 | 711.9 | 24.51 |
| | #4 | 738.3 | 23.13 |
| | #5 | 758.2 | 25.37 |
| **Average #3** | | **745.0 ± 27.7** | **23.73 ± 1.19** |
| #4 | #1 | 970.5 | 27.53 |
| | #2 | 917.5 | 28.32 |
| | #3 | 897.6 | 29.62 |
| | #4 | 924.1 | 28.24 |
| | #5 | 944.0 | 30.48 |
| **Average #4** | | **930.7 ± 27.7** | **28.84 ± 1.19** |
| #5 | #1 | 665.3 | 23.35 |
| | #2 | 612.3 | 24.14 |
| | #3 | 592.4 | 25.44 |
| | #4 | 618.9 | 24.06 |
| | #5 | 638.8 | 26.30 |
| **Average #5** | | **625.5 ± 27.7** | **24.66 ± 1.19** |
| **Total average** | | **748.9 ± 121.9** | **23.56 ± 3.42** |

**Table S7:** Mechanical properties of SCG extracted by comparing 80 nm SCG-PC and PC films.

| 80 nm SCG-PC | 80 nm PC | $E$ (GPa) | $\sigma$ (GPa) |
|---|---|---|---|
| #1 | #1 | 254.4 | 9.84 |
|  | #2 | 249.6 | 9.86 |
|  | #3 | 283.9 | 10.11 |
|  | #4 | 274.1 | 9.79 |
|  | #5 | 247.1 | 10.0 |
| **Average #1** |  | **261.8 ± 16.3** | **9.92 ± 0.13** |
| #2 | #1 | 441.7 | 11.60 |
|  | #2 | 436.8 | 11.62 |
|  | #3 | 471.1 | 11.87 |
|  | #4 | 461.3 | 11.55 |
|  | #5 | 434.3 | 11.76 |
| **Average #2** |  | **449.0 ± 16.3** | **11.68 ± 0.13** |
| #3 | #1 | 454.0 | 10.85 |
|  | #2 | 449.1 | 10.86 |
|  | #3 | 483.4 | 11.12 |
|  | #4 | 473.6 | 10.80 |
|  | #5 | 446.6 | 11.01 |
| **Average #3** |  | **461.3 ± 16.3** | **10.93 ± 0.13** |
| #4 | #1 | 330.8 | 10.58 |
|  | #2 | 325.9 | 10.60 |
|  | #3 | 360.3 | 10.85 |
|  | #4 | 350.5 | 10.53 |
|  | #5 | 323.5 | 10.74 |
| **Average #4** |  | **338.2 ± 16.3** | **10.66 ± 0.13** |
| #5 | #1 | 355.5 | 10.06 |
|  | #2 | 350.6 | 10.07 |
|  | #3 | 384.9 | 10.33 |
|  | #4 | 375.1 | 10.01 |
|  | #5 | 348.1 | 10.22 |
| **Average #5** |  | **362.8 ± 16.3** | **10.14 ± 0.13** |
| **Total average** |  | **374.7 ± 76.8** | **10.67 ± 0.65** |

**Table S8:** Mechanical properties of SCG extracted by comparing 150 nm SCG-PC and PC films.

| 150 nm SCG-PC | 150 nm PC | $E$ (GPa) | $\sigma$ (GPa) |
|---|---|---|---|
| #1 | #1 | 656.3 | 15.41 |
|  | #2 | 674.0 | 15.97 |
|  | #3 | 665.1 | 15.66 |
|  | #4 | 642.9 | 15.10 |
|  | #5 | 629.6 | 15.92 |
| **Average #1** |  | **653.6 ± 17.7** | **15.61 ± 0.36** |
| #2 | #1 | 616.1 | 13.62 |
|  | #2 | 633.9 | 14.18 |
|  | #3 | 625.0 | 13.88 |
|  | #4 | 602.8 | 13.32 |
|  | #5 | 589.5 | 14.13 |
| **Average #2** |  | **613.5 ± 17.7** | **13.83 ± 0.36** |
| #3 | #1 | 625.1 | 15.16 |
|  | #2 | 642.8 | 15.72 |
|  | #3 | 633.9 | 15.41 |
|  | #4 | 611.7 | 14.85 |
|  | #5 | 598.4 | 15.67 |
| **Average #3** |  | **622.4 ± 17.7** | **15.36 ± 0.36** |
| #4 | #1 | 665.2 | 12.33 |
|  | #2 | 683.0 | 12.89 |
|  | #3 | 674.1 | 12.58 |
|  | #4 | 651.8 | 12.02 |
|  | #5 | 638.5 | 12.84 |
| **Average #4** |  | **662.5 ± 17.7** | **12.53 ± 0.36** |
| #5 | #1 | 691.9 | 18.03 |
|  | #2 | 709.7 | 18.59 |
|  | #3 | 700.8 | 18.28 |
|  | #4 | 678.6 | 17.72 |
|  | #5 | 665.2 | 18.53 |
| **Average #5** |  | **689.2 ± 17.7** | **18.23 ± 0.36** |
| **Total average** |  | **648.2 ± 32.4** | **15.11 ± 1.98** |

**Table S9:** Mechanical properties of SCG extracted by comparing 300 nm SCG-PC and PC films.

| 300 nm SCG-PC | 300 nm PC | $E$ (GPa) | $\sigma$ (GPa) |
|---|---|---|---|
| #1 | #1 | 618.4 | 28.11 |
|    | #2 | 591.6 | 25.57 |
|    | #3 | 404.3 | 19.96 |
|    | #4 | 654.1 | 23.20 |
|    | #5 | 466.7 | 19.50 |
| **Average #1** | | **547.0 ± 106.5** | **23.27 ± 3.67** |
| #2 | #1 | 663.0 | 30.41 |
|    | #2 | 636.3 | 27.86 |
|    | #3 | 448.9 | 22.26 |
|    | #4 | 698.7 | 25.49 |
|    | #5 | 511.3 | 21.80 |
| **Average #2** | | **591.6 ± 106.5** | **25.56 ± 3.67** |
| #3 | #1 | 627.3 | 32.67 |
|    | #2 | 600.5 | 30.13 |
|    | #3 | 413.2 | 24.52 |
|    | #4 | 663.0 | 27.76 |
|    | #5 | 475.7 | 24.07 |
| **Average #3** | | **555.9 ± 106.5** | **27.83 ± 3.67** |
| #4 | #1 | 734.5 | 33.99 |
|    | #2 | 707.7 | 31.45 |
|    | #3 | 520.4 | 25.83 |
|    | #4 | 770.2 | 29.07 |
|    | #5 | 582.8 | 25.38 |
| **Average #4** | | **663.1 ± 106.5** | **29.14 ± 3.67** |
| #5 | #1 | 663.0 | 30.57 |
|    | #2 | 636.3 | 28.03 |
|    | #3 | 448.9 | 22.41 |
|    | #4 | 698.7 | 25.65 |
|    | #5 | 511.4 | 21.96 |
| **Average #5** | | **591.7 ± 106.5** | **25.72 ± 3.67** |
| **Total average** | | **589.9 ± 105.8** | **26.31 ± 3.94** |

Table S10: Mechanical properties of 200 nm SCG-PC films along armchair and zigzag direction.

| Sample | Sample number | $E_{Gr-PC}$ (GPa) | $\sigma_{Gr-PC}$ (MPa) | Strain at max stress (%) | Failure Strain (%) |
|---|---|---|---|---|---|
| ZZ | #1 | 3.76 | 119.7 | 4.55 | 10.90 |
|  | #2 | 3.66 | 103.6 | 6.34 | 7.78 |
|  | #3 | 3.70 | 109.6 | 5.84 | 5.94 |
|  | #4 | 3.64 | 115.7 | 7.05 | 7.14 |
|  | #5 | 3.63 | 104.6 | 6.27 | 7.17 |
|  | average | **3.68 ± 0.05** | **110.6 ± 7.0** | **6.01 ± 0.92** | **7.79 ± 1.86** |
| AC | #1 | 3.47 | 102.3 | 3.91 | 10.10 |
|  | #2 | 3.32 | 106.4 | 4.15 | 8.97 |
|  | #3 | 3.57 | 94.4 | 3.71 | 7.98 |
|  | #4 | 3.78 | 95.3 | 3.13 | 9.98 |
|  | #5 | 3.50 | 100.6 | 3.56 | 13.47 |
|  | average | **3.53 ± 0.17** | **99.8 ± 5.0** | **3.69 ± 0.38** | **10.10 ± 2.07** |

**Table S11:** Mechanical properties of SCG with tensile loading along zigzag direction.

| 200 nm SCG-PC | 200 nm PC | $E$ (GPa) | $\sigma$ (GPa) |
|---|---|---|---|
| #1 | #1 | 1203 | 32.10 |
| | #2 | 1150 | 32.89 |
| | #3 | 1130 | 34.20 |
| | #4 | 1156 | 32.81 |
| | #5 | 1176 | 35.05 |
| Average #1 | | **1163 ± 27.7** | **33.41 ± 1.19** |
| #2 | #1 | 1136 | 21.43 |
| | #2 | 1083 | 22.21 |
| | #3 | 1063 | 23.52 |
| | #4 | 1090 | 22.13 |
| | #5 | 1110 | 24.37 |
| Average #2 | | **1096 ± 27.8** | **22.73 ± 1.19** |
| #3 | #1 | 1163 | 25.41 |
| | #2 | 1110 | 26.19 |
| | #3 | 1090 | 27.50 |
| | #4 | 1117 | 26.11 |
| | #5 | 1136 | 28.35 |
| Average #3 | | **1123 ± 27.7** | **26.71 ± 1.19** |
| #4 | #1 | 1123 | 29.45 |
| | #2 | 1070 | 30.24 |
| | #3 | 1050 | 31.54 |
| | #4 | 1077 | 30.16 |
| | #5 | 1097 | 32.40 |
| Average #4 | | **1083 ± 27.8** | **30.76 ± 1.19** |
| #5 | #1 | 1116 | 22.09 |
| | #2 | 1063 | 22.88 |
| | #3 | 1044 | 24.18 |
| | #4 | 1070 | 22.80 |
| | #5 | 1090 | 25.04 |
| Average #5 | | **1077 ± 27.5** | **23.40 ± 1.19** |
| **Total average** | | **1109 ± 41.0** | **27.40 ± 4.36** |

**Table S12:** Mechanical properties of SCG with tensile loading along armchair direction.

| 200 nm SCG-PC | 200 nm PC | $E$ (GPa) | $\sigma$ (GPa) |
|---|---|---|---|
| #1 | #1 | 1010 | 20.56 |
|  | #2 | 957.3 | 21.35 |
|  | #3 | 937.4 | 22.66 |
|  | #4 | 963.9 | 21.27 |
|  | #5 | 983.8 | 23.51 |
| Average #1 |  | **970.5 ± 27.6** | **21.87 ± 1.19** |
| #2 | #1 | 910.7 | 23.28 |
|  | #2 | 857.8 | 24.07 |
|  | #3 | 837.9 | 25.38 |
|  | #4 | 864.4 | 23.99 |
|  | #5 | 884.3 | 26.23 |
| Average #2 |  | **871.0 ± 27.7** | **24.59 ± 1.19** |
| #3 | #1 | 1077 | 15.32 |
|  | #2 | 1024 | 16.11 |
|  | #3 | 1004 | 17.41 |
|  | #4 | 1030 | 16.03 |
|  | #5 | 1050 | 18.27 |
| Average #3 |  | **1037 ± 27.7** | **16.63 ± 1.19** |
| #4 | #1 | 1216 | 15.92 |
|  | #2 | 1163 | 16.71 |
|  | #3 | 1143 | 18.01 |
|  | #4 | 1170 | 16.63 |
|  | #5 | 1189 | 18.87 |
| Average #4 |  | **1176 ± 27.7** | **17.23 ± 1.19** |
| #5 | #1 | 1030 | 19.43 |
|  | #2 | 977.2 | 20.22 |
|  | #3 | 957.3 | 21.53 |
|  | #4 | 983.8 | 20.14 |
|  | #5 | 1004 | 22.38 |
| Average #5 |  | **990.5 ± 27.7** | **20.74 ± 1.19** |
| **Total average** |  | **1009 ± 105** | **20.21 ± 3.22** |

**Table S13:** Comparison of SCG's mechanical properties with reported theoretical and experimental results.

| Ref | Method | $E$ (TPa) | | $\sigma$ (GPa) | | $\varepsilon_f$ (%) | |
|---|---|---|---|---|---|---|---|
| | | ZZ | AC | ZZ | AC | ZZ | AC |
| This Work | Float on Water (CVD grown SCG) | 1.11 | 1.01 | 27.40 | 20.21 | 6.01 | 3.69 |
| **Theoretical Results** | | | | | | | |
| A(*34*) | DFPT | 1.05 | | 121 | 110 | 26.6 | 19.4 |
| B(*35*) | MD/LAMMPS/AIREBO | 1.01 | | 129 | 102 | 20.0 | 13.0 |
| C(*51*) | LAMMPS/AIREBO (100 K) | - | | 112 | 137 | 15.5 | 23.0 |
| | LAMMPS/AIREBO (300 K) | - | | 106 | 126 | 14.0 | 19.5 |
| D(*52*) | MD/LAMMPS/LCBOP | 0.939 | | - | | - | |
| E(*53*) | MD/LAMMPS/AIREBO | 1.31 | 1.19 | 162 | 137 | 14.0 | 10.0 |
| F(*54*) | MD/LAMMPS | 0.99 | | 121 | 99 | - | |
| **Experimental Results** | | | | | | | |
| G(*6*) | Nanoindentation (Exfoliated graphene) | 1.0 | | 130 | | 25 | |
| H(*7*) | Nanoindentation (CVD graphene) | 1.0 | | 90-99 | | - | |
| I(*13*) | PTP Tensile Test (microscale) | 0.9-1.0 | | 50-60 | | 5.8 | |
| J(*21*) | DMA Tensile Test (macroscale) | 0.74 | | 3.33 | | 0.6 | |

**DFPT**: Density Functional Perturbation Theory; **MD**: Molecular Dynamics; **LAMMPS**: Large-scale Atomic/Molecular Massively Parallel Simulator; **AIREBO**: Adaptive Intermolecular Reactive Empirical Bond Order; **LCBOP:** Long range Carbon Bond Order Potential.

**References:**

1. I. Toray Composite Materials America. (2023), vol. 2024.
2. F. Tanaka, Pioneering the carbon fiber frontier: A half-century of industry leadership and the road ahead. *Composites Part B: Engineering* **281**, (2024).
3. I. Toray Composite Materials America. (2024), vol. 2024.
4. H. Okuda, R. J. Young, F. Tanaka, J. Watanabe, T. Okabe, Tensile failure phenomena in carbon fibres. *Carbon* **107**, 474-481 (2016).
5. Y. Bai *et al.*, Carbon nanotube bundles with tensile strength over 80 GPa. *Nat Nanotechnol* **13**, 589-595 (2018).
6. C. Lee, X. Wei, J. W. Kysar, J. Hone, Measurement of the elastic properties and intrinsic strength of monolayer graphene. *Science* **321**, 385-388 (2008).
7. Gwan-Hyoung Lee *et al.*, High-Strength Chemical-Vapor – Deposited Graphene and Grain Boundaries. *Science* **340**, 1073-1076 (2013).
8. D. Krajcinovic, *Damage mechanics*. (Amsterdam, New York : Elsevier, Amsterdam, New York, 1996).
9. M. L. Kachanov, *Handbook of Elasticity Solutions*. B. Shafiro, I. Tsukrov, Eds., (Dordrecht : Springer Netherlands : Imprint: Springer, ed. 1st 2003.. 2003).
10. T. L. Anderson, *Fracture mechanics : fundamentals and applications*. (Boca Raton, FL : Taylor & Francis, Boca Raton, FL, ed. 3rd . 2005).
11. T. Zhu, J. Li, Ultra-strength materials. *Progress in Materials Science* **55**, 710-757 (2010).
12. R. B. Abernethy, *The New Weibull Handbook*. (R.B. Abernethy, 1996).
13. K. Cao *et al.*, Elastic straining of free-standing monolayer graphene. *Nat Commun* **11**, 284 (2020).
14. Xuesong Li *et al.*, Large-Area Synthesis of High-Quality and Uniform Graphene Films on Copper Foils. *Science* **324**, 1312-1314 (2009).
15. Yi Zhang, Luyao Zhang, C. Zhou, Review of Chemical Vapor Deposition of Graphene and Related Applications. *Acc Chem Research* **46**, 2329–2339 (2012).
16. W. C. Xuesong Li, Luigi Colombo, and Rodney S. Ruof, Evolution of Graphene Growth on Ni and Cu by Carbon Isotope Labeling. *Nano Lett.* **9**, 4268-4272 (2009).
17. M. Wang *et al.*, Single-crystal, large-area, fold-free monolayer graphene. *Nature* **596**, 519-524 (2021).
18. D. Luo *et al.*, Adlayer-Free Large-Area Single Crystal Graphene Grown on a Cu(111) Foil. *Adv Mater* **31**, e1903615 (2019).
19. D. Luo *et al.*, Folding and Fracture of Single-Crystal Graphene Grown on a Cu(111) Foil. *Adv Mater* **34**, e2110509 (2022).
20. S. Nie, J. M. Wofford, N. C. Bartelt, O. D. Dubon, K. F. McCarty, Origin of the mosaicity in graphene grown on Cu(111). *Physical Review B* **84**, (2011).
21. B. Wang *et al.*, Camphor-Enabled Transfer and Mechanical Testing of Centimeter-Scale Ultrathin Films. *Adv Mater* **30**, e1800888 (2018).
22. F. Liu *et al.*, Achievements and Challenges of Graphene Chemical Vapor Deposition Growth. *Advanced Functional Materials* **32**, (2022).
23. J. Zhang *et al.*, Clean Transfer of Large Graphene Single Crystals for High-Intactness Suspended Membranes and Liquid Cells. *Adv Mater* **29**, (2017).
24. J. H. Kim *et al.*, Tensile testing of ultra-thin films on water surface. *Nat Commun* **4**, 2520 (2013).
25. W. J. Choi, R. K. Bay, A. J. Crosby, Tensile Properties of Ultrathin Bisphenol-A Polycarbonate Films. *Macromolecules* **52**, 7489-7494 (2019).

26. A. Miyagawa, S. Korkiatithaweechai, S. Nobukawa, M. Yamaguchi, Mechanical and Optical Properties of Polycarbonate Containing p-Terphenyl. *Industrial & Engineering Chemistry Research* **52**, 5048-5053 (2013).
27. P. M. Toor, A review of some damage tolerance design approaches for aircraft structures. *Engineering Fracture Mechanics* **5**, 837-880 (1973).
28. R. O. Ritchie, The conflicts between strength and toughness. *Nat Mater* **10**, 817-822 (2011).
29. E. Cadelano, P. L. Palla, S. Giordano, L. Colombo, Nonlinear elasticity of monolayer graphene. *Phys Rev Lett* **102**, 235502 (2009).
30. X. Wei, B. Fragneaud, C. A. Marianetti, J. W. Kysar, Nonlinear elastic behavior of graphene:Ab initiocalculations to continuum description. *Physical Review B* **80**, (2009).
31. Y. Wei, R. Yang, Nanomechanics of graphene. *Natl Sci Rev* **6**, 324-348 (2019).
32. Chunxiao Cong, Ting Yu, H. Wang, Raman Study on the G Mode of Graphene for Determination of Edge Orientation. *ACS Nano* **4**, 3175–3180 (2010).
33. Q. Yu *et al.*, Control and characterization of individual grains and grain boundaries in graphene grown by chemical vapour deposition. *Nat Mater* **10**, 443-449 (2011).
34. F. Liu, P. Ming, J. Li, Ab initiocalculation of ideal strength and phonon instability of graphene under tension. *Physical Review B* **76**, (2007).
35. K. M. H. Zhao, N. R. Aluru, Size and Chirality Dependent Elastic Properties of Graphene Nanoribbons under Uniaxial Tension. *Nano Lett.* **9**, 3012-3015 (2009).
36. Traian Dumitrica, Ming Hua, B. I. Yakobson, Symmetry-, time-, and temperature-dependent strength of carbon nanotubes. *Proceedings of the National Academy of Sciences* **103**, 6105–6109 (2006).
37. N. M. Pugno, R. S. Ruoff, Nanoscale Weibull statistics. *J. Appl. Phys.* **99**, 024301 (2006).
38. N. M. Pugno, R. S. Ruoff, Quantized fracture mechanics. *Philosophical Magazine* **84**, 2829-2845 (2004).
39. A. Carpinteri, N. Pugno, Fracture instability and limit strength condition in structures with re-entrant corners. *Engineering Fracture Mechanics* **72**, 1254-1267 (2005).
40. S. Jaddi *et al.*, Definitive engineering strength and fracture toughness of graphene through on-chip nanomechanics. *Nat Commun* **15**, 5863 (2024).
41. T.-M. C. J. W. Ju, Effective Elastic Moduli of Two-Dimensional Brittle Solids With Interacting Microcracks, Part I Basic Formulations. *J. Appl. Mech.* **61**, 349-357 (1994).
42. K. F. W.A. Curtin, Microcrack toughening? *Acta Metallurgica* **38**, 2051-2058 (1990).
43. J. W. Hutchinson, Crack tip shielding by micro-cracking in brittle solids. *Acta Metallurgica* **35**, 1605-1619 (1986).
44. R. K. Bay, S. Shimomura, Y. Liu, M. Ilton, A. J. Crosby, Confinement Effect on Strain Localizations in Glassy Polymer Films. *Macromolecules* **51**, 3647-3653 (2018).
45. A. Alesadi, W. Xia, Understanding the Role of Cohesive Interaction in Mechanical Behavior of a Glassy Polymer. *Macromolecules* **53**, 2754-2763 (2020).
46. R. Song *et al.*, Unveiling the Stress–Strain Behavior of Conjugated Polymer Thin Films for Stretchable Device Applications. *Macromolecules* **53**, 1988-1997 (2020).
47. L. A. Galuska *et al.*, SMART transfer method to directly compare the mechanical response of water-supported and free-standing ultrathin polymeric films. *Nat Commun* **12**, 2347 (2021).
48. R. K. n. Bay *et al.*, Decoupling the Impact of Entanglements and Mobility on the Failure Properties of Ultrathin Polymer Films. *Macromolecules* **55**, 8505-8513 (2022).
49. G. Wang *et al.*, Mechanical Size Effect of Freestanding Nanoconfined Polymer Films. *Macromolecules* **55**, 1248-1259 (2022).

50. J. E. Mark, *Physical Properties of Polymers Handbook*. (New York, NY: Springer-Verlag, New York, NY, ed. 2. Aufl., 2007).
51. Y. I. Jhon, Y. M. Jhon, G. Y. Yeom, M. S. Jhon, Orientation dependence of the fracture behavior of graphene. *Carbon* **66**, 619-628 (2014).
52. S. Thomas, K. M. Ajith, S. U. Lee, M. C. Valsakumar, Assessment of the mechanical properties of monolayer graphene using the energy and strain-fluctuation methods. *RSC Adv* **8**, 27283-27292 (2018).
53. S. Thamaraikannan, M. R. Sunny, S. C. Pradhan, Chirality dependent mechanical properties of carbon nano-structures. *Materials Research Express* **6**, (2019).
54. B. Mortazavi *et al.*, First-Principles Multiscale Modeling of Mechanical Properties in Graphene/Borophene Heterostructures Empowered by Machine-Learning Interatomic Potentials. *Adv Mater* **33**, e2102807 (2021).

**Acknowledgements:** This work was supported by the Institute for Basic Science (IBS-R019-D1). We thank J. H. Lee of the UNIST Center Research Facilities for the TEM imaging. We thank Han Gi Chae and Fumihiko Tanaka for discussions about strength/modulus of carbon fibers. We thank Ben Cunning who refined an initial draft [Claude/ChatGPT].